\documentclass[12pt]{article} 
\usepackage{theorem}

\newtheorem{theorem}{Theorem}[section]

\newtheorem{lemma}[theorem]{Lemma}

\def\indirizzo#1{{}}
\def\email#1{{}}
\def\qed{\raise1pt\hbox{\vrule height5pt width5pt depth0pt}}
\def\qedd{{\qed}}
\def\brm#1\erm{\vskip.8em{\bf\0Remark.}\ #1\vskip.8em}
\def\bpr{{\it\0Proof.}\ }
\def\epr{\vskip.8em}
\def\bac#1\eac{\vskip.8em{\bf\0Acknowledgements}\ #1\vskip.8em}

\usepackage{amsmath}
\usepackage{amssymb}
\usepackage{amsfonts}
\usepackage{graphicx}

\numberwithin{equation}{section}

\newcount\driver
\newcount\bozza

\def\a{\alpha}        \def\b{\beta}         \def\d{\delta}     
\def\e{\varepsilon}   \def\z{\zeta}         \def\h{\eta}     
        \def\l{\lambda}       \def\m{\mu}           
\def\n{\nu}           \def\p{\pi}           \def\r{\rho}
\def\s{\sigma}        \def\t{\tau}                
          \def\ps{\psi}              
    \def\g{\gamma}        \def\G{\Gamma}        
\def\D{\Delta}        \def\L{\Lambda}       
           \def\Si{\Sigma}                
\def\th{\vartheta}    \def\O{\Omega}        
\def\x{\xi}           \def\f{\varphi}       \def\o{\omega}

\def\PP{{\mathcal P}}\def\EE{{\mathcal E}}\def\MM{{\mathcal M}}\def\VV{{\CMcal V}}
\def\CC{{\mathcal C}}\def\FF{{\mathcal F}}\def\WW{{\mathcal W}}
\def\TT{{\mathcal T}}\def\NN{{\mathcal N}}\def\BB{{\mathcal B}}
\def\RR{{\mathcal R}}\def\LL{{\mathcal L}}
\def\SS{{\mathcal S}}
\def\KK{{\mathcal K}}\def\UU{{\mathcal U}}

\def\hK{{\widehat  K}}

\def\hA{{\widehat  A}}
\def\hD{{\widehat \D}}

\def\hGG{{\widehat  \G}}

\def\mce{{\widehat e}}

\def\RRR{\mathbb{R}}

\def\WWW{\mathbb{W}}
\def\CCC{\mathbb{C}}
\def\EEE{\mathbb{E}}
\def\ZZZ{\mathbb{Z}}
\def\UU{\mathcal{U}}  
\def\VV{\mathcal{V}} 

\let\dpr=\partial                  \let\bs=\backslash           
\let\io=\infty                     \let\==\equiv

\def\lft{\left}                \def\rgt{\right}

       \def\tilde#1{{\widetilde #1}}

\def\cfr{\hbox{\it cfr.\ }} 
\def\ie{\hbox{\it i.e.\ }}

\def\sms{{\SS\hskip-.6em /}}

\def\wh#1{\widehat{#1}}
\def\hat#1{\wh{#1}}
\def\bar#1{\overline {#1}}

\def\lb#1{\label{#1}}

\def\pref#1{(\ref{#1})}

\def\*{{\hfill\break\null\hfill\break}}
\let\0=\noindent

\def\be{\begin{equation}}    \def\ee{\end{equation}}
\def\bea{\begin{eqnarray}}   \def\eea{\end{eqnarray}}
\def\bean{\begin{eqnarray*}} \def\eean{\end{eqnarray*}}
\def\bfr{\begin{flushright}} \def\efr{\end{flushright}}
\def\bc{\begin{center}}      \def\ec{\end{center}}
\def\bal#1\eal{\begin{align}#1\end{align}}
\def\ba#1{\begin{array}{#1}} \def\ea{\end{array}}
\def\bd{\begin{description}} \def\ed{\end{description}}
\def\bv{\begin{verbatim}}    \def\ev{\end{verbatim}}

\def\ins#1#2#3{\vbox to0pt{\kern-#2 \hbox{\kern#1 #3}\vss}\nointerlineskip}

\newdimen\xshift \newdimen\xwidth \newdimen\yshift
\newcount\griglia

\def\insertplot#1#2#3#4#5#6{%
\xwidth=#1pt \xshift=\hsize \advance\xshift by-\xwidth \divide\xshift by 2%
\begin{figure}[ht]
\vspace{#2pt}
\hspace{\xshift}
\begin{minipage}{#1pt}
#3
\ifnum\driver=1 \griglia=#6
\includegraphics{#4.ps}\fi%
\ifnum\driver=2 \fi
\end{minipage}
\caption{#5}
\end{figure}
}

\begin{document}
\title
{Kosterlitz-Thouless Transition Line for the\\ 
Two Dimensional  Coulomb Gas}
\author{Pierluigi Falco}
\indirizzo{School of Mathematics\\
Institute for Advanced Study, Princeton, New Jersey 08540}
\email{falco@math.ias.edu}
\maketitle
\begin{abstract}
With a rigorous  renormalization group approach, 
we  study the pressure of the two dimensional Coulomb Gas 
along a small piece of the Kosterlitz-Thouless transition line, 
i.e. the boundary of the dipole region in the activity-temperature
phase-space.  
\end{abstract}
\setcounter{tocdepth}{2}
\tableofcontents
\section{Introduction}
Occasionally in Statistical Mechanics the study of 
a special model inspired
major advances of the general theory.
This was the case,  for example, 
of the exact solutions of the two dimensional 
Ising and  Six-Vertex models  
in the ambit  of the theory of universality;   
and was certainly the case of the Renormalization Group (RG) 
analysis of two dimensional Coulomb Gas
as prototype of the {\it Kosterlitz-Thouless} (KT) 
transition for systems with long range 
interactions.

Two dimensional Coulomb Gas is the statistical system 
of point particles on a plane,
carrying a charge $\pm 1$, 
and interacting through the two-dimensional electrostatic potential
that, for large distances, is  
$$
V(x-y)\sim -{1\over 2\p} \ln |x-y|\;.
$$
An ideal realization of the model is a system of 
infinite, parallel, uniformly charged wires in thermal equilibrium; 
historically it was proposed as a model     
of strongly magnetized plasma (see introduction of
\cite{[DL]} and references therein). Very soon, anyways, 
the Coulomb Gas  acquired a
great theoretical importance,  for Berezinskii, \cite{[Be]}, and 
Kosterlitz and Thouless, \cite{[KT]},
found in it the solution of a  puzzling dichotomy  
in the theory of the two dimensional {\it XY model}:
the {\it spin-wave} approximation  
(quite reliable for low temperatures)
predicted, in agreement with the general Mermin-Wagner argument, 
absence of order and power law fall-off of the spin correlation;  on the
other hand, high temperature expansion clearly demonstrated exponential decay
of the correlations. The two scenarios were merged together 
by the fundamental observation
that the former picture did not take into account the spin
configurations with {\it vortex excitations}, which
interacted through the same logarithmic potential written above.

Using RG ideas,  Kosterlitz and Thouless, \cite{[KT]},
\cite{[Ko]}, were able to obtain the  diagram of phases 
of the Coulomb Gas (and so of the XY model) depicted in 
Fig.\ref{fig1}, where $z$ is the activity  
and $\b$ the inverse temperature.
\insertplot{420}{150}
{\ins{312pt}{77pt}{$\b$}
\ins{190pt}{78pt}{$8\p$}
\ins{110pt}{78pt}{$0$}
\ins{110pt}{150pt}{$z$}
}%
{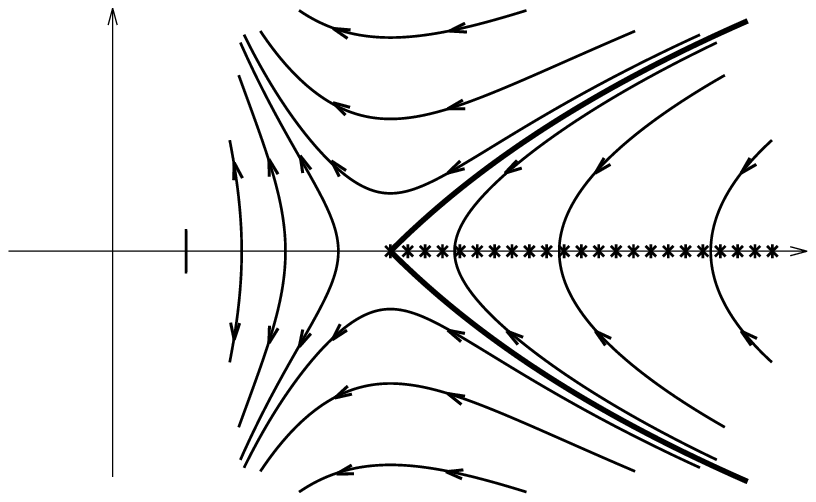}{\lb{fig1}Diagram of phases: the separatrix is the KT line.}{0}  
At high temperatures, the gas is in the {\it plasma phase} 
- or {\it Debye screening phase}: 
the configurations of significant probability are those 
in which the long-range electrostatic interactions generated by
a uniform density of particles of opposite charges almost cancel each others, 
resulting in an effective, short range potential; 
as consequence, the correlation length is expected to be finite, and
certain {\it screening sum rules} are conjectured.  
Low temperatures, on the contrary, favor the formation of
pairs of opposite charges, the dipoles: the effective range
of the interactions remains long, and the correlations length is 
infinite; this regime is the {\it dipole phase} - or 
{\it KT phase}.
In between the two phases, there is  (at least) one critical curve,
that is called
{\it KT transition line}, and was found to 
intersect the $z=0$ axis at $\b_c(0)=8\p$, \cite{[Ko]}. 
This picture represented a {\it new kind} of phase transition, because
all the temperatures below $\b_c(z)$ are, in a sense, critical; in fact,
approaching $\b_c(z)$ from higher temperatures the correlations length
is expected to diverge as $\x\sim \exp[c(z)|\b-\b_c(z)|^{-\frac12}]$, as
opposed to the $\x\sim |\b-\b_c|^{-\n}$ (or $\x\sim\ln |\b-\b_c|$) of
the second-order phase transitions.

After the pioneering analysis of Kosterlitz and Thouless, 
efforts of many authors were addressed to the topic, 
in search of stronger evidence of the KT phase transition: 
the reader interested in theoretical physics works can find useful
discussions and a good selection of references in \cite{[AGG]}, 
\cite{[Ka2]} and  \cite{[DL]} (see also the conjecture in \cite{[GN]}
of an infinite series of  KT transition lines,  
intersecting the $z=0$ axis at $\b_{c,n}(0)=8\p(1-1/2n)$, $n=1,2,\ldots
\io$; and  criticism in 
\cite{[Te]}).  

The first major rigorous result was the 
proof of Fr\"ohlich and Park, \cite{[FP]},  
of the existence of the thermodynamic
limit for pressure and correlations.
Later on,  Fr\"ohlich and Spencer, \cite{[FS]}, \cite{[FS2]}, proved, 
for $\b$ large enough, an upper and lower  power-law bound 
for the correlations of {\it fractional} (i.e. non-integer) charges; 
then, refinements of the
same technique allowed Marchetti, Klain and Peres, \cite{[MKP]}, \cite{[Ma]}
\cite{[MK]}, to cover increasing regions in the dipole phase 
that eventually included the point 
$(\b,z)=(8\p,0)$, but not the  rest of the KT transition 
line. Despite its  fast improvement, 
Fr\"ohlich-Spencer method seemed to have  some unavoidable limitations:
it could not provide the exact power  of the correlations fall-off, nor
could exclude logarithmic corrections to such decay (which actually were
expected along the KT transition line); and it did not provide any useful
bound for correlations of  {\it integer} charges. 
For this reasons, different authors started developing an RG approach to the
model - at the beginning in some approximate form:  hierarchical metric, or
order by order in perturbation theory;  see
\cite{[BGN]}, \cite{[MP]}, \cite{[Di2]}, \cite{[NP]}, 
\cite{[KPW]}, \cite{[BR]}.  
Later,  Dimock and Hurd, \cite{[DH]},  achieved  a rigorous 
construction, under no approximation,  of the pressure in a region of
the dipole phase that  included $(\b,z)=(8\p,0)$ but not the rest of 
the KT transition line; they could not discuss charge correlations, though. 
Finally, the only rigorous result on the plasma phase is the work of
Yang, \cite{[Ya]}, that extended to dimension two the proof of the
dynamical  mass generation for small $\b$ obtained in 
\cite{[Br1]}, \cite{[BF]}, for higher dimensional cases.

The objective of this paper is to study the Coulomb Gas  {\it along}
the  KT transition line, for small activity.  
Using the general RG approach of \cite{[Br]} and some model-specific ideas 
in \cite{[DH]}, we are able to give a constructive proof
of the existence of the pressure.  We shall not discuss here the
critical exponents of the correlation functions; but in view of the
bounds of this paper,  the task should not be difficult and we plan to pursue
it in a possible forthcoming work.

Besides, we shall not consider in these pages more complicated models
studied by  Fr\"ohlich and Spencer, such as  the XY, 
Villain, discrete 
Gaussian, $Z_n$-Clock and  the solid-on-solid models, because they
require (perhaps) a large-activity framework.  
Even more interesting - and more difficult to treat - 
is the surprising (formal) equivalence between the KT transition
and the second-order phase transitions of certain two-dimensional,  
lattice models, such as Ashkin-Teller, Six-Vertex and 
 Eight-Vertex,  $Q$-states and anti-ferromagnetic Potts models, 
$O(n)$ models (including the $n=0$ case, i.e. the Self Avoiding Walks)
 etc. (see \cite {[Ka]}, \cite{[Ni]} and references therein).
Future efforts should be  addressed 
to these appealing  applications of the Coulomb Gas.

\section{Definitions and Results}

A possible microscopic realization  of the two-dimensional 
Coulomb Gas  is the following. 
Consider a system of point particles labeled with numbers
$1, 2, 3\ldots$ and a finite square lattice
$\L\subset\ZZZ^2$ endowed with periodic boundary conditions: 
a configuration is the assignment   
to each of the particles, say the $j$-th,  of a charge
$\s_j=\pm1$ and  a position $x_j\in \L$.
Particles interact through a two-body electrostatic potential
$W_\L(x_i,x_j)\=W_\L(x_i-x_j)$, so that,  if  $\O_n$ is the set of all the
possible configurations
of $n$ particles,
the total energy  ({\it self-energy} included) of the configuration
$\o\in\O_n$ is
\be\lb{ene}
H_\L(\o)={1\over 2}\sum_{i,j=1}^n\s_i\s_j W_\L(x_i-x_j)\;.
\ee
If $z$ is the activity and $\b>0$ the inverse temperature, 
the Grand Canonical partition function is
\be\lb{gf}
Z_\L(\b,z)=\sum_{n\ge 0}
{z^n\over n!}
\sum_{\o\in \O_n}
e^{-\b H_\L(\o)}\;.
\ee
Before stating the main theorem, we have to give a  precise 
definition of  electrostatic potential $W_\L$.
It is to be the inverse
of $-\D_\L$, the Laplacian operator on $\L$; anyways, 
because of the periodic boundary conditions on $\L$, $-\D_\L$ has
zero modes, and  we have to assign a  regularization procedure 
to make sense of $(-\D_\L)^{-1}$. 
Define the Yukawa interaction on $\L$ with mass $m$ 
\be\lb{yuk}
W_\L(x;m)={1\over |\L|}\sum_{k\in \L^*}{e^{ikx}\over m^2-\hat\D(k)}
\ee
where $\L^*$ is the reciprocal lattice of $\L$ and
$-\hat\D(k)=2\sum_{j=0,1}(1-\cos k_j)$ is the Fourier transform of  $-\D_\L$;
then, if $H_\L(\o;m)$ is the Yukawa energy of the configuration $\o$,
re-define \pref{ene}  such that
\be\lb{eneb}
e^{-\b H_\L(\o)}:=\lim_{m\to 0}e^{-{\b\over 2}\sum_{i,j=1}^n\s_i\s_j W_\L(x_i-x_j;m)}\;.
\ee
Of course in the massless limit $m\to 0$ the Yukawa potential,  by
itself, is ill
defined; though we shall see in Sec. \ref{a1} that the above
definition makes sense and, in fact, assignes weight zero to
configurations of non-neutral total charge - because of that, the
system is symmetric under $z\to -z$.
Finally, for $R$ and $L$ positive integers, $L$ odd and bigger than one,
assume that $\L$ is a square with a side of $L^R$ lattice
sites; then the {\it thermodynamic limits} of the pressure is
\be\lb{td}
p(\b,z)=\lim_{R\to \io}{1\over \b|\L|}\ln Z_\L(\b,z)\;.
\ee
The main result of the paper is the following theorem.
\begin{theorem}
\lb{main} 
For $\e>0$ small enough, there exists a function 
$\Sigma(z)\ge\Sigma(0)=8\p$ such that,
if  $|z|\le \e$ and $\b=\Sigma(z)$, the limit \pref{td}
exists.
\end{theorem} 
As stated at this point, the Theorem is slightly imprecise, because 
we have not found a way to
characterize the Kosterlitz-Thouless transition line in terms of the
pressure only;  though it will become apparent in the next section that
the curve $\b=\Sigma(z)$ is exactly such line. 
Lengthier computations, simple but not explicitly 
pursued in this paper, would  prove that
$\Sigma(z)$ is a smooth function.  
\section{Strategy of the Proof}
The starting point of this analysis is the functional integral
representation of the partition function  that
allows for a more standard RG approach. In Sec. \ref{a1}
we shall prove an equivalent formula for the  
partition function for finite lattice $\L$:
\bal\lb{fif}
Z_\L(\b,z)=
&
\lim_{m\to0}\int\!dP_{\ge R}(\z^{(R)};m) 
\int\!dP_{R-1}(\z^{(R-1)})\cdots
\int\!dP_{1}(\z^{(1)})\;\cdot
\cr
&\qquad\qquad\cdot\int\!dP_{0}(\z^{(0)})\; 
e^{\VV\big(\z^{(0)}+\z^{(1)}+\cdots+\z^{(R-1)}+\z^{(R)}\big)}
\eal
where $dP_{\ge R}(\z;m)$ is a Gaussian with {\it massive}
covariance, 
$$
\int dP_{\ge R}(\z;m)\; \z_x\z_y=\G_{\ge R}(x-y;m)\;,
$$
while $dP_j$, for $j=0,\ldots,R-1$, is a
Gaussian measure with {\it massless} covariance 
$$
\int dP_j(\z)\; \z_x\z_y=\G_j(x-y)\;.
$$ 
Explicit
definitions of  $\G_{\ge R}$ and $\G_j$ (and precise meaning of 'massive'
and 'massless') are given in Sec. \ref{a1}. Here we
stress that, besides being positive-definite functions, 
$\G_0, \G_1, \ldots, \G_{R-1}$
satisfy the following properties
\bal
\lb{p1}&\G_j(x)=0
\qquad{\rm for\ }|x|\ge L^{j+1}/2\;,
\\ \cr
\lb{p2}&|\dpr^a\G_j(x)|\le C_a L^{-j|a|}
\qquad {\rm for\ }|x|\le
L^{j+1}/2\;,\quad {\rm if\ }|a|>0\;,
\\\cr
\lb{p3}&\G_j(0)={1\over 2\p}\ln L + c_j(L)
\qquad
{\rm for\ } |c_j(L)|\le c L^{-\frac{j}{4}}\;; 
\eal
where $C_a$ and $c$ are independent of $L$. Namely,  
$\G_j$ has  compact support $O(L^{j+1})$ 
and typical momentum $O(L^j)$.
Finally, to complete the explanation
of \pref{fif},  
\bal\lb{pot}
\VV(\f):= |\L|{1\over 2}\ln (1-s)
+{s\over 2}\sum_{x\in \L\atop \m\in\hat e}(\dpr^\m\f_x)^2+
z\sum_{x\in\L\atop\s=\pm}e^{i\s\a\f_x}\;,
\eal
where $\hat e=\{(1,0),(0,1),(-1,0),(0,-1)\}$ and  
$\dpr^\m\f_x=\f_{x+\m}-\f_{x}$; the notation $\sum_{\m\in\hat e}$ implies also
a factor $1/2$ that we do not write explicitly (so that the
Fourier transform of  $\sum_{\m\in\hat e}\dpr^{-\m}\dpr^\m$ coincides
with $\hD(k)$ defined after \pref{yuk}).  
The parameter $s$ is in $[0,1/2)$ and will be chosen as function of
$z$: in the final limit $R\to \io$, it
will fix the relation between $\a$ 
and the inverse temperature $\b$ through the formula
$\a^2= (1-s)\b$. 

The RG approach consists in computing the integrals in \pref{fif} 
progressively from the random variable with highest momentum  to the
one with  lowest. After each integration  we
define $\VV_j$, the effective potential on scale $j$, 
such that $\VV_0=\VV$, 
\bal\lb{itr}
&
e^{\VV_{j+1}(\f)}=\int\!dP_{j}(\z)\ \;e^{\VV_{j}(\f+\z)}\;,
\qquad j=0,1\ldots,R-1\;;
\eal
and, at last, 
\bal\lb{0itr}
&Z_\L(\b,z)=\lim_{m\to 0}\int\!dP_{\ge R}(\z;m)\ \;e^{\VV_{R}(\z)}\;.
\eal
In this way the evaluation of the partition function
is transformed into the flow of 
a dynamical system of effective interactions $\VV_0,
\VV_1,\ldots, \VV_{R}$. To have control
on it, we must distinguish  the {\it irrelevant part} of $\VV_j$, namely
the terms that, along the flow, 
become smaller and smaller by simple 'power counting'
arguments, 
from the {\it relevant part}, namely the terms that require a
more careful study. In order to do that, it is important 
to introduce some special kind of lattice domains: 
{\it blocks} and {\it polymers} (\cite{[Br]}, \cite{[BS]}). 
Define $|x|:=\max\{|x_0|,|x_1|\}$.
Recall that $L$ was chosen odd; and, for $j=0,1,\ldots,R$,  
pave the periodic lattice $\L$ with $L^{2(R-j)}$ disjoint 
squares of size $L^{j}$ in a natural way: there is the central square,
$$
\big\{x\in \L:|x|\le L^{j}/2 \big\}
$$
and all the other squares are translations of this one by 
vectors in $L^{j}\ZZZ$. We call these squares $j$-blocks, and
we denote the set of $j$-blocks by $\BB_j\=\BB_j(\L)$. $0$-blocks
are made of  single points, so, for example, $\BB_0(\L)=\L$. 
A union of $j-$blocks is called $j-$polymer, 
and the set of all $j-$polymers 
in $\L$ is denoted $\PP_j\=\PP_j(\L)$. Suppose $X$ is a  $j-$polymer:  
$\dpr X$ is the set of sites in $X$ with a nearest neighbor outside $X$; 
$\BB_j(X)$ is the set of the $j-$blocks in $X$;
$|X|_j$ is the cardinality of $\BB_j(X)$; 
and $\bar X$ is the smallest polymer in $\PP_{j+1}(\L)$ 
that contains $X$. A polymer made of  two blocks, $B,D\in \BB_j(\L)$, 
is connected
if there exist $x\in B$ and $y\in D$ s.t. $\|x-y\|=1$; the
definition extends to connected polymers of more blocks in the usual
way. 
We call $\PP_j^c\=\PP^c_j(\L)$ and $\SS_j\=\SS_j(\L)$ 
the set of the connected $j$-polymers 
and the set of the connected $j$-polymers that are made of no more than $4$
$j-$blocks -  the ``small polymers'' -  respectively. 
$\sms_j\=\sms_j(\L)$ is the set
of the connected polymers that are
not small; and $S$ is the ($j$-independent) number of 
small $j-$polymers that contain a given $j-$block. 
Given a $j-$polymer $X$, 
the collection of its maximal connected parts (each of which is a $j-$polymer
by construction) is called 
$\CC_j(X)$; while its {\it small set neighborhood} is 
$X^*=\cup\{Y\in \SS_j(\L):Y\cap X\neq \emptyset\}$. 
The empty set is considered as an element of $\PP_j(\L)$, 
but not of $\PP^c_j(\L)$.
\insertplot{420}{210}
{\ins{240pt}{97pt}{$x_0$}
\ins{129pt}{210pt}{$x_1$}

\ins{335pt}{62pt}{2-block}
\ins{335pt}{108pt}{1-block}
\ins{335pt}{128pt}{0-block}

\ins{238pt}{204pt}{$\L$}
}%
{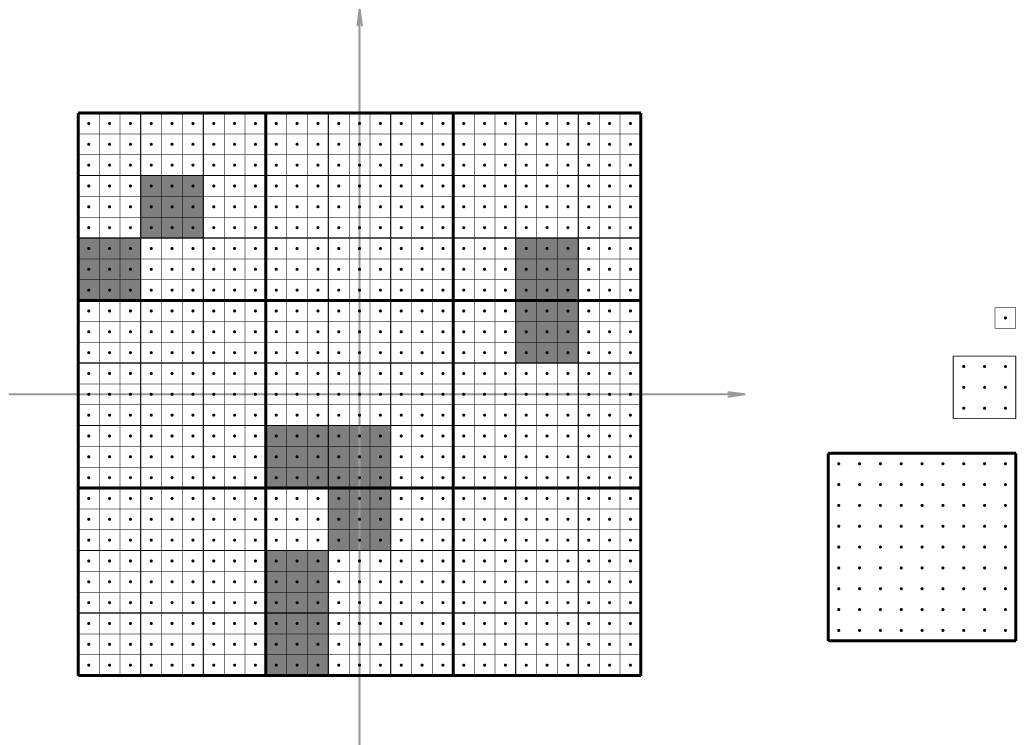}{\lb{fig2} Lattice paving with blocks of different sizes
  in the  case $L=3$ and $R=3$}{0}
\*
Given a block  $B\in \BB_j$, we define the interaction
\be\lb{uu}
U_j(s,z,\f,B)=V_j(s,z,\f,B)+W_j(s,z,\f,B)
\ee
where $W_j(s,z,\f,B)$ is quadratic in $s,z$ and
will be defined precisely below; while $V_j$ is basically given by
the original interaction  
\be \lb{vv}
V_j(s,z,\f,B)={s\over 2}\sum_{x\in B\atop \m\in\hat e}
(\dpr^\m\f_x)^2+zL^{-2j}\sum_{x\in B\atop \s=\pm}e^{i\s\a\f_x}\;.
\ee
Notice anyways the factor $L^{-2j}$ that makes $V_j$ explicitly
$j-$dependent; besides, it   
actually depends upon $\{\f_x\}_x$ for $x$ in a domain slightly bigger that $B$
(because includes the outer boundary sites).  
We assume the $W_j(s,z,\f,B)$ depends upon $\{\f_x\}_{x\in B^*}$.
We extend these definitions to
polymers $X\in \PP_j$ additively:
\be\lb{nt}
U_j(s,z,\f,X):=
\sum_{B\in \BB_j(X)}U_j(s,z,\f,B)\;;
\ee
 $V_j(s,z,\f,X)$ and $W_j(s,z,\f,X)$ are defined in the same way.
If we drop the first two variable of $U_j$, $V_j$ and $W_j$,  it
means they are $s_j$ and $z_j$ - i.e. their have the same label $j$ of
the potential. 
Finally, inductively assume the following formula 
for the effective potential
\bal\lb{pfr}
\VV_j(\f)=E_j|\L|
+\ln\Bigg[\sum_{X\in \PP_j(\L)} e^{U_j(\f,\L\bs X)} 
\prod_{Y\in \CC_j(X)} K_j(\f,Y)\Bigg]
\eal
where the {\it polymer activity}, 
$K_j(\f,Y)$, is a function of $\{\f_x\}_{x\in Y^*}$.
For example, the first terms in the sum inside 
the square brackets are
$$
e^{U_j(s_j,z_j,\f,\L)}+
\sum_{Y\in \PP^c_j(\L)} e^{U_j(s_j,z_j,\f,\L\bs Y)} 
K_j(\f,Y) +\cdots
$$
 Finally,
from \pref{0itr} and  \pref{pfr} with  $j=R$
\be\lb{0pfr}
Z_\L(\b,z)=e^{E_R|\L|} 
\lim_{m\to 0}\int\!dP_{\ge R}(\z;m)\lft[e^{U_{R}(\z, \L)}+K_R(\z, \L)\rgt]
:=e^{E_{R+1}|\L|}\;.
\ee
The details of these constructions are in Sec. \ref{RGM}; as opposed
to the original RG method in \cite{[BY]}, we will not
need any  'cluster expansion' 
to handle the tails of the covariances, for in
our setting they have compact support (this idea was devised in
\cite{[MS]}, \cite{[Br]}).
We shall be more specific on the regularity of $U_j$ and
$K_j$ later. At this stage, we just mention 
that $U_j$ is going to contain the second order part of 
{\it marginal} terms of the
iteration, whereas 
the {\it irrelevant} terms, as well higher order marginal terms, 
will be stored in $K_j$. 

Assumption \pref{pfr} holds at $j=0$, for $(E_0,s_0,z_0)=(\frac12\ln(1-s),s,z)$,
$W_0\=0$ and  $K_0\=0$. In Sec. \ref{RGM} we shall prove that, 
iterating \pref{itr}, assumption \pref{pfr} holds at any scale  
$j=1,\ldots R$, with  
$(s_j, z_j, K_j)$ recursively given by: for $j=0$,  
\bal\lb{0lk}
&s_{1}=s
\cr
&z_{1}=L^2e^{-\frac{\a^2}{2}\G_0(0)} z
\cr
&K_{1}=\RR_0(z, s)\;;
\eal
for $j=1, 2, \ldots, R-1$, 
\bal\lb{lk}
&s_{j+1}=s_j-a_j z_j^2+\FF_j(K_j)
\cr
&z_{j+1}=L^2e^{-\frac{\a^2}{2}\G_j(0)}\Big[z_j-b_j s_jz_j+\MM_j(K_j)\Big]
\cr
&K_{j+1}=\LL_j (K_j) + \RR_j(z_j, s_j, K_j)\;.
\eal
$\FF_j$ and $\MM_j$ are real functions 
of the polymer activity;  $\LL_j$ and $\RR_j$ are a
linear and high order maps of the polymer activities (and functions 
of $z_j$ and $s_j$). Note that   $a_j\=a_j(L)$, $b_j\=b_j(L)$. 
\begin{lemma}\lb{t3.0}
If $\a^2=8\p$, set $a:=8\p^2e^c\ln L$ and $b:=2 \ln L$,
where $c$ is a constant introduced in \pref{smpt}; then there exist $C$ and
$C(L)$ such that 
\be\lb{r0}
|L^2e^{-\frac{\a^2}{2}\G_j(0)}-1|\le C L^{-\frac j4}\;,\quad
|a_j-a|,\; 
|b_j-b|\le C(L) L^{-\frac j4}\;.
\ee
\end{lemma}
The former inequality is a consequence of \pref{p3}; the latter in
proven in Sec. \ref{aec}.
The energy parameters, $E_0, E_1,\ldots E_{R}$, are recursively
defined by 
\be\lb{sk}
E_{j+1}=E_j+L^{-2j}\lft[s^2_j\mce_{3,j}+
z^2_j\mce_{4,j}+s_j\mce_{2,j}+\mce_{1,j}(K_j)\rgt]\;,
\ee
where $\mce_{1,j}(K_j)$ is linear in $K_j$, while the other 
$\mce_{m,j}$'s are independent of  $s_j$, $z_j$ and $K_j$.
$E_{R+1}$ is defined in  \pref{0pfr}. 
For $j=0,1,\ldots, R$, 
let $\EE_j:=E_{j+1}-E_j$.
\begin{lemma}\lb{t3.5}
There exists $C(\a,L)$ such that,  for any given $j=0,1,\ldots, R$, 
if  $|s_j|, |z_j|,\|K_j\|_{h,T_j}\le \e_0$,
\be\lb{r2}
|\EE_j|\le C(\a,L) L^{-2j}\e_0 \;.
\ee
Besides, $\EE_0, \ldots, \EE_{R-1}$ (but not $\EE_{R}$) 
are the same on $\L$ and on
$\ZZZ^2$. 
\end{lemma}
The proof is in Sec. \ref{s6.1}.
\pref{lk} is the {\it RG map}. The solution of \pref{lk}, for initial data
$(s_0,z_0, K_0)=(s,z,0)$ will be called the {\it RG flow}; the
sequence of energies given by  \pref{sk} has no influence 
on the RG flow, hence can be seen as the history of an
observable.
Consider the case $\a^2>8\p$:  by \pref{p3}, for $L$ large enough, 
$L^2e^{-\frac{\a^2}{2}\G_j(0)}<1$; consequently, if also $|z|$ is
small enough, the flow goes to zero as was
discussed in \cite{[DH]} (in the sub-case of $|z|$ small w.r.t
the value of $\b$). This paper 
is focused on the more complicated case $\a^2=8\p$, when
$L^2e^{-\frac{\a^2}{2}\G_j(0)}$ is basically  1 and then the flow is 
determined by the second order terms (see Lemma \ref{t3.0}). 
In fact, neglecting higher orders and neglecting the RG map for  
$(K_j)_j$, \pref{lk} is the equation that Kosterlitz obtained 
with a (formal) RG technique in coordinate space, \cite{[Ko]}. 
The corresponding approximate solution is, for $j\ge 1$, 
\be \lb{kos}
s_j=\frac{1}{b} |q_j|\;, \qquad
z_j=\frac{1}{\sqrt{ab}}q_j\;, 
\ee
with  $q_1$ proportional to $z$ and 
\bal\lb{dq}
q_j:=\frac {q_1}{1+|q_1| (j-1)}\;.
\eal  
\pref{kos} describes a line, \ie the first approximation of  
the separatrix in Fig. \ref{f1}, 
which is the Kosterlitz-Thouless transition line. 
Our goal is a proof that the solution of the full RG map
\pref{0lk}, \pref{lk}, is, qualitatively, not too different from
\pref{kos}. In order to achieve that, we need to know the smoothness
of the remainder terms. In Sec. \ref{nr}, 
we will set up a norm $\|\cdot\|_{h,T_j}$ 
for polymer activities on scale $j$ depending on two parameters $h$
and $A$: if properly chosen, the following lemmas hold. 

\begin{lemma}\lb{t3.1}
There exist $C,\; C(\a)>1$ such that, if $L$ is large
enough,   
\be\lb{e3.1}
|\FF_j(K_j)|\le C A^{-1} \|K_j\|_{h,T_j}\;,
\qquad
|\MM_j(K_j)|\le C(\a) A^{-1} \|K_j\|_{h,T_j}\;.
\ee
Besides, $\FF_1, \ldots, \FF_{R-1}$ and 
$\MM_1, \ldots, \MM_{R-1}$ are the same on $\L$ and on
$\ZZZ^2$. 
\end{lemma}
\begin{lemma}\lb{t3.2} 
There exist $C(\a)>1$ and $\h,\;\th >0$ such that, for $L$ large
enough,  
\be\lb{e3.2}
\|\LL_j(K_j)\|_{h,T_{j+1}}\le C(\a)\lft(L^{-\th}+ A^{-\h}\rgt)
\|K_j\|_{h,T_j}\;.
\ee
\end{lemma}
\begin{lemma}\lb{t3.4}
If $\e_j>0$ is small enough, there exists $C\=C(A,L,\a)>1$ such that, for any 
$(s_j,z_j,K_j)$ and $(\dot s_j, \dot z_j,\dot K_j)$ 
satisfying  $|s_j|,|z_j|, |\dot s_j|,|\dot z_j|\le \e_j$
and $\|K_j\|_{h,T_j}, \|\dot K_j\|_{h,T_j}\le  \e_j^2$: 
for $j=0$
\bal\lb{0e3.4}
\|\RR_0(z, s)-
\RR_0(\dot z, \dot s)\|_{h,T_{1}}
\le 
C \e_0\lft[|s-\dot s|+ |z-\dot z|\rgt]\;; 
\eal
while, for $j=1,2, \ldots, R-1$, 
\bal\lb{e3.4}
&\|\RR_j(z_j, s_j, K_j)-
\RR_j(\dot z_j, \dot s_j, \dot K_j)\|_{h,T_{j+1}}
\cr
&\qquad\le 
C \lft[\e_j^2|s_j-\dot s_j|+ \e_j^2|z_j-\dot z_j|
+\e_j \|K_j-\dot K_j\|_{h,T_j}\rgt]. 
\eal
\end{lemma}
The proofs are in Sec. \ref{s6}. 
In particular, Lemma \ref{t3.2} is crucial: 
to prove the contraction of $\LL_j$ we
have to show that $\LL_j(K_j)$ is made of irrelevant terms.
The role of constants and parameters so far introduced is the
following: $\a^2=8\p$, although in many sub-results  of the paper
we will just assume $\a^2\ge 8\p$; 
$h$ is a numerical constant related to the propagator, 
see \pref{hcon}; $L$ will be taken large; 
$A$ is to be large enough w.r.t. $L$; finally, $\e_0$,  the 
size of $z$, is to be small enough w.r.t. $A$ and $L$ (and $\a$).
In a sense, the major improvement w.r.t. \cite{[DH]}, crucial
for studying the Kosterlitz-Thouless line, is
that in this paper $h$ is independent of $L$.    

Thanks to Lemma \ref{t3.1}, \ref{t3.2} and \ref{t3.4}, 
we have the following result on the  RG flow.
\begin{theorem}\lb{t.sm}
Given  $L$, $A$ large enough and an $\e\=\e(A,L,\a)$, 
in correspondence of any $z$,  $|z|\le \e$,
there exists
an $s=\Si(z)$ such that the solution of \pref{lk} with initial
data $(z, s)$ is 
\be\lb{slk}
s_j=\frac{|q_j|}{b}+O(j^{-\frac32})\;,\quad
z_j=\frac{q_j}{\sqrt {ab}}+O(j^{-\frac32})\;,\quad
\|K_j\|_{h,T_j}=O(j^{-3})\;,
\ee
for  $j=0, \ldots, R$  and $q_1=O(z)$. Besides, $\Si(z)$  can be
chosen 
independent of $\L$.
\end{theorem} 
The proof is given in Sec.\ref{s.7}, and uses the fixed point theorem
for Banach spaces. Since the flow of $(s_j,z_j,K_j)$ remains bounded, Lemma
\ref{t3.5} applies, and  
$$
\b p(\b,z)=\lim_{R\to \io}\frac{\ln Z_\L(\b,z)}{|\L|}
=\lim_{R\to \io} E_{R+1}=E_0+\sum_{j\ge0}\EE_j\;, 
$$
where the series is convergent because of \pref{r2} 
(valid also for $\EE_{R+1}$).
This completes the proof of the main result, Theorem \ref{main};  
in the remaining sections, we shall take
up the task of proving all the above sub-results. Note that 
$C$, $C(\a)$,  $C(L,\a)$ and $C(A,L,\a)$, 
will  indicate (possibly) different values in different equations.     

\section{Renormalization Group Map}\lb{RGM}
In this and the following section we adopt an abridged
notation for the fields.
In general, 
we remove the labels $j$ because they will be clear from the context, 
and label the sum of the fields on higher
scales with a prime, so that $\z_x:=\z^{(j)}_x$ and 
$\f'_x:=\z^{(R)}_x+\z^{(R-1)}_x+\cdots+\z^{(j+1)}_x$; besides,   
$\f_x:=\f'_x+\z_x$. We also use $\EEE_j[\;\cdot\;]$ 
for the expectations
w.r.t. the measure $dP_j(\z^{(j)})$. Therefore the RG map for the
effective potential is
\be\lb{rgm}
e^{\VV_{j+1}(\f')}=\EEE_j\lft[e^{\VV_j(\f'+\z)}\rgt]\;,
\ee
for $\VV_j(\f)$ given by \pref{pfr}.
As function of the fields, $\VV_0$ is made of a periodic term of
period $2\p/\a$, and a derivative term: then
$\VV_0$ is invariant under $\f_x\to \f_x+ \frac{2\p}{\a}t$ for the
a constant, integer field $t$. The latter property remains true,
by induction, for $\VV_j$ and then for $K_j$. As consequence, 
we shall prove in
appendix \ref{abCG} the following decomposition of $K_j(\f,Y)$ into
{\it charged components}. 
\begin{lemma}\lb{ldec3}
For any $j=0,1,\ldots,R$, there exists a decomposition
\be\lb{dec3}
K_j(\f,X)=\sum_{q\in \ZZZ} \hK_j(q,\f,X)
\ee
such that, if $\th$ is a constant field, 
\be\lb{dec4}
\hK_j(q,\f+\th,X)= e^{i q\a\th}\hK_j(q,\f,X)\;.
\ee
\end{lemma}
This simple result is borrowed from \cite{[DH]} and for completeness
is reviewed in Sec. \ref{abCG}. The 'power counting' argument -
that we shall make rigorous in the rest of the paper - implies: 
a) terms with charge $q$  contract by a factor 
$L^{-\frac{\a^2}{4\p}q^2}$; b) terms proportional to $(\dpr \f')^n$
contract by a factor $L^{-n}$; c) all terms are increased by a
volume factor $L^2$ (the ratio of  volumes of $j+1$- and $j$-blocks). 
Therefore, at $\a^2=8\p$, RG reduces the size of the  
components with charge $|q|\ge 2$;  of the components with charge
$|q|=1$, if the 0-th order Taylor expansion in $\dpr \f'$ has been
taken away; and also the of neutral charge component, if the
2-th order Taylor expansion in $\dpr \f'$ has been
taken away. The terms so removed are absorbed in 
$E_j, s_j,z_j$ to give  $E_{j+1}, s_{j+1},z_{j+1}$. 
Guided by these ideas, we pass to a technical description of RG;
because of technical reasons, it is more convenient to define
the first RG steps in a different (in fact simpler) 
way. 
\subsection{First RG step} 
After integrating the field $\z^{(0)}$ in  $V_0$, 
we want to recast the effective potential $\VV_1$ into the form
\pref{pfr}.
There are many ways to achieve this.
\begin{lemma}\lb{llin0}
Given $s$ and $z$, define $s_1$ and $z_1$ as in \pref{0lk}, and 
\bal\lb{lk1}
E_1=-\frac{s}{2}\sum_{\m\in\hat e}(\dpr^\m\dpr^\m \G_0)^2(0)\;.
\eal
There exists a choice of $K_1$
 and $U_1$
such that \pref{pfr} holds with $K_1={\rm O}(V_0^2)$. 
\end{lemma}
{\bf\0Proof of Lemma \ref{llin0} - } 
For $D\in \BB_1$ and $Y\in \PP^c_1$ define
\bal\lb{spl}
&P_0(\f',\z,D):=e^{V_0(\f,D)}
-e^{V_1(\f',D)+E_1|D|}\;,
\cr\cr&
P_0^Y(\f',\z):=\prod_{D\in \BB_{1}(Y)}P_0(\f',\z,D)\;.
\eal
(recall $V_0(\f,D)\=V_0(s,z,\f,D)$ and $V_1(\f',D)\=V_1(s_1,z_1,\f',D)$).
Expand formula \pref{itr} at $j=0$ w.r.t. $P_0$, namely 
\be
\EEE_0\lft[e^{\VV_0(\f,\L)}\rgt]=
\EEE_0\lft[\prod_{D\in \BB_1(\L)}\lft(P_0(\f', \z,D)+e^{V_1(\f',D)+E_1|D|}\rgt)\rgt]\;,
\ee
to obtain \pref{pfr} for  $U_1\=V_1$ (i.e. $W_1\=0$) and 
\bal\lb{5.16no}
K_1(\f',Y)=e^{-E_1|Y|}
\EEE_0\lft[P^Y_0(\f',\z)\rgt]\;.
\eal
The linear order in $V_0$ of \pref{5.16no} may not be zero
only when  $Y$ is a $1-$block $D$:
\bal\lb{5.25bis}
 \lft[\LL_0K_{0}\rgt](\f',D)
=\EEE_0[V_0(\f,D)]-
V_1(\f',D)- E_1|D|\;.
\eal 
But this expression is zero by the choice 
of $s_1$, $z_1$ and $E_1$. \qedd
 
\subsection{$j$-th RG Step}\lb{s4.2} Consider now 
$j=1,2,\ldots,R-1$.
After integrating the field $\z$, we want to recast the
effective potential $V_{j+1}$ into the form
\pref{pfr}:
\be\lb{pfr2}
e^{\VV_{j+1}(\f')}=e^{E_{j+1} |\L|}
\sum_{X\in \PP_{j+1}(\L)}e^{U_{j+1}(\f',\L\bs X)}
\prod_{Y\in \CC_{j+1}(X)} K_{j+1}(\f',Y)\;.
\ee
Again, this formula does not determine $K_{j+1}$ and $U_{j+1}$ 
in a unique way; we want to take advantage of this
freedom in order to make  $K_{j+1}$ either 
irrelevant or third order in 
$V_j$. To achieve that, we have to extract a $\bar
Q_j(\f',X)$ from $\EEE_j [K_j]$  and a $Q_j(\f',Y)$ 
from $\EEE^T_j[V_j;V_j]$: the next lemma, purely
combinatorial, furnishes such a $K_{j+1}$ and 
the corresponding  
$U_{j+1}$. 
\begin{lemma}
\lb{l5.3}
Suppose to be given two activities: 
$\bar Q_j(\f',X)={\rm O}(K_j)$ 
with support on sets $X\subset\SS_{j}$, and  
$Q_j(\f',Y)={\rm O}(V^2_j)$ with support 
on sets $Y\subset\SS_{j+1}$; and suppose 
the difference between $(s_j,z_j, E_j)$ and  
$(s_{j+1},z_{j+1}, E_{j+1})$ is such
that 
\bal\lb{cond1}
(E_{j+1}-E_j)|D|+V_{j+1}(\f',D)-\EEE_j\lft[V_j(\f,D)\rgt]
={\rm O}(K_j, V_j^2)\;.
\eal 
Then there exists a choice of  
$K_{j+1}$ and $U_{j+1}$
such that  \pref{pfr2} holds  
with
\be\lb{lnr}
K_{j+1}(\f',Y'):=\lft[\LL_j K_j\rgt](\f',Y')+\RR_j(\f',Y')\;, 
\ee
where 
$\RR_j(\f',Y')={\rm O}(K^2_j, K_j V_j, V^3_j)$
and,  
\bal\lb{5.25}
&
[\LL_j K_{j}](\f',Y')
=\sum_{X\in \PP^c_j(Y')}^{\overline X=Y'}
\Big[\EEE_j[K_j(\f,X)]-\bar Q_j(\f',X)\Big]
\cr
&
+
{1\over 2}\sum_{B_0, B_1\in \BB_{j}(Y')}^{\overline{B_0\cup B_1}=Y'}
\EEE^T_j\lft[V_j(\f,B_0);V_j(\f,B_1)\rgt]-Q_j(\f',Y')
\cr
&
-\sum_{D\in \BB_{j+1}}^{D=Y'}
\Bigg[W_{j+1}(\f',D)-
\EEE_j\lft[W_j(s_{j+1},z_{j+1},\f,D)\rgt]
-\sum_{Y\in \SS_{j+1}}^{Y\supset D} {Q_j(\f',Y)\over
  |Y|_{j+1}}\Bigg]
\cr
&
-\sum_{B\in \BB_{j}}^{\bar B=Y'}
\Bigg[\EE_{j}|B|+V_{j+1}(\f',B)
-\EEE_j\lft[V_j(\f,B)\rgt]
-\sum_{X\in \SS_j}^{X\supset B}
{\bar Q_j(\f',X)\over
  |X|_{j}}\Bigg]
\eal
(recall the short notations  for $V_j$, 
$V_{j+1}$ and $W_{j+1}$ defined after \pref{nt}). 
\end{lemma}
In the third line there is 
$W_j(s_{j+1},z_{j+1},\f,D)$ as opposed to $W_j(s_{j},z_{j},\f,D)$: the
former is more convenient for our analysis and  the difference with
the latter is $O(V_j^3, K_j V_j^2, K_j^2)$, 
which can be left inside $\RR_j$. Besides, 
\pref{lnr} gives  the third line of \pref{lk}, with  
$\RR_j(\f',Y')\=[\RR_j(s_j,z_j,K_j)](\f',Y')$.

Before giving the proof we elaborate on \pref{5.25}.
As planned before, 
in the first and second line of the r.h.s. member 
we read the extraction of 
$\bar Q_j$  and $Q_j$ from $\EEE_j[K_j]$ and $\EEE^T_j[V_j;V_j]$;  
the same terms are re-absorbed into 
$E_{j}$, $s_{j}$ $z_{j}$ 
in the third line, 
so obtaining   $E_{j+1}$, $s_{j+1}$,
$z_{j+1}$. Therefore, to determine the latter parameters, 
we have to choose $\bar Q_j$  and $Q_j$ first.

The choice of  $\bar Q_j$ requires Taylor expansion in $\nabla \f'$, 
that we now define. 
Let $F(\x,X)$ be a smooth
function of the field $(\x_x)_{x\in X^*}$; the $n$-order
Taylor expansion of $F(\x,X)$ at $\x=0$ is
\bal
&({\rm Tay}_{n} F)(\x,X)
:=\sum_{m=0}^n {1\over m!}
\sum_{x_1\ldots, x_m\in X^*}\x_{x_1}\cdots\x_{x_m}
{\dpr^m F\over \dpr \x_{x_1}\cdots \dpr \x_{x_m}}(0,X)\;;
\eal
correspondingly, the $n$-order remainder is
\bal
&({\rm Rem}_{n} F)(\x,X):= F(\x,X)-({\rm Tay}_{n} F)(\x,X)\;.
\eal
The Taylor expansion of  $K_j$ is to be in  the field $\x\sim\nabla
\f'$, namely
we have to single out the part of the activity that is 
{\it purely dipolar}. 
That is accomplished  
by \pref{dec4}: for any point $x_0\in X$, 
if $(\d \f')_x:= \f'_x-\f'_{x_0}$ (which is 
a sum of $\nabla \f'$'s),
$$
\hK_j(q,\f,X)= e^{i\a q \f'_{x_0}} \hK_j(q,\d\f'+\z,X)\;.
$$
Therefore, our choice for $\bar Q_j$ is 
\bal\lb{qbr}
&\bar Q_j(\f',X)
=\frac{1}{|X|}\sum_{x_0\in X}{\rm Tay}_{2} 
\EEE_j\lft[\hK_j(0,\d\f'+\z,X)\rgt]
\cr
&\qquad
+\frac{1}{|X|}\sum_{x_0\in X}\sum_{\s=\pm1}e^{i\s\a\f'_{x_0}}{\rm Tay}_{0}
\EEE_j\lft[\hK_j(\s,\d\f'+\z,X)\rgt]
\cr
&=\EEE_j\lft[\hK_j(0,\z,X)\rgt]
+\frac{1}{|X|}\sum_{x_0\in X}\sum_{\s=\pm1}
e^{i\s\a\f'_{x_0}}\EEE_j\lft[\hK_j(\s,\z,X)\rgt]
\cr
&\qquad+\frac{1}{|X|}\sum_{x_0\in X\atop x_1,x_2\in X^*}
\EEE_j\lft[\frac{\dpr^2\hK_j}
{\dpr \f_{x_1}\dpr \f_{x_2}}(0,\z,X)\rgt]
\frac{(\f'_{x_1}-\f'_{x_0})(\f'_{x_2}-\f'_{x_0})}{2}\;
\eal
where Taylor expansions are in $\x=\d\f'$
(the special point $x_0$ is averaged over $X$); in \pref{qbr}
we also used the fact that $K_j$ is even in $\z$, therefore the
expectation of one derivative of $K_j$ is zero.
We shall prove in Sec. \ref{s6.4}  
that \pref{qbr} makes irrelevant the first line of \pref{5.25}. 
Then, define intermediate parameters  
$\bar \EE_{j}=\bar E_j-E_j$, $\bar s_{j}$ and 
$\bar z_{j}$  such that
\bal\lb{lk2}
&\bar s_j=s_j +\FF_j(K_j)\;,\qquad
\bar z_j=L^{2}e^{-\frac{\a^2}{2}\G_j(0)}\lft[z+\MM_j(K_j)\rgt]
\cr\cr
&\bar \EE_j= L^{-2j}\lft[\mce_{1,j}(K_j) + s_j \mce_{2,j}\rgt]\;,
\eal
where 
\bal\lb{dfm1} 
&\mce_{1,j}(K_j)=\sum_{X\in \SS_j}^{X\ni0}
\frac{1}{|X|_{j}}\EEE_j\lft[\hK_j(0,\z, X)\rgt]\;,
\qquad
\mce_{2,j}=-{L^{2j}\over 2}\sum_{\m\in \hat e} (\dpr^\m\dpr^\m\G_j)(0)\;.
\cr\cr
&\FF_j(K_j)=\sum_{X\in \SS_j}^{X\ni0}
\frac{L^{-2j}}{|X|_{j}|X|}\sum_{x_0\in X\atop x_1,x_2\in X^*}
\EEE_j\lft[\frac{\dpr^2\hK_j}
{\dpr \f_{x_1}\dpr \f_{x_2}}(0,\z,X)\rgt]
\sum_{\m\in\hat e}(x_1-x_0)^\m(x_2-x_0)^\m
\cr\cr
&\MM_j(K_j)=\frac{e^{\frac{\a^2}{2}\G_j(0)}}{2}\sum_{X\in \SS_j}^{X\ni0}
\frac{1}{|X|_{j}}\sum_{q=\pm1}
\EEE_j\lft[\hK_j(q,\z,X)\rgt]\;.
\eal 
In Sec. \ref{s6.4} we shall also prove that, with the above choices,  
the following quantity
(compare it with the third line of \pref{5.25}) 
is irrelevant: 
\be\lb{irb}
\sum_{B\in \BB_{j}}^{\bar B=Y'}
\Bigg[\bar \EE_{j}|B|+V_{j+1}(\bar s_j, \bar z_j,\f',B)
-\EEE_j\lft[V_{j}(\f,B)\rgt]
-\sum_{X\in \SS_j}^{X\supset B}
{\bar Q_j(\f',X)\over
  |X|_{j}}\Bigg]
\ee
Then set $Q_j$ so that  the second line
of \pref{5.25} is vanishing:
\be\lb{qqq}
Q_j(\f',Y'):={1\over 2}\sum_{B_0, B_1\in \BB_{j}(Y')}^{\overline{B_0\cup B_1}=Y'}
\EEE^T_j\lft[V_j(\f,B_0);V_j(\f,B_1)\rgt]\;;
\ee
and choose $\EE_{j}$, $s_{j+1}$,
$z_{j+1}$ so that the remaining part of \pref{5.25} vanish:
\bal\lb{eW}
&\sum_{D\in \BB_{j+1}}^{D=Y'}
\Bigg[\lft(\EE_{j}-\bar \EE_{j}\rgt)|D|
+V_{j+1}(s_{j+1}-\bar s_j, z_{j+1}-\bar z_j,\f',D)
\cr
&+
W_{j+1}(\f',D)-\EEE_j\lft[W_j(s_{j+1}, z_{j+1},\f,D)\rgt]
-\sum_{Y\in \SS_{j+1}}^{Y\supset D} {Q_j(\f',Y)\over
  |Y|_{j+1}}
\Bigg]=0\;.
\eal
Because of the simple identity
\bal\lb{srp}
&\sum_{Y\in \SS_{j+1}}^{Y\supset D}{Q_j(\f',Y)\over |Y|_{j+1}}
={1\over 2}
E^T\lft[V_j(s_j,z_j,\f,D);V_j(s_j,z_j,\f,D^*)\rgt]
\eal 
and computations in Sec. \ref{sc2},  
cancellation \pref{eW} is obtained if    
\bal\lb{lk3}
&s_{j+1}=\bar s_j-a_jz^2_j\;,\qquad
z_{j+1}=\bar z_j-L^2 e^{-\frac{\a^2}{2}\G_j(0)} b_js_j z_j
\cr\cr
&\EE_{j}=\bar \EE_{j}+L^{-2j}\lft[s^2_j\mce_{3,j}+
z^2_j\mce_{4,j}\rgt]
\eal
and if $W_j$ is given by    
\bal\lb{wj}
s^2 W_{a,j}(\f,B)
+z^2\Big[W_{b,j}(\f,B)+
W_{c,j}(\f,B)\Big]+zs
W_{d,j}(\f,B)\;,
\eal
where, setting $\G_{j,n}(x):=\G_{n}(x)+\G_{n+1}(x)+\cdots+\G_{j}(x)$ 
and $f(0|x):= f(0)-f(x)$,
the definitions of the functions 
used in \pref{lk3} and \pref{wj} are
\bal\lb{dfm2} 
&a_j:={\a^2\over 2} \sum_{y\in \ZZZ}|y|^2
\lft[
w_{b,j}(y)\lft(e^{-\a^2\G_j(0|y)}-1\rgt)
+
e^{-\a^2\G_j(0)}\lft(e^{\a^2\G_j(y)}-1\rgt)L^{-4j}\rgt]
\notag\\
&b_j:={\a^2 \over 2}\sum_{y\in \ZZZ\atop\m\in \hat e}
\lft[\lft(\dpr^\m\G_{j}\rgt)^2(y) + 
2\sum_{n=0}^{j-1}\lft(\dpr^\m\G_{n}\rgt)(y)\lft(\dpr^\m\G_{j}\rgt)(y)
e^{-{\a^2\over2}\G_{j-1,n}(0)} L^{2(j-n)}\rgt]
\notag\\
&\mce_{3,j}:={L^{2j}\over 4}\sum_{y\in \ZZZ^2}\sum_{\m\in\hat e\atop\n\in\hat e}
\Big[(\dpr^\m\dpr^\n\G_{j,0})(y) + 2(\dpr^\m\dpr^\n\G_{j-1,0})(y)\Big]
(\dpr^\m\dpr^\n\G_{j})(y|0)\;,
\notag\\
&
\mce_{4,j}:=2L^{2j}\sum_{y}w_{b,j}(y)\lft[e^{-\a^2\G_j(0|y)}-1-{\a^2\over 2}
\sum_{\m,\n}\dpr^\m\dpr^\n \G_j(0)y^\m y^\n\rgt]
\notag\\
&\qquad\qquad+L^{-2j}\sum_{y}e^{-\a^2\G_j(0)}
\lft(e^{\a^2\G_j(y)}-1\rgt)\;,
\eal
and $W_{m,0}(s,z,\f,B)=0$ for $m=a,b,c,d,e$, while for $j\ge 1$, 
\bal\lb{dfW}
W_{a,j}(\f,B)
=&-
\sum_{\m\in\hat e\atop\n\in\hat e}
\sum_{y\in \ZZZ^2} w_{a,j}^{\m\n}(y)
\sum_{x\in B}(\dpr^\m\f_x)
\Big[(\dpr^\n\f_{x+y})- (\dpr^\n\f_x)\Big]
\cr
W_{b,j}(\f,B)
=&\sum_{y\in \ZZZ^2}w_{b,j}(y)
\sum_{x\in B\atop\s=\pm} 
\Bigg[ e^{i\s\a(\f_{x}-\f_{x+y})}
-1
+{\a^2\over 2}\sum_{\m\n}(\dpr^{\m}\f_x)(\dpr^{\n}\f_x)y^\m y^\n
\Bigg]
\cr
W_{c,j}(\f,B)
=&\sum_{y\in \ZZZ^2} 
w_{c,j}(y)
\sum_{x\in B\atop\s=\pm}
e^{i\s\a(\f_{x}+\f_{x+y})}
\cr
W_{d,j}(\f,B)
=&
\sum_{\m}\sum_{y\in \ZZZ^2} 
w^\m_{d,j}(y)
\sum_{x\in B\atop\s=\pm}
i\s\lft[
e^{i\s\a\f_{x}}(\dpr^\m\f_{x+y})-
e^{i\s\a\f_{x+y}}(\dpr^\m\f_x)\rgt]
\cr
&-
\sum_{y\in \ZZZ^2}
w_{e,j}(y)\sum_{x\in B\atop\s=\pm}
\lft(e^{i\s\a\f_{x+y}}-e^{i\s\a\f_x}\rgt)
\eal
for
\bal\lb{lw}
&w_{a,j}^{\m\n}(y)={1\over 2}
(\dpr^\m\dpr^\n \G_{j-1,0})(y)\;;
\cr
&w_{b,j}(y)=\sum_{n=0}^{j-1}
e^{-\a^2\G_{j-1,n+1}(0|y)}
e^{-\a^2\G_n(0)}\lft(e^{\a^2\G_n(y)}-1\rgt)L^{-4n}\;;
\cr
&w_{c,j}(y)={1\over 2}
\sum_{n=0}^{j-1} e^{-\a^2[\G_{j-1,n+1}(0)+\G_{j-1, n+1}(y)]}
e^{-\a^2\G_n(0)}\lft(e^{-\a^2\G_n(y)}-1\rgt)L^{-4n}\;;
\cr
&w^\m_{d,j}(y)={\a\over 2}\sum_{n=0}^{j-1}
e^{-{\a^2\over 2}\G_{j-1, n}(0)}
(\dpr^\m\G_{n})(y)
L^{-2n}
\cr
&w_{e,j}(y)={\a^2\over 4}
\sum_{n=0}^{j-1} e^{-{\a^2\over 2}\G_{j-1,n}(0)}
\sum_\m \lft[\lft(\dpr^\m\G_{j-1,n}\rgt)^2(y)-
\lft(\dpr^\m\G_{j-1,n+1}\rgt)^2(y)\rgt]
L^{-2n}\;.
\cr
\eal
By \pref{p1}, $W_j(\f, B)$ depends on the field $\f_x$ for $x$ in a
neighborhood of $B$ of diameter $L^j/2$, which is a subset of $B^*$. 
Finally, joining \pref{lk3} with \pref{lk2} we obtain
\pref{lk} and fulfill condition \pref{cond1}.
\*
{\bf\0Proof of Lemma \ref{l5.3}. -}
The {\it re-blocking} operation used here 
is different from the one in \cite{[Br]}, partly because the extraction of
$Q_j$ (that for the dipoles of \cite{[Br]} is not required), partly
because of a different {\it large field regulators} introduced below.

Starting from \pref{pfr}, reblock the polymers 
on scale $j+1$ and obtain:
\bal\lb{pfj+1}
e^{V_j(\f)}=e^{E_j|\L|}\sum_{X\in \PP_{j+1}} e^{U_{j}(\f,\L\bs X)} 
\prod_{Y\in \CC_{j+1}(X)} \tilde K_j(\f, Y)
\eal
for
\bal
\tilde K_j(\f, Y):=\sum_{X'\in \PP_{j}(Y)}^{\overline  {X'}=Y} 
e^{U_{j}(\f,Y\bs X')} 
\prod_{Y'\in \CC_{j}(X')} K_j(\f, Y')\;.
\eal
If $D$ is a $j+1$-block, and $Z$ a $j+1$-polymer,
define
\bal\lb{4.19bis}
&P_j(\f',\z,D):=e^{U_j(\f,D)}-e^{U_{j+1}(\f',D)+\EE_{j}|D|}\;,
\notag\\
&P^Z_j(\f',\z):=\prod_{D\in\BB_{j+1}(Z)}P_j(\f',\z,D)\;;
\eal
besides, if  $Y$ and $X$ are $j+1$-polymers, define
\bal
\lb{5.9bis}
&R_j(\f',\z,Y):=\tilde K_j(\f,Y)-\sum_{D\in \BB_{j+1}(Y)}J_j(\f',D,Y)\;,
\cr
&R^X_j(\f',\z):=\prod_{Y\in\CC_{j+1}(X)}R_j(\f',\z,Y)
\eal
where $J_j(\f',D,Y)$ contains the extracted terms $\bar Q_j$ and
$Q_j$ (which have support on $j+1$-small and
$j$-small polymers, respectively): if $Y\not\in \SS_{j+1}$, or 
$D\not\subset Y$, $J_j(\f',D,Y)\=0$;
otherwise 
\bal\lb{dfj}
&J_j(\f',D,Y):=\frac{Q_j(\f',Y)}{|Y|_{j+1}}
+\sum_{B\in \BB_j(D)}
\sum_{X\in \SS_j\atop X\supset B}^{\bar X=Y}
\frac{\bar Q_j(\f',X)}{|X|_j}
\cr
&\qquad-\d_{D,Y}\sum_{Y'\in \SS_{j+1}}^{Y'\supset D}
\lft[
\frac{Q_j(\f',Y')}{|Y'|_{j+1}}+\sum_{B\in \BB_j(D)}
\sum_{X\in \SS_j\atop X\supset B}^{\bar X=Y'}
\frac{\bar Q_j(\f',X)}{|X|_j}
\rgt]\;;
\eal
hence $J_j(\f',D,Y)$ is a function of $\{\f'_x\}_{x\in Y^*}$; if
$Y\not\in \SS_{j+1}$; and, by construction, 
\bal\lb{5.23}
\sum_{Y\in \PP^c_{j+1}} J_j(\f',D,Y)=0\;.
\eal
Plugging decompositions \pref{4.19bis} and \pref{5.9bis} 
in \pref{pfj+1} and expanding we find
\pref{pfr2}, for
\bal\lb{j+1}
K_{j+1}(\f', Y')=&\sum_{X_0,X_1\atop  Z, (D) }^{\to Y'}
e^{-\EE_{j}|W|+U_{j+1}(\f',Y'\bs W)}\;\cdot
\cr
&\qquad\cdot   
\EEE_j\lft[P^Z_j(\f',\z)R^{X_1}_j(\f',\z)\rgt]
J_j^{X_0,(D)}(\f')\;.
\eal
In this formula, we abridged $X_0\cup X_1\cup Z$ into $W$. 
And we  
labeled $\to Y'$ the sum  
over three $j+1$-polymers,  $X_0$, $X_1$, $Z$,
and over one $j+1$-block, $D_Y$, for each polymer 
$Y\in \CC_{j+1}(X_0)$, such
that: a) $X_0$ and $X_1$ are separated by at least 
by one $j+1$-block, so that $\CC_{j+1}(X_0\cup X_1)=
\CC_{j+1}(X_0)+ \CC_{j+1}(X_1)$; b) $Z$ is a subset 
of $Y'\bs (X_0\cup X_1)$;
c) each connected 
component of $X_0$ is $j+1$-small; 
d) $\cup_{Y}D^*_Y\cup Z\cup X_1=Y'$.
Finally, given $X_0$ and a
$D_Y$ for each $Y\in \CC_{j+1}(X_0)$, we defined  
$$
J^{X_0,(D)}_j(\f'):=\prod_{Y\in \CC_{j+1}(X_0)} J_j(\f', D_Y, Y)\;.
$$
For \pref{j+1} we also used the factorization of $\EEE_j$ over
sets that are in two different connected components of a $j+1$-polymer
(hence at distance not smaller than
$L^{j+1}$, while the range of $\G_j$ is $L^{j+1}/2$). 
Besides, by construction, $W\subset Y'$
so that that $K_{j+1}(\f',Y)$ depends 
on the fields $(\f'_y)_{y\in Y^*}$ as required. 

Finally, we have to prove that the linear part in $K_j$ and second
order part in $V_j$ of this choice of
$K_{j+1}$ is \pref{5.25}: this follows from 
two simple identities,
\bal\lb{*id1}
&\sum_{D\in \BB_{j+1}(Y')}J_j(\f',D,Y')=Q_j(\f',Y')
+\sum_{X\in\SS_j}^{\bar X=Y'} \bar Q_j(\f',X)
\cr
&\qquad\qquad-
\sum_{D\in \BB_{j+1}}^{D=Y'}\sum_{Y\in \SS_{j+1}}^{Y\supset D}
\frac{Q_j(\f',Y)}{|Y|_{j+1}}
-\sum_{B\in \BB_j}^{\bar B=Y'}
\sum_{X\in \SS_j}^{X\supset B}
\frac{\bar Q_j(\f',X)}{|X|_j}\;;
\eal
and, by \pref{5.23}, 
\bal\lb{*id2}
&\sum_{Y\in \SS_{j+1}}
\sum_{D\in \BB_{j+1}(Y)}^{D^*=Y'}\; J_j(\f',D,Y)
=\sum_{D\in \BB_{j+1}}^{D^*=Y'}
\sum_{Y\in \SS_{j+1}}^{Y\supset D}\; J_j(\f',D,Y)=0\;.
\eal
This completes the proof.
\qedd

%

\section{Estimates}\lb{s5}

\subsection{Norms and Regulators}\lb{nr}
Unless otherwise stated, $j=0,1,\ldots, R-1$.
Given $X\in \PP^c_j$, let  $\CC^2_j(X)$ be 
the vector space of the 
bounded functions $\f:X^*\to \CCC$.
For $n=0,1,2$, define (for $\dpr^\m$ the discrete derivative defined
after \pref{pot})
\be\lb{ln}
\|\nabla^n_j\f\|_{L^\io (X)}:=\max_{\m_1,\ldots,\m_n}\max_{x\in X}
L^{nj}\big|\dpr^{\m_1}\cdots \dpr^{\m_n} \f_x\big|
\ee
and the  norm  
\bal
\|\f\|_{\CC^2_j(X)}:=\max_{n=0,1,2}\|\nabla_j^n\f\|_{L^\io(X^*)}.
\eal
Also, consider $L^2$ norms:
\bal\lb{ln2}
&\|\nabla^n_j\f\|^2_{L^2_j(X)}=
L^{-2j}\sum_{x\in X}\sum_{\m_1,\ldots,\m_n}
L^{2nj}\big|\dpr^{\m_1}\cdots \dpr^{\m_n} \f_x\big|^2\;,
\cr
&\|\nabla^n_j\f\|^2_{L^2_j(\dpr X)}=
L^{-j}\sum_{x\in \dpr X}\sum_{\m_1,\ldots,\m_n}
L^{2nj}\big|\dpr^{\m_1}\cdots \dpr^{\m_n} \f_x\big|^2\;.
\eal
Bounds on expectations of $\CC^2_j$-norms can be 
obtained by these $L^2_j$-norms.
\footnote{By {\it lattice Sobolev lemma} 
((6.136) of \cite{[Br]})  
there exists $c>0$ such that, for any $B\in \BB_j$, 
$$
\|\f\|^2_{L^\io(B)}\le c\sum_{p=0}^2 \|\nabla^p_j\f\|^2_{L^2_j(B)}\;;
$$
therefore, for any $X\in \PP^c_j$ and $n=1,2$, 
$$
\EEE_j\lft[\|\nabla_j^n\z^{(j)}\|^2_{L^\io(X)}\rgt]
\le C |X|_j\;,
\qquad
\EEE_j\lft[\|\z^{(j)}\|^2_{\CC^2_j(X)}\rgt]\le C|X|_j \log L\;.
$$
The powers $L^j$ in definition \pref{ln} are
are chosen to have $C$ independent of $L$ and $j$.}

Let $\NN_j(X)$ be the space of the smooth complex activities of the
polymer $X^*$,
i.e. the set of  $C^\io$ functions $F(\f,X):\CC^2_j(X)\rightarrow \CCC$.
The $n$-order derivative of  $F$ along the directions 
$f_1,\ldots,f_n\in \CC^2_j(X)$ 
is 
$$
D^n_{\f} F(\f,X)\cdot (f_1,\ldots,f_n)
=\sum_{x_1,\ldots,x_n\in X^*}(f_1)_{x_1}\cdots (f_n)_{x_n}
{\dpr^n F\over \dpr \f_{x_1}\cdots\dpr \f_{x_n}}(\f,X)\;.
$$
The size of the differential of order $n$ is given by
\be\lb{tjn}
\|F(\f,X)\|_{T^n_j(\f,X)}=\sup_{\|f_i\|_{\CC^2_j(X)}\le 1}
\big|D^n_{\f} F(\f,X)\cdot (f_1,\ldots,f_n)\big|\;.
\ee
Then, given any  $h>1$, we define the norm
\bal\lb{sn}
&\|F(\f,X)\|_{h,T_j(\f,X)}
=\sum_{n\ge 0}{h^n\over n!}\|F(\f,X)\|_{T^n_j(\f,X)}\;. 
\eal
In order to control the norm of the activities as function of the field $\f$,
for any scale $j$ and any $X\in \PP^c_j$
introduce the {\it field regulators}, $G_j(\f,X)\ge 1$, 
that  depends upon  derivatives of $\f$ only. 
Accordingly, define 
\bal\lb{rn}
\|F(X)\|_{h,T_j(X)}
&=\sup_{\f\in \CC^2_j(X)}{\|F(\f,X)\|_{h,T_j(\f,X)}\over G_j(\f,X)}
\;.
\eal
Finally, we have to weight the polymer activity w.r.t. the size of the
set. Given a parameter $A>1$, define 
\bal\lb{an}
\|F\|_{h,T_j}&=\sup_{X\in \PP_j^c} A^{|X|_j}\|F(X) \|_{h,T_j(X)}\;.
\eal
As already announced,  in this paper, as opposed to \cite{[DH]},
$h$ is  independent  $L$ and of the scale
$j$ (as well as independent of  $\a$, $s$ and  $z$). 
Consider the function 
$h_j(x):=\G_j(x)-\G_j(0)$:
by \pref{p2}, there exists a numerical constant $C_0>0$ such that,
for any $X\in \SS_j$ that contains $0$,  
$\|h_j\|_{\CC^2_j(X)}\le C_0$; then any fixed  $h>1$ that satisfies
\be\lb {hcon}
h\ge 2\a C_0
\ee
will work for our purposes.
 
A convenient choice for the functions $G_j$, 
that guarantees the integrability of the polymer
activities at any scale, is inspired by \cite{[Br]}: 
for $X$ a j-polymer,  define 
\be\lb{6.109}
W_j(\f, X)^2:=\sum_{B\in \BB_j(X)} \|\f\|^2_{L^{\io}(B^*)}\;;
\ee
then, 
given two positive constants $c_1$, $c_3$, 
and a positive function of $L$, $\kappa_L$, 
if $X\in \PP_j^c$
\be\lb{6.69}
\ln G_j(\f,X)=c_1\kappa_L
\|\nabla_j\f\|^2_{L^2_j(X)}+c_3\kappa_L\|\nabla_j\f\|^2_{L^2_j(\dpr X)}
+c_1\kappa_L W_j(\nabla^2_j\f,X)^2\;.
\ee
$\NN_j(X)$ with the norm $\|\cdot\|_{h,T_j(X)}$ is a
Banach space.
For the field dependence of $U_j$ we shall
use the {\it strong field regulator}, $G^{\rm str}_j$: for
$B\in \BB_j$ and $X\in \PP_j$, 
\be\lb{6.66}
\ln G^{\rm str}_j(\f,B)=\kappa_L
\max_{n=1,2}\|\nabla^n_j\f\|^2_{L^\io(B^*)}\;,
\quad
G_j^{\rm str}(\f,X)=\prod_{B\in \BB_{j}(X)}G_j^{\rm str}(\f,B)\;.
\ee
These definitions suits the  Coulomb Gas,  
more than those ones in \cite{[Br]} (Sec. 6.2.4-5.),
designed for the Dipole Gas; the reason is the need of a
{\it finer decomposition} of the covariance suggested in
\cite{[Br2]}; we will discuss this in Sec. \ref{d}.
Anyways the basic structure of the regulators is unchanged,
therefore we still refer to \cite{[Br]} in most of 
the proofs below.

We now list some useful features of the field regulator. 
As apparent from the definition, if $X\in P_{j+1}$, 
\be\lb{nw}
G^{\rm str}_j(\f',X)\le G^{\rm str}_{j+1}(\f',X)\;. 
\ee
For any polymer  $X\in \PP_j$, 
the sets $Y$'s in $\CC_j(X)$
have disjoint boundaries, then the norm $L^2_j(\dpr Y)$ (besides 
the norm $L^2_j(Y)$) is additive in $Y$; hence 
\be\lb{6.51}
\prod_{Y\in \CC_j(X)}G_j(\f,Y)=G_j(\f,X)\;.
\ee
In the following results, borrowed or inspired by \cite{[Br]}, 
$c_3$ and $c_1$ must be large enough, but independent of the
scale $j$ and of  the parameters $L$, $s$ and $z$.  
\begin{lemma}\lb{l6.100b} 
For  $X\in \PP_j$,
\bal\lb{6.100b}
G^{\rm str}_j(\f,X)\le G_j(\f,X)\;.
\eal
For $X\in \PP_j$ and $B$ a $j-$block that is not inside $X$,
\bal\lb{6.52}
G^{\rm str}_j(\f,B)G_j(\f,X)\le G_j(\f,B\cup X)\;.
\eal
\end{lemma} 
\begin{lemma}\lb{l6.53} 
If $\kappa_L=c(\log L)^{-1}$ with $c>0$  and small enough:
\bd
\item{a)} for $X\in \PP_j^c$,
\be\lb{6.54}
\EEE_j \lft[G_j(\f,X)\rgt]\le  2^{|X|_j} G_{j+1}(\f',\bar X)\;;
\ee
\item{b)} if $X\in \SS_j$, for a $C>1$,  
\bal\lb{6.58}
\lft(1+\max_{n=1,2}\|\nabla^n_{j+1}\f'\|_{L^\io(X^*)}\rgt)^3
\EEE_j
\lft[G_{j}(\f,X)\rgt]\le
C {2^{|X|_j}\over \kappa_L^{3/2}}G_{j+1}(\f',\bar X)\;.
\eal
\ed
Besides \pref{6.58} holds also if $G_{j}(\f,X)$ is replaced by 
$\sup_{t\in[0,1]} G_{j}(t\f'+\z,X)$.
\end{lemma}
For the proof of Lemma 
\ref{l6.100b} see Lemma 6.21 and formula (6.52) 
in \cite{[Br]}; for the former we need $c_1>4$, in the latter we
have to assume $c_3<c c_1$, for a certain geometrical
constant $c$.  Lemma \ref{l6.53} has analogies with formulas (6.53) and
(6.58) in \cite{[Br]}, but the proof is a substantial re-shuffling of
the one in that paper, and so
is given in Sec. \ref{d}. It means that 
$\kappa_L={\rm O}(1/\ln L)$,  as opposed to  $\kappa_L={\rm O}(1/L^2)$
of \cite{[DH]}: this makes an important difference, because the
r.h.s. of \pref{6.58} is then proportional to 
${\rm O}(\ln L)^{3/2}$, and is
beaten, for large $L$, by any power of $L$. 

Lemma \ref{l6.53} holds for $j=0,1,\ldots,R-1$. 
Let $\EEE_R$ be the expectation with covariance  $\G_{\ge R}(x;m)$,
followed by the limit $m\to 0$. 
\begin{lemma}\lb{l6.53bis}
If $\kappa_L=c(\log L)^{-1}$ with $c>0$  and small enough,
\be\lb{ex6.53bis}
\EEE_R \lft[G_R(\f,\L)\rgt]\le 2\;.
\ee 
\end{lemma}
We conclude this section with two useful bounds in
\cite{[Br]}. For $\l\in (0,1)$, define
\bal\lb{gld}
&k_s(A,\l)=\sup_{V\in P^c_{j+1}}A^{|V|_{j+1}}
\sum_{Y\in \SS_j}^{\bar Y=V}(\l A)^{-|Y|_j}
\eal
and 
\bal\lb{gld2}
k_l(A,\l)=\sup_{V\in P^c_{j+1}}A^{|V|_{j+1}}
\sum_{Y\in \not\SS_j}^{\bar Y=V}(\l A)^{-|Y|_j}\;.
\eal
\begin{lemma}
There exists $c_s(\l)>0$ such that
\be\lb{6.90}
k_s(A,\l)\le c_s(\l)L^2\;.
\ee
There exist $\h>0$ and  $A_l(\l,L)$ such that, 
if $A> A_l(\l,L)$ then 
\be\lb{6.87}
k_l(A,\l)\le A^{-\h}\;.
\ee
\end{lemma}
In brief, when the sum is over small sets the bound is proportional
to a volume factor $L^2$; when the sum is over large sets, 
the bound is finite in $L$ and vanishing for large $A$. 
For the proof  see Lemma 6.19  and Lemma 6.18 in \cite{[Br]}.

\subsection{Power Counting Analysis}
For the definition set up so far, we can easily derive some
fundamental bounds (mostly introduced in \cite{[Br]}), 
that will be repeatedly used in the rest of the
paper. 
From the definitions one can easily verify the two following facts: 
first, for any $\f\in \CC^2_{j+1}(X)$, we have 
$\|\f\|_{\CC^2_j(X)}\le \|\f\|_{\CC^2_{j+1}(X)}$, so that, 
for any $F\in \NN_j(X)$
\be\lb{6.8}
\|F(\f,X)\|_{h,T_{j+1}(\f,X)}
\le\|F(\f,X)\|_{h,T_{j}(\f,X)} \;;
\ee
second, if $Y\subset X$, for any $\f\in \CC^2_j(X)$ we have
$\|\f\|_{\CC^2_j(Y)}\le \|\f\|_{\CC^2_j(X)}$, so that 
$\CC^2_j(X)\subset \CC^2_j(Y)$ and
\be\lb{6.36}
\|F(\f,X)\|_{h,T_{j}(\f,X)}
\le\|F(\f,X)\|_{h,T_{j}(\f,Y)}\;.
\ee
For any two  polymers $Y_1, Y_2$ not
necessarily disjoint and such
that $Y_1\cup Y_2\subset X$, and any two polymer activities, 
$F_{1}\in\NN_j(Y_1)$ and $F_{2}\in\NN_j(Y_2)$, we have
the {\it triangular inequality}
\be\lb{6.5}
\|F_{1}(\f,Y_1)+F_{2}(\f,Y_2)\|_{h,T_{j}(\f,X)}
\le\|F_{1}(\f,Y_1)\|_{h,T_{j}(\f, Y_1)} 
+ \|F_{2}(\f, Y_2)\|_{h,T_{j}(\f, Y_2)} \;,
\ee
and the {\it factorization property}
\be\lb{6.37}
\|F_{1}(\f,Y_1)F_{2}(\f,Y_2)\|_{h,T_{j}(\f,X)}
\le\|F_{1}(\f,Y_1)\|_{h,T_{j}(\f, Y_1)} 
\|F_{2}(\f, Y_2)\|_{h,T_{j}(\f, Y_2)}\;.
\ee
(The former is a consequence of \pref{6.36} and 
of the triangular inequality for norms; 
the latter follows from
differentiation rules.)
By \pref{6.8}  and \pref{6.54},  
for any $X\in \PP_j^c$ 
\bal\lb{6.4}
\|\EEE_j\lft[K_j(\f,X)\rgt]\|_{h,T_{j+1}(\f',X)}
\le \|K_j\|_{h,T_j}\lft({A\over 2}\rgt)^{-|X|_j}
G_{j+1}(\f',\bar X)\;; 
\eal
and similarly, for each charged component 
\bal\lb{6.4bis}
\|\EEE_j\lft[\hK_j(q,\f,X)\rgt]\|_{h,T_{j+1}(\f',X)}
\le \|K_j\|_{h,T_j}\lft({A\over 2}\rgt)^{-|X|_j}
G_{j+1}(\f',\bar X)\;.
\eal
We passed from scale $j+1$ to scale $j$ by
the bound \pref{6.8} which is of  general validity.
But, under special circumstances, this step can
be done in a more efficient way and the above bounds can be considerably 
improved.

\begin{theorem}\lb{l6.12}
If $L$ is large enough, there exists $\th>0$ such
that, for any $X\in \SS_j$,  
\be\lb{6.62}
\|{\rm Rem}_{2}\;
\EEE_j\lft[\hK_j(0,\f,X)\rgt]\|_{h,T_{j+1}(\f',X)}\le
L^{-(2+\th)}
\|K_j\|_{h,T_j} \lft({A\over 2}\rgt)^{-|X|_j} G_{j+1}(\f',\bar X)\;.
\ee
\end{theorem}
\begin{theorem}\lb{l13}
If $L$ is large enough, $\a^2\ge 8\p$ and $|q|\ge 1$,
then, for any  $X\in \SS_j$,  
\be\lb{119}
\|\EEE_j\lft[\hK_j(q,\f,X)\rgt]\|_{h,T_{j+1}(\f',X)}\le
C(\a) L^{-2|q|}
\|K_j\|_{h,T_j} \lft({A\over 2}\rgt)^{-|X|_j} G_{j+1}(\f',\bar X)\;. 
\ee
\end{theorem}
\begin{theorem}\lb{l13b} 
If $L$ is large enough, $\a^2\ge 8\p$ and  $q=\pm1$, there exists
$\th>0$ such that, for  any $X\in \SS_j$, 
\be\lb{119b}
\|{\rm Rem}_0\ 
\EEE_j\lft[\hK_j(q,\f,X)\rgt]\|_{h,T_{j+1}(\f',X)}\le
L^{-(2+\th)}
\|K_j\|_{h,T_j} \lft({A\over 2}\rgt)^{-|X|_j} G_{j+1}(\f',\bar X) 
\ee
\end{theorem} 
These results make rigorous power counting arguments mentioned below
Lemma \ref{ldec3}. 
Theorem \ref{l13} is borrowed from \cite{[DH]} 
and means that the charged terms - on small sets - 
are {\it irrelevant} as soon as  $L^{-2|q|}$ beats the volume factor 
$L^2$, i.e. for $|q|\ge 2$.
Theorem \ref{l6.12} is taken from \cite{[Br]} and means that 
the second order Taylor remainder - again, on small sets only - 
is an {\it irrelevant term} because the
factor $L^{-3}(\ln L)^{3/2}$ on the r.h.s. member 
beats the volume factor. Theorem \ref{l13b} is a fusion of the
first two: it was missing in \cite{[DH]} and prevented that paper to
cover the case of the Kosterlitz-Thouless line. It means that 
a charged component with $|q|=1$ is irrelevant if the
zeroth order Taylor term has been taken away. 
Although we do not need any essential change respect 
to the formulations in \cite{[DH]} (even in the case of Theorem
\ref{l13b}, once the $L$-independence of $h$,  found in this paper,  
is taken into account)  
we reviewed the  proofs of these fundamental results 
in Sec. \ref{A6}.

\section{Smoothness of RG Map}\lb{s6}
\subsection{Proof of Lemma \ref{t3.1} and \ref{t3.5} }\lb{s6.1}
Given $X\in \SS_j$ and $x_0\in X$, 
set $f^\m_x:=(x-x_0)^\m$. 
\bal\lb{ffj}
|\FF_j(K_j)|&\le 
C \sum_{X\in \SS_j}^{X\ni0}
\frac{L^{-2j}}{|X|_{j}}
\|\EEE_j\lft[\hK_j(0,\z,X)\rgt]\|_{h,T_{j+1}(\f',X)}
\sum_{\m\in\hat e}\|f^\m\|^2_{\CC^2_j(X)}\;;
\eal
as  $\|f^\m\|_{\CC^2_j(X)}\le  CL^j$
and by \pref{6.4bis}, \pref{ffj} is bounded by $C A^{-1}\|K_j\|_{h,T_j}$.
\bal\lb{mmj}
|\MM_j(K_j)|
&\le\sum_{X\in \SS_j}^{X\ni0}
\frac{e^{\frac{\a^2}{2}\G_j(0)}}{|X|_{j}}\sum_{\s=\pm1}
\|\EEE_j\lft[\hK_j(\e,\z,X)\rgt]\|_{h,T_{j+1}(\f',X)}\;,
\eal
which, by \pref{p3} and \pref{119}, is bounded by 
$C(\a) A^{-1}\|K_j\|_{h,T_j}$. Lemma \ref{t3.1} is proven.

By \pref{p1}, \pref{p2}, 
\pref{p3} and as $\dpr^\m\G_j(0)\=0$, for $ |y|\le C(L) L^j$, 
$$
|e^{-\a^2\G_j(0|y)}-1-{\a^2\over 2}
\sum_{\m,\n\in \hat e}(\dpr^\m\dpr^\n \G_j)(0)y^\m y^\n|\le C(L) L^{-3j} |y|^3\;,
$$
$$
e^{-\a^2 \G_j(0)}|e^{\pm\a^2 \G_j(y)}-1|\le C(L,\a) |\G_j(y)|\;,\qquad
|e^{-\a^2\G_j(0|y)}-1|\le C(L,\a) |y| L^{-j}\;,
$$
$$
|(\dpr^\m\dpr^\n \G_j)(0|y)|\le C(L) |y| L^{-3j}\;,
$$
so that (using also the bounds \pref{c13}, \pref{c14})
\be\lb{sD}
|\mce_{3,j}|\le C(L)\;,
\qquad |\mce_{2,j}|\le C\;, \qquad
|\mce_{4,j}|\le C(L,\a)\;,
\ee
\be\lb{sab}
|a_j|\le C(L,\a)\;,\qquad |b_j|\le C(L,\a)\;.
\ee
Finally
\bal\lb{sA}
|\mce_{1,j}(K_j)|&\le  \sum_{X\in \SS_j}^{X\ni0}
\frac{1}{|X|_{j}}
\|\EEE_j\lft[\hK_j(0,\z,X)\rgt]\|_{h,T_{j+1}(\f',X)}\;,
\eal
which, by \pref{6.4bis}, is bounded by $C A^{-1}\|K_j\|_{h,T_j}$.
Finally, a bound for  $\EE_{R}$ is
$$
|\EE_R|\le \frac{1}{|\L|} \ln \EEE_{R}
\lft[1+\|e^{U_R(\z,\L)}-1\|_{h, T_R(\z,\L)}+\|K_R(\z,\L)\|_{h,T_R(\z,\L)}\rgt]\;;
$$
therefore \pref{r2} with $j=R$ follows from \pref{ex6.53bis}.
Proof of Lemma \ref{t3.5} is complete.

\subsection{Proof of Theorem \ref{t3.2}}\lb{s6.4}
Taking into account cancellation \pref{eW}, 
$$
(\LL_j K_j)(\f',V)
=\sum_{p=1}^5(\LL^{(p)}_j K_j)(\f',V)
$$
for (Taylor expansions are in $\d \f'$)
\bal
(\LL^{(1)}_j K_j)(\f',V)
&:=\sum_{Y\in\not\SS_j(V)}^{\bar Y=V}\EEE_j[K_j(\f,Y)]
\lb{el1}
\\
(\LL^{(2)}_j K_j)(\f',V)
&:=\sum_{q:|q|\ge 2}
\sum_{Y\in \SS_j(V)}^{\bar Y=V}
\EEE_j\Big[\hK_j(q,\f,Y)\Big]
\lb{el2}
\\
(\LL^{(3)}_j K_j)(\f',V)
&:=\sum_{Y\in \SS_j(V)}^{\bar Y=V}{\rm Rem}_2\;
\EEE_j\Big[\hK_j(0,\f,Y)\Big]
\\
(\LL^{(4)}_j K_j)(\f',V)
&:=\sum_{q=\pm1}\sum_{Y\in \SS_j(V)}^{\bar Y=V}
{\rm Rem}_{0}\;
\EEE_j\Big[\hK_j(q,\f,Y)\Big]
\\
(\LL^{(5)}_j K_j)(\f',V)
&:=-\sum_{B\in \BB_j(V)}^{\bar B=V}
\Big[V_{j+1}(\bar s_j,\bar z_j,\f',B)
-\EEE_j\lft[V_{j}(\f,B)\rgt]
\cr
&\qquad\qquad\qquad\qquad\qquad
+\bar \EE_{j}|B| 
-\sum_{X\in \SS_j}^{X\supset B}
\frac{\bar Q(\f',X)}{ |X|_j}\Big]
\eal
Consider each term separately:
we want to prove that 
\be\lb{contr}
\|\LL^{(p)}_j K_j\|_{h,T_{j+1}}\le \r(A,L) \|K_j\|_{h,T_j}
\ee
for $p=1,\ldots ,5$; the constant  $\r(A,L)$ is to be  
small enough if  $L$ and $A$ are large enough, 
uniformly in  the scale $j$.
\subsubsection{Norm of $\LL^{(1)}$} Use \pref{6.5}, 
\pref{6.4} and definition \pref{gld2} to find
\bal
&\|(\LL^{(1)}_j K_j)(\f',V)\|_{h,T_{j+1}(\f',V)}
\le  \sum_{Y\in\not\SS_j}^{\bar Y=V} 
\|\EEE_j\lft[K_j(\f,Y)\rgt]\|_{h,T_{j+1}(\f',Y)}
\cr
&\qquad\qquad\qquad\qquad
\le  G_{j+1}(\f',V)\|K_j\|_{h,T_j}
\sum_{Y\in\not\SS_j}^{\bar Y=V} A^{-|Y|_j}2^{|Y|_j}
\cr
&\qquad\qquad\qquad\qquad
\le  G_{j+1}(\f',V)A^{-|V|_{j+1}}\;\|K_j\|_{h,T_j}\; 
k_l(A,1/2)\;;
\eal
by \pref{6.87}, we find \pref{contr} for 
$\r(A,L)\ge A^{-\h}$.
\subsubsection{Norm of $\LL^{(2)}$} Use \pref{119} 
 and definition \pref{gld} to find
\bal
&\|(\LL^{(2)}_j K_j)(\f',V)\|_{h,T_{j+1}(\f',V)}
\le \sum_{q:|q|\ge 2}
\sum_{Y\in \SS_j(V)}^{\bar Y=V}
\|\EEE_j\Big[\hK_j(q,\f,Y)\Big]\|_{h,T_{j+1}(\f',Y)}
\cr
&\qquad\qquad
\le   C(\a)G_{j+1}(\f',V)A^{-|V|_{j+1}}
\;\|K_j\|_{h,T_j} 
\;k_s(A,1/2)\sum_{q:|q|\ge 2}
L^{-2|q|}\;;
\eal
by \pref{6.90} we obtain \pref{contr} for $\r(A,L)\ge C(\a)L^{-2}$.
\subsubsection{Norm of $\LL^{(3)}$} By \pref{6.62}
\bal
&\|(\LL^{(3)}_j K_j)(\f',V)\|_{h,T_{j+1}(\f',V)}
\le \sum_{Y\in \SS_j(V)}^{\bar Y=V}
\|{\rm Rem}_{2}\EEE_j\Big[\hK_j(0,\f,Y)\Big]\|_{h,T_{j+1}(\f',Y)}
\cr
&\qquad\qquad
\le  G_{j+1}(\f',V)A^{-|V|_{j+1}}\;\|K_j\|_{h,T_j}\; 
L^{-(2+\th)}k_s(A,1/2)\;;
\eal
by \pref{6.90} we obtain \pref{contr} for $\r(A,L)\ge CL^{-\th}$.
\subsubsection{Norm of $\LL^{(4)}$} By \pref{119b},
\bal
&\|(\LL^{(4)}_j K_j)(\f',V)\|_{h,T_{j+1}(\f',V)}
\le 
\sum_{q=\pm1}\sum_{Y\in \SS_j(V)}^{\bar Y=V}
\|{\rm Rem}_{0}\EEE_j\Big[\hK_j(q,\f,Y)\Big]\|_{h,T_{j+1}(\f',Y)}
\cr
&\qquad\qquad
\le   G_{j+1}(\f',V)A^{-|V|_{j+1}}\;\|K_j\|_{h,T_j}\; 
L^{-(2+\th)}k_s(A,1/2)\;.
\eal 
by \pref{6.90} we obtain \pref{contr} for $\r(A,L)\ge CL^{-\th}$.
\subsubsection{Norm of $\LL^{(5)}$}
Here we prove that \pref{irb} is irrelevant as claimed before 
\pref{irb}. Further decompose this term as 
$$
(\LL^{(5)}_jK_j)(\f',V):=(\LL^{(5,a)}_jK_j)(\f',V)
+(\LL^{(5,b)}_jK_j)(\f',V)\;,
$$
where 
\bal
&(\LL^{(5,a)}_jK_j)(\f',V)=\sum_{B\in \BB_j(V)}^{\bar B=V}
\Bigg[
\sum_{X\in \SS_j}^{X\supset B}\sum_{x_0\in X}
\frac{{\rm Tay}_{2}\EEE_j\lft[\hK_j(0,\f,X)\rgt]}{|X| |X|_j}
\cr
&\qquad\qquad\qquad\qquad\qquad\qquad\quad
-V_{j+1}(\FF_j(K_j),0,\f',B) -\mce_{1,j}(K_j)\Bigg]
\eal
\bal
&(\LL^{(5,b)}_jK_j)(\f',V)=\sum_{B\in \BB_j(V)}^{\bar B=V}
\Bigg[
\sum_{X\in \SS_j}^{X\supset B}\sum_{q=\pm}\sum_{x_0\in X}
\frac{e^{iq\a \f'_{x_0}}\EEE_j\lft[\hK_j(q,\z,X)\rgt]}{|X| |X|_j}
\cr
&\quad\qquad\qquad\qquad\qquad\qquad\qquad\quad
-V_{j+1}(0,L^2 e^{-{\a^2\over 2}\G_j(0)}\MM_j(K_j),\f',B) \Bigg]
\eal
Consider $\LL^{(5,a)}_j$. The term proportional to $\mce_{1,j}$ 
exactly cancels the one proportional to the 
zero order of ${\rm Tay}_2\EEE_j\lft[\hK_j(0,\f,X)\rgt]$. We are left 
with the difference between $V_{j+1}(\FF_j(K_j),0,\f',B)$ 
and the 2-nd order of  ${\rm Tay}_2\EEE_j\lft[\hK_j(0,\f,X)\rgt]$ (the
first being zero by symmetry),
that, by invariance under space translations 
of the problem, yields 
\bal
(\LL^{(5,a)}_jK_j)(\f',V)=\sum_{B\in \BB_j(V)}^{\bar B=V}
\sum_{X\in\SS_j}^{X\supset B}
\sum_{x_0\in X,\atop x_1,x_2\in X^*}
\frac{R^{x_0}_{x_1,x_2}(\f')}{2|X||X|_j}
\EEE_j\lft[{\dpr^2 \hK_j\over \dpr\f_{x_1}\dpr\f_{x_2}}(0,\z,X)\rgt]
\eal
where $R^{x_0}_{x_1,x_2}(\f')$ is given by
\bal
&
(\f'_{x_2}-\f'_{x_0})(\f'_{x_1}-\f'_{x_0})
-L^{-2j}\sum_{x\in B}\sum_{\m,\n\in\hat e}
(\dpr^\m\f')_x(\dpr^\n\f')_x(x_2-x_0)^\m(x_1-x_0)^\n
\notag\\
&=
L^{-2j}
\sum_{x\in B}\Big[u_{x_1,x_0}(x,\f')u_{x_2,x_0}(x,\f')
+v_{x_1,x_0}(x,\f')u_{x_2,x_0}(x,\f')
\notag\\
&\qquad\qquad\qquad\qquad\qquad\qquad\qquad\qquad\qquad
+ u_{x_1,x_0}(x,\f')v_{x_2,x_0}(x,\f')\Big]
\eal
and  
\bal
&u_{x_1,x_0}(x,\f):=\f_{x_1}-\f_{x_0}-
\sum_{\m\in\hat e}(\dpr^\m\f)_x(x_1-x_0)^\m
\notag\\
&v_{x_1,x_0}(x,\f):=\sum_{\m\in\hat e}(\dpr^\m\f)_x(x_1-x_0)^\m\;.
\eal
Notice that $|x_0-x_1|,\; |x_1-x_2|\le C L^{j}$; it implies
\bal
\|u_{x_1,x_0}(x,\f')\|_{h,T_{j+1}(\f,X)}\le C
L^{-2}(1+\max_{n=1,2}\|\nabla_{j+1}^n\f'\|_{L^\io(X^*)})
\notag\\
\|v_{x_1,x_0}(x,\f')\|_{h,T_{j+1}(\f,X)}\le C
L^{-1}(1+\max_{n=1,2}\|\nabla_{j+1}^n\f'\|_{L^\io(X^*)})
\eal
and then a bound for
$\|(\LL^{(5,a)}_jK_j)(\f',V)\|_{h,T_{j+1}(\f',V)}$ is 
\bal
&\qquad\le C \|K_j\|_{h,T_j}\sum_{B\in \BB_j(V)}^{\bar B=V}
\sum_{X\in\SS_j}^{X\supset B}
\lft({A\over 2}\rgt)^{-|X|_j}
\sup_{x_0,x_1,x_2}\|R^{x_0}_{x_0,x_1,x_2}(\f')\|_{h,T_{j+1(\f',V)}}
\notag\\
&\qquad\le C'   \|K_j\|_{h,T_j}
A^{-|V|_{j+1}}
L^{-1}
\lft(1+\max_{n=1,2}\|\nabla_{j+1}^n\f'\|^2_{L^\io(V^*)}\rgt)
\notag\\
&\qquad\le C''  \kappa_L^{-1}L^{-1} \|K_j\|_{h,T_j} G^{\rm str}_{j+1}(\f',V)
A^{-|V|_{j+1}}
\eal
(the sums over $B$ and $X$ have no more than $CL^2$ terms, in
all of which $|V|_{j+1}=1$).   $G^{\rm str}_{j+1}(\f',V)$ can be
replaced by $G_{j+1}(\f',V)$ by \pref{6.100b}.
Next consider $\LL^{(5,b)}_j$. Simple computations and symmetry 
$\hat K_j(-1,\z,X)=\hat K_j(1,\z,X)$, give
\bal
&(\LL^{(5,b)}_jK_j)(\f',V)=\sum_{B\in \BB_j(V)}^{\bar B=V}
L^{-2j}\sum_{x_0\in B\atop q=\pm1}e^{iq\a \f'_{x_0}}\cdot
\cr
&\qquad\qquad
\cdot\sum_{X\in \SS_j}^{X\supset B}
\sum_{y_0\in X}
\frac{\lft(e^{iq\a (\f'_{y_0}-\f'_{x_0})}-1\rgt)}
{2|X| |X|_j}\sum_{\s=\pm1}\EEE_j\lft[\hK_j(\e,\z,X)\rgt]\;.
\eal
Notice that by construction $|x_0-y_0|\le C L^{j}$ so that, 
by \pref{6.100b}, 
\bal
\|e^{iq\a (\f'_{y_0}-\f'_{x_0})}-1\|_{h,T_{j+1}(\f',X)}
&\le C(\a)
L^{-1}(1+\max_{n=1,2}\|\nabla_{j+1}^n\f'\|_{L^\io(X^*)})
\cr
&\le 
C'(\a)\kappa_L^{-1}L^{-1}G_{j+1}(\f',V)
\eal
and then, using \pref{119}, a bound for
$\|(\LL^{(5,b)}_jK_j)(\f',V)\|_{h,T_{j+1}(\f',V)}$ is 
\bal
&C(\a)\kappa_L^{-1}L^{-1}G_{j+1}(\f',V)
\sum_{\s=\pm}
\sum_{B\in \BB_j(V)}^{\bar B=V}
\sum_{X\in \SS_j}^{X\supset B}
\frac{1}{ |X|_j}
\|\EEE_j\lft[\hK_j(\e,\z,X)\rgt]\|_{h,T_{j+1}(\f',V)}
\notag\\
&\qquad\qquad\le 
C'(\a)\kappa_L^{-1}L^{-1}G_{j+1}(\f',V) 
A^{-|V|_{j+1}}\|K_j\|_{h,T_j}\;.
\eal
Again we obtain \pref{contr} 
for $\r(A,L)\ge C(\a)\kappa_L^{-1}L^{-1}$.

\subsection{Preparation Bounds} We begin with 
bound (6.74) of \cite{[Br]}. 
\begin{lemma}\lb{l6.74}
There exists $\h>0$
such that, for any $X\in P_j$, 
\be\lb{6.74}
(1+2\h)|\bar X|_{j+1}\le |X|_j+8(1+2\h)|\CC_j(X)|\;.
\ee
\end{lemma}
In the next section we shall need 
bounds on the ``building  blocks'' 
of the RG map.
Let us introduce a vector 
notation and a  norm with weight  $\m\ge 0$, 
\bal\lb{vec}
\KK_j:=(s_j,z_j, K_j)\;, 
\qquad
|\KK_j|_{\m,j}:=\m^{-1}|s_j|+ \m^{-1}|z_j|+\m^{-2}\|K_j\|_{h,T_j}\;.
\eal
\begin{lemma}\lb{l6.1}
There exist $\h>0$,  $C\=C(A,L,\a)>1$  
such that, for $\e_0$ small enough, 
any $\KK_j$ and $\dot \KK_j$ with 
$|\KK_j|_{1,j}, |\dot \KK_j|_{1,j}\le \e_0$ 
and  $D\in \BB_{j+1}$ or $Y\in \PP_{j+1}^c$ 
\bal\lb{dif-}
&\|e^{U_{j}(\f,D)}-e^{\dot U_{j}(\f,D)}\|_{h,T_j(\f',D)}
\le C \e_0|\KK_j-\dot{\KK}_j|_{\e_0,j}
G_j^{\rm str}(\f,D)\;,
\\\cr
\lb{dif0}
&\|e^{U_{j+1}(\f',D)}-e^{\dot U_{j+1}(\f',D)}\|_{h,T_j(\f',D)}
\le C \e_0|\KK_j-\dot \KK_j|_{\e_0,j}
G_{j+1}^{\rm str}(\f',D)\;,
\\\cr
\lb{dif1}
&\|P_j(\f',\z,D)-\dot P_j(\f',\z,D)\|_{h,T_j(\f',D)}
\cr\cr
&\qquad
\le
C\e_0|\KK_j-\dot \KK_j|_{\e_0,j} A^{-(1+\h)}
\lft[G_{j+1}^{\rm str}(\f',D)+G_j^{\rm str}(\f,D)\rgt]\;,
\\\cr
\lb{dif3}
&\|J_j(\f',D,Y)-\dot J_j(\f',D,Y)\|_{h,T_j(\f',Y)}
\cr\cr
&\qquad
\le C \e_0^2
|\KK_j-\dot \KK_j|_{\e_0,j} 
A^{-(1+\h)|D^*|_{j+1}} G^{\rm str}_{j+1}(\f',D)\;,
\\\cr
\lb{dif2}
&\|R_j(\f',\z,Y)-\dot R_j(\f',\z,Y)\|_{h,T_j(\f',Y)}
\cr\cr
&\quad
\le C 
\e_0^2|\KK_j-\dot \KK_j|_{\e_0,j}  
A^{-(1+\h)|Y|_{j+1}}\lft[G^{\rm str}_{j+1}(\f',Y)+G_j(\f,Y)\rgt]\;,
\eal
where $\dot U_{j}$, $\dot U_{j+1}$, $\dot P_j$, $\dot R_j$ and $\dot J_j$
contains parameters $\dot \KK_j=(\dot s_j, \dot z_j, \dot K_j)$ instead of 
$\KK_j=(s_j, z_j, K_j)$.
\end{lemma}
\brm
Although the $T_{j+1}(\f',Y)$ norm requires a decay in the size 
of $Y$ as  $A^{-|Y|_{j+1}}$, in the r.h.s. of 
\pref{dif1}, \pref{dif3} and \pref{dif2}
the decay is (basically) $A^{-(1+\h)|Y|_{j+1}}$; this is because 
the extra factor $A^{-\h|Y|_{j+1}}$ 
is needed to control 
the sum in \pref{j+1} (\cfr Sec. \ref{pt3}). In 
\pref{dif1} and \pref{dif3},  $|Y|_{j+1}$ is bounded and  
a $A^{-\h|Y|_{j+1}}$ factor 
can be extracted from $C(A,L,\a)$; 
in \pref{dif2}, instead, we shall need Lemma \ref{l6.74} 
to have such an improvement.   
\erm
 
\bpr
Let  $B\in \BB_j(D)$ and use \pref{6.5} and \pref{6.37} to obtain
\bal\lb{bvv}
&\|V_j(\f,B)-\dot V_j(\f,B)\|_{h,T_j(\f',B)}
\le |s_j-\dot s_j| \lft(h+\|\nabla_j\f\|_{L^\io(B)}\rgt)^2
+2|z_j-\dot z_j| e^{\a h}
\cr
&\qquad \le C(\a) \Big[|s_j-\dot s_j|+ |z_j-\dot z_j|\Big] 
\lft(1+\max_{n=1,2}\|\nabla_j^n\f\|^2_{L^\io(B^*)}\rgt)\;.
\eal
A similar bound holds for $W_j$ that, by support property of the
propagator, depends on fields at $B^*$ 
(more details are in Appendix \ref{sC.3}): 
\bal\lb{bww}
&\|W_{j}(\f,B)-\dot W_{j}(\f,B)\|_{h,T_j(\f',B)}
\cr
&\qquad \le C(\a) \Big[|s_j-\dot s_j|+ |z_j-\dot z_j|\Big] A^{-1}_* 
\lft(1+\max_{n=1,2}\|\nabla_j^n\f\|^2_{L^\io(B^*)}\rgt)\;.
\eal
From the definition of $ G_j^{\rm str}$
\be\lb{id1}
\sum_{B\in \BB_{j}(D)}
\lft(1+\max_{n=1,2}\|\nabla_j^n\f\|^2_{L^\io(B^*)}\rgt)
\le C(L) \;G_j^{\rm str}(\f,D)^{\frac13}\;;
\ee
and for $\e_0$ smaller than a $\e_0(L,\a)$,
\be\lb{id2}
\prod_{B\in \BB_{j}(D)}
\exp\lft\{C(\a) \e_0\lft(1+\max_{n=1,2}\|\nabla_j^n\f\|^2_{L^\io(B^*)}\rgt)\rgt\}
\le 2G_j^{\rm str}(\f,D)^{\frac13}\;.
\ee
Then \pref{dif-} follows
by factorization property \pref{6.52}  
and these inequalities. 
\pref{dif0} is obtained by repeating \pref{bvv} and \pref{bww} on
scale $j+1$; and by taking into account that, 
by \pref{sk}, \pref{lk} and the bounds in 
Sec. \ref{s6.1},  if  $A$ is large enough,
\bal\lb{inc0}
\max\lft\{|\EE_{j}-\dot \EE_{j}|L^{2j+2}\;,
|s_{j+1}-\dot s_{j+1}|\;,
|z_{j+1}-\dot z_{j+1}|\rgt\}
\le C(L,\a) |\KK_j-\dot \KK_j|_{1,j}&\;.
\eal
Accordingly, for $\e_0$ smaller than a $\e_0(L,\a)$,
\bal\lb{4bvv}
\|e^{U_{j+1}(\f',D)+\EE_{j}|D|}
-e^{\dot U_{j+1}(\f',D)+\dot\EE_{j}|D|}\|_{h,T_j(\f',D)}
\cr
\le 
C(L,\a) 
|\KK_j-\dot \KK_j|_{1,j}
G_j^{\rm str}(\f',D)\;,
\eal
which, together to  \pref{dif-}, proves \pref{dif1}. 
Next, consider \pref{dif3}. 
By \pref{qbr},
for any $X\in \SS_j$ that contains a given 
$B\in \BB_j(D)$ (if $L>8$ then $X^*\subset D^*$)
\bal\lb{bq}
\|\overline Q_j(\f',X)-\dot{\overline Q}_j(\f',X)\|_{h,T_j(\f',X)}
\le C(\a) A^{-|X|_{j}} \|K-\dot K\|_{h,T_j}G^{\rm str}_{j+1}(\f',D)\;;
\eal
then, as $|D^*|\le 4S$,  
replace  $C(\a) A^{-|X|_{j}}$ with 
$C(A,\a) A^{-(1+\h)|D^*|_{j+1}}$. 
Besides, for any $Y\in \SS_{j+1}$ that contains a given block 
$D\in \BB_{j+1}$, by the second order computation in 
in Appendix \ref{sc2},
\bal\lb{bbq}
&\|Q_j(\f',Y)-\dot{Q}_j(\f',Y)\|_{h,T_j(\f',Y)}
\cr
&\le C(L,\a)\e_0\Big[|s_j-\dot s_j|+|z_j-\dot z_j|\Big] 
G^{\rm str}_{j+1}(\f',D)\;;
\eal
again we replace $C(L,\a)$ with  
$C(A,L,\a) A^{-(1+\h)|D^*|_{j+1}}$ 
as $|D^*|_{j+1}\le 4S$.
Together, \pref{bq} and \pref{bbq} prove \pref{dif3}.
Bounds so far have been straightforward consequences of the properties
of the norms. As announced, bound \pref{dif2} is
slightly more sophisticated because requires Lemma \ref{l6.74}.
\bal
&\tilde K_j(\f,Y)-\dot {\tilde K}_j(\f,Y)
=\sum_{X\in \PP^c_j}^{\overline X=Y}
\lft[U_j(\f,Y\bs X)-\dot U_j(\f,Y\bs X)\rgt]\prod_{Z\in  \CC_j(X)} K_j(\f,Z)
\cdot\notag\\
&\qquad\qquad\qquad\qquad\qquad\qquad\qquad\qquad
\cdot \int_0^1dt\; \lft[e^{tU_j(\f,Y\bs X)+(1-t)\dot U_j(\f,Y\bs X)}\rgt]
\notag\\
&
+ \sum_{X\in \PP^c_j}^{\overline X=Y}
e^{\dot U_j(\f,Y\bs X)}
\sum_{X_0\in ((X))_j}
\prod_{Z\in \CC_j(X_0)}\lft[K_j(\f,Z)-\dot K_j(\f,Z)\rgt]
\prod_{Z\in  \CC_j(X\bs X_0)} \dot K_j(\f,Z)
\eal
where $ ((X))_j$ is the family of sets made of unions of connected
parts of $X$, empty set excluded. Therefore,  using the factorization
property \pref{6.52} and previous bounds
\bal\lb{pb}
&\|\tilde K_j(\f,Y)-\dot {\tilde K}_j(\f,Y)\|_{h,T_j(\f,Y)}
\notag\\
&\quad
\le
\Big[|s_j-\dot s_j|+ |z_j-\dot z_j|\Big]
G_j(\f,Y)
\sum_{X\in \PP^c_j}^{\overline X=Y}
C(\a)^{|Y\bs X|_j} A^{-|X|_j}\e_0^{|\CC_j(X)|}
\notag\\
&\quad\qquad
+
\e_0^{-1}\|K_j-\dot K_j\|_{h,T_j}
G_j(\f,Y)
\sum_{X\in \PP^c_j}^{\overline X=Y}
2^{|Y\bs X|_j}4^{|C_j(X)|} A^{-|X|_j}\e_0^{|\CC_j(X)|}
\notag\\
&\quad
\le \e_0 |\KK_j-\dot \KK_j|_{\e_0,j}\; G_j(\f,Y)\;
C'(\a)^{|Y|_j}\sum_{X\in \PP^c_j}^{\overline X=Y}
A^{-|X|_j}\lft(4\e_0\rgt)^{|\CC_j(X)|}\;
\eal
In order to extract the factor $A^{-(1+\h)|Y|_{j+1}}$ and, at the same
time, to control the sum over $X$, which is made of no more than 
$2^{|Y|_j}=2^{L^2|Y|_{j+1}}$ terms, use
\pref{6.74} and obtain
\bal\lb{ineq}
A^{-|X|_j}A^{-8(1+2\h)|C_j(X)|}
\le A^{-(1+2\h)|Y|_{j+1}}\;.
\eal
Therefore, for $4A^{8(1+2\h)}\e_0<1$, bound \pref{pb} 
is smaller than   
\bal 
A^{-(1+\h)|Y|_{j+1}} \e_0 |\KK_j-\dot \KK_j|_{\e_0,j}\; 
G_j(\f,Y)\;
\lft(C(\a)^{L^2}A^{-\h}\rgt)^{|Y|_{j+1}}
4A^{8(1+2\h)}\e_0\;;
\eal
finally, 
for $A$ large enough, $C(\a)^{L^2}A^{-\h}\le 1$ and
choosing 
$C(A,\a,L)\ge 4A^{8(1+2\h)}$, 
\bal\lb{bk}
&\|\tilde K_j(\f,Y)-\dot {\tilde K}_j(\f,Y)\|_{h,T_j(\f',Y)} 
\notag\\
&\qquad\qquad
\le
C(A,L,\a) A^{-(1+\h)|Y|_{j+1}} 
\e^2_0|\KK_j-\dot \KK_j|_{\e_0,j}\;
G_j(\f,Y)\;.
\eal
To conclude, \pref{dif2} follows from \pref{dif3} and \pref{bk}.
This completes the proof. \qedd 
\epr
\subsection{Proof of Theorem \ref{t3.4}}\lb{pt3}
We need an explicit formula for $\RR_j(s_j,z_j,K_j)$. 
Expanding \pref{j+1} we obtain: 
\bal\lb{j+1xp}
&K_{j+1}(\f', Y')=  
\EEE_j\lft[P^{Y'}_j(\f',\z)\rgt]+ 
\lft(e^{-\EE_{j}|Y'|}-1\rgt) 
\EEE_j\lft[P^{Y'}_j(\f',\z)\rgt] 
\notag\\
&\qquad\qquad+\sum_{\CC_{j+1}(X_0\cup X_1)\ge 1\atop }^{\to Y'}  
\EEE_j\lft[P^Z_j(\f',\z)R^{X_1}_j(\f',\z)\rgt]
J_j^{X_0,(D)}(\f') 
\notag\\
&\qquad\qquad
+\sum_{\CC_{j+1}(X_0\cup X_1)\ge 1\atop }^{\to Y'}
\lft(e^{-\EE_{j}|W|+U_{j+1}(\f',W\bs Y')}-1\rgt)\;   
\cdot 
\notag\\
&\qquad\qquad\qquad\qquad\qquad\qquad
\cdot\EEE_j\lft[P^Z_j(\f',\z)R^{X_1}_j(\f',\z)\rgt]
J_j^{X_0,(D)}(\f')\;.
\eal
Next, we have to remove the part $\LL_j K_j$: to this purpose, 
further expand the first term in the  r.h.s. of \pref{j+1xp}
\bal
\sum_{D\in \BB_{j+1}}^{D= Y'}  
\EEE_j\lft[P_j(\f',\z,D)\rgt]
&+{1\over 2}
\sum_{D_0, D_1\in \BB_{j+1}\atop D_0\neq D_1}^{D_0\cup D_1= Y'}  
\EEE_j\lft[P_j(\f',\z,D_0)P_j(\f',\z,D_1)\rgt]
\cr\cr
&+\sum_{Z\in P_{j+1}\atop |Z|_{j+1}\ge 3}^{Z=Y'} 
\EEE_j\lft[P^{Z}_j(\f',\z)\rgt]\;;
\eal
and also further expand the second line of the r.h.s. member 
of \pref{j+1xp}
\bal
&\EEE_j\lft[R_j(\f',\z,Y')\rgt]
+\sum_{Y\in \SS_{j+1}}\sum_{D\in \BB_{j+1}(Y)}^{D=Y'}  
J_j(\f',D,Y)
\cr\cr
&
+\sum_{\CC_{j+1}(X_0\cup X_1)\ge 1
\atop|Z|_{j+1}+\CC_{j+1}(X_0\cup X_1)\ge 2}
^{\to Y'}  
\EEE_j\lft[P^Z_j(\f',\z)R^{X_1}_j(\f',\z)\rgt]
J_j^{X_0,(D)}(\f')\;.
\eal
Grouping together the above decompositions and definition \pref{5.25},
we obtain
\bal
&\RR_j(\f',Y'):=K_{j+1}(\f',Y')-(\LL_j K_j)(\f',Y')=
\sum_{s=1}^6\RR^{(s)}_{j}(\f',Y')
\eal
where the eight summands are 
\bal
&\RR^{(1)}_{j}(\f',Y'):=\sum_{D\in \BB_{j+1}}^{D= Y'}  
\Bigg[\EEE_j\lft[P_j(\f',\z,D)\rgt]
+\EE_{j}|D|+U_{j+1}(\f',D)
\cr
&\qquad\qquad\qquad\qquad\qquad\qquad\quad\qquad
-\EEE_j\lft[U_j(\f,D)\rgt]-{1\over 2}
\EEE^T_j\lft[V_j(\f,D);V_j(\f,D)\rgt]\Bigg]
\notag\\\cr
&
\RR^{(2)}_{j}(\f',Y'):=\sum_{D\in \BB_{j+1}}^{D= Y'} 
\Bigg[\EEE_j\lft[W_{j}(s_{j+1},z_{j+1},\f,D)\rgt]
-
\EEE_j\lft[W_{j}(s_j,z_j,\f,D)\rgt]\Bigg]
\notag\eal
\bal
&
\RR^{(3)}_{j}(\f',Y'):={1\over 2}
\sum_{D_0, D_1\in \BB_{j+1}\atop D_0\neq D_1}^{D_0\cup D_1= Y'}  
\Bigg[\EEE_j\lft[P_j(\f',\z,D_0)P_j(\f',\z,D_1)\rgt]
\notag\\
&\qquad\qquad\qquad\qquad\qquad\qquad\qquad\qquad
\qquad\qquad\qquad
-
\EEE^T_j\lft[V_j(\f,D_0);V_j(\f,D_1)\rgt]\Bigg]
\notag\eal
\bal
&
\RR^{(4)}_{j}(\f',Y'):=\sum_{\CC_{j+1}(X_0\cup X_1)\ge 1
\atop|Z|_{j+1}+\CC_{j+1}(X_0\cup X_1)\ge 2}
^{\to Y'}  
\EEE_j\lft[P^Z_j(\f',\z)R^{X_1}_j(\f',\z)\rgt]
J_j^{X_0,(D)}(\f')
\notag\eal
\bal
&
\RR^{(5)}_{j}(\f',Y'):=\sum_{\CC_{j+1}(X_0\cup X_1)\ge 1\atop }^{\to Y'}
\lft(e^{-\EE_{j}|W|+U_{j+1}(\f',Y'\bs W)}-1\rgt)\;   \cdot
\notag\\
&\qquad\qquad\qquad\qquad\qquad\qquad\qquad\qquad
\cdot\EEE_j\lft[P^Z_j(\f',\z)R^{X_1}_j(\f',\z)\rgt]
J_j^{X_0,(D)}(\f')
\notag\\\cr
&
\RR^{(6)}_{j}(\f',Y'):=
\lft(e^{-\EE_{j}|Y'|}-1\rgt) 
\EEE_j\lft[P^{Y'}_j(\f',\z)\rgt]
+\sum_{Z\in P_{j+1}\atop |Z|_{j+1}\ge 3}^{Z=Y'} 
\EEE_j\lft[P^{Z}_j(\f',\z)\rgt]
\notag\\\cr
&
\RR^{(7)}_{j}(\f',Y'):=
\sum_{X'\in P_j\atop \CC_j(X')\ge 2}^{\bar {X'}=Y'}
\EEE_j\lft[e^{U_j(\f, Y'\bs X')}\prod_{Y\in \CC_j(X')}K_j(\f,Y)\rgt]
\notag\\\cr
&
\RR^{(8)}_{j}(\f',Y'):=
\sum_{X'\in P_j}^{\bar {X'}=Y'}
\EEE_j\lft[\lft(e^{U_j(\f, Y'\bs X')}-1\rgt)
\prod_{Y\in \CC_j(X')}K_j(\f,Y)\rgt]\;.
\eal
Notice that, for $j=0$, we instead have 
\bal
&\RR_{0}(\f',Y'):=\sum_{D\in \BB_{1}}^{D= Y'}  
\Bigg[\EEE_0\lft[P_0(\f',\z,D)\rgt]
+\EE_{1}|D|+V_{1}(\f',D)-\EEE_1\lft[V_1(\f,D)\rgt]\Bigg]
\cr&\qquad\qquad\qquad
+
\lft(e^{-\EE_{1}|Y'|}-1\rgt) 
\EEE_j\lft[P^{Y'}_0(\f',\z)\rgt]
+\sum_{Z\in P_{1}\atop |Z|_{1}\ge 2}^{Z=Y'} 
\EEE_j\lft[P^{Z}_j(\f',\z)\rgt]
\eal
Then, Lemma \ref{t3.4} is a direct consequence of the following result.
\begin{lemma}
There exists $C(A,L,\a)>1$  
such that, for $\e_0$ small enough, 
any $\KK_j$ and $\dot \KK_j$ with 
$|\KK_j|_{\e_0,j}, |\dot \KK_j|_{\e_0,j}\le 1$ 
and  $Y'\in \PP_{j+1}^c$: if j=0,  
\bal\lb{0pr}
\|\RR_0-\dot {\RR}_0\|_{h,T_0}
\le C(A,L,\a) \e_0
\Big[|s-\dot s|+|z-\dot z|\Big]\;;
\eal
while, if $j=1, \ldots, R-1$, 
\bal\lb{pr}
\|\RR^{(m)}_{j}-\dot {\RR}^{(m)}_{j}\|_{h,T_j}
\le C(A,L,\a) \e_0^3
|\KK_j-\dot{\KK}_j |_{\e_0,j}\;
\eal
for any $m=1,2,\ldots,8$.
\end{lemma}
\brm
Many details of the construction of $K_{j+1}$ out of $K_j$ and $U_j$, 
\pref{j+1},  are designed with the goal of having \pref{0pr}, \pref{pr}.
In particular, the preliminary  re-blocking of $K_j$, with support on
$j$-polymers, into $\tilde K_j$,  with support on  $j+1$-polymers - a
step not done in \cite{[Br]} -  seems necessary here to obtain 
\pref{ggg2} form \pref{ggg} through the new integration property 
\pref{6.54}.
\erm
\bpr
First consider $\RR^{(4)}_{j}$ that is the most
instructive term.
By Lemma \ref{l6.1}, its bound is made of 
three kind of factors: a product of field regulators; a product of
$A^{-1}$'s;
a product of $\|\KK_j\|_{h,T_j}$'s 
and $C(A,L,\a)$'s. The field regulators stemming from
the $P_j$ and $R_j$ factors can be merged together by Lemma
\ref{l6.100b}: 
given disjoint $Z, X_1\in \PP_{j+1}$, 
\bal\lb{ggg}
&
\prod_{D\in \BB_{j+1}(Z)} \lft[G^{\rm str}_j(\f,D)
+ G^{\rm str}_{j+1}(\f',D)\rgt]
\prod_{Y\in \CC_{j+1}(X_1)} \lft[G_j(\f,Y)
+ G^{\rm str}_{j+1}(\f',Y)\rgt]
\cr\cr
&\le 
\sum_{W_1\in \PP_{j+1}(Z)\atop
W_2\in ((X_1))_{j+1}} 
G^{\rm str}_{j+1}(\f',Z\bs W_1)\; 
G^{\rm str}_{j+1}(\f',X_1\bs W_2)\;
G_j(\f,W_1\cup W_2)
\eal
where $((X_1))_{j+1}$ is the collections of subsets of $X_1$ made of
unions of connected parts of $X_1$. There are no more than 
$2^{|\CC_{j+1}(X_1)|+|Z|_{j+1}}$ terms in the sum above.
Then take the expectation 
$\EEE_j$ in 
\pref{ggg} using \pref{6.54}, and 
merge also the (strong) field regulators stemming from the bounds 
for the $J_j$'s factors, to obtain the upper 
bound for \pref{ggg}
\be\lb{ggg2}
2^{|\CC_{j+1}(X_1)|+|Z|_{j+1}} 
2^{L^2|X_1\cup Z|_{j+1}}G_{j+1}(\f',\cup_{Y}D_Y\cup X_1\cup Z) \;.
\ee
Next, collect all the  $A^{-1}$ factors from the bounds of 
$P_j$'s,  $R_j$'s and $J_j$'s to obtain
$A^{-(1+\h)|Y'|_{j+1}}$. Finally, since the variation in the vector
$\KK_j$ can occur in each of the factors, that are no more than
$|Y'|_{j+1}$,  we obtain:  
\bal\lb{r4r}
&\|\RR_{j}^{(4)}(\f',Y')-
\dot \RR_{j}^{(4)}(\f',Y')\|_{h,T_j(\f',Y')}
\cr\cr
&
\le  G_{j+1}(\f',Y') A^{-(1+\h)|Y'|_{j+1}}
|\KK_j-\dot \KK_j|_{\e_0,j}
|Y'|_{j+1}2^{L^2|Y'|_{j+1}}
\;\cdot
\cr\cr
&\quad
\cdot
\sum_{\CC_{j+1}(X_0\cup X_1)\ge 1
\atop|Z|_{j+1}+\CC_{j+1}(X_0\cup X_1)\ge 2}^{\to Y'}  
\lft[2C(A,L,\a)\e_0\rgt]^{|Z|_{j+1}+2|\CC_{j+1}(X_0\cup X_1)|}\;.
\eal 
Notice the following facts: 
the number of term in the sum is not bigger 
than $4^{|Y'|_{j+1}} 4^{|\CC_{j+1}(X_0)|}$; if $\e_0$ is small enough, 
$8 C(A,L,\a)\e_0<1$; by the constraint on 
the sum there is a least factor $[8C(A,L,\a)\e_0]^3$ 
in each summand; finally, 
for $A$ large enough, $(A^{-\h}2^{(L^2+2)})^{|Y'|_{j+1}} 
|Y'|_{j+1}<1$. Therefore, for $C'(A,L,A)>[8C(A,L,\a)]^3$,
 an upper bound for \pref{r4r} is 
\bal
 C'(A,L,\a)  \e_0^3|\KK_j-\dot \KK_j|_{\e_0,j}
G_{j+1}(\f',Y')A^{-|Y'|_{j+1}}\;.
\eal
This proves \pref{pr} for $m=4$. 
Now consider $\RR^{(5)}_j$. It  can be recasted as 
\bal\lb{r5}
&\quad\sum_{\CC_{j+1}(X_0\cup X_1)\ge 1\atop}^{\to Y'}
\sum_{W_0\in \PP_j(W)\atop W_1\in \PP_j(Y'\bs W)}
^{\atop |W_0\cup W_1|_{j+1}\ge 1}
\lft(e^{-\EE_{j}L^{2j+2}}-1\rgt)^{|W_0|_{j+1}}
J_j^{X_0,(D)}(\f')\;\cdot   
\cr\cr
&
\cdot\EEE_j\lft[P^Z_j(\f',\z)R^{X_1}_j(\f',\z)\rgt]
\prod_{D_1\in \BB_{j+1}(W_1)}\lft(e^{U_{j+1}(\f',D_1)}-1\rgt)\;.
\eal
Bound \pref{pr} for $m=5$ follows, mostly,  
as for $m=4$; so we just  stress   one detail: 
after field regulators are multiplied and integrated,  
we have to merge strong field regulators originated by
the bound on (possibly, the variation of) $J_j$'s,  but also  
$e^{-U_{j+1}(\f',D_1)}-1$'s;  
still, by construction,  the final bound for the field regulators 
is $G_{j+1}(\f', Y')$, as wanted. 

Bounds \pref{pr} for $m=7,8$ can be obtained in the
very same fashion as \pref{dif2}; while cases $m=1,2,3,6$ are even 
simpler than the previous ones. This completes the proof of Lemma
\ref{t3.4}.  \qedd
\epr

\section{Stable Manifold Theorem} \lb{s.7}
In this discussion,  $\a^2=8\p$. 
Then, a natural way to recast \pref{lk} is
\bal\lb{blk}
&s_{j+1}=s_j-a z_j^2+\FF_j(s_j, z_j,K_j)
\cr
&z_{j+1}=z_j-b s_jz_j+\MM_j(s_j, z_j, K_j)
\cr
&K_{j+1}=\LL_j (K_j) + \RR_j(z_j, s_j, K_j)
\eal
for $a\=a(L)$ and $b\=b(L)$ given in Lemma \ref{t3.0}, and 
\bal
&\FF_j(s_j,z_j, K_j):=\FF_j(K_j)-(a_j-a) z_j^2\;,
\cr
&\MM_j(s_j,z_j,K_j):=L^2e^{-\frac{\a^2}{2}\G_j(0)}
\MM_j(K_j)
+\lft[L^2e^{-\frac{\a^2}{2}\G_j(0)}-1\rgt]z_j
\cr
&\qquad\qquad\qquad\qquad
-\lft[L^2e^{-\frac{\a^2}{2}\G_j(0)}b_j-b\rgt] s_jz_j\;.
\eal
The flow equation for $j=0$ is simpler and given in \pref{0lk}. 
We want to redefine the  
dynamical system so that it begins at step $1$ instead of step
$0$ (and so to absorb the constants $a$ and $b$). 
To this purpose, define $(x_j, y_j, W_j)_{j\ge1}$
such that: $(x_1, y_1, W_1):=(b s_1, \sqrt{ab}z_1, 0)$, while, 
for $j\ge2$,   $(x_j, y_j, W_j):=(b s_j,\sqrt{ab}z_j,K_j)$. Accordingly 
the  flow equation  becomes
\bal\lb{lk8}
&x_{j+1}-x_j= -y_j^2+\tilde \FF_j(x_j,y_j,W_j)
\cr
&y_{j+1}-y_j= -x_jy_j+\tilde \MM_j(x_j,y_j,W_j)
\cr
&W_{j+1}=\LL_j (W_j) + \tilde \RR_j(x_j, y_j, W_j)\;.
\eal
with, for $j=1$ and $\l=\lft[L^2e^{-\frac{\a^2}{2}\G_0(0)}\rgt]^{-1}$, 
\bal
&\tilde \FF_1(x_1,y_1, W_1)
:=b\FF_1(s_1,z_1,\RR_0(s_1,\l z_1))\;,
\cr
&\tilde \MM_j(x_1,y_1, W_1)
:=\sqrt{ab}\MM_1(s_1,z_1,\RR_0(s_1,\l z_1))\;,
\cr
&\tilde \RR_1(x_1,y_1, W_1):=\LL_1(\RR_0(s_1,\l z_1))
+\RR_1(s_1,z_1,\RR_0(s_1,\l z_1))\;;
\eal
while,  for $j\ge2$,  
\bal
&\tilde \RR_j(x_j,y_j,W_j)
:=\RR_j(s_j,z_j,K_j)\;, 
\cr
&\tilde \FF_j(x_j,y_j,W_j)
:=b\FF_j(s_j,z_j,K_j)\;,
\cr
&\tilde \MM_j(x_j,y_j,W_j)
:=\sqrt{ab}\MM_j(s_j,z_j,K_j)\;.
\eal
Then, equations \pref{0lk} and \pref{lk} for the RG flow 
are equivalent to \pref{lk8}. 
Note the following consequences of Lemmas \ref{t3.0}, \ref{t3.1},
\ref{t3.2} and \ref{t3.4}: if  
$|x_j|,|y_j|, |\dot x_j|,|\dot y_j|\le \e_j$
and $\|W_j\|_{h,j}, \|\dot W_j\|_{h,j}\le  \e_j^2$, 
\bal\lb{aaF}
&|\tilde \FF_j(x_j,y_j,W_j)-\tilde \FF_j(\dot x_j,\dot y_j,\dot W_j)|
\le C(L) L^{-\frac j4}|y^2_j-\dot y_j^2|
\cr&\qquad\qquad
+ C(L)A^{-1}\Big[\|W_j-\dot W_j\|_{h,T_j}+\d_{j,1}\e_1(|x_1-\dot x_1|
+|y_1-\dot y_1|)\Big]
\\\cr\lb{aaM}
&|\tilde \MM_j(x_j,y_j,W_j)-\tilde \MM_j(\dot x_j,\dot y_j,\dot W_j)| 
\le c L^{-\frac j4}|y_j-\dot y_j|+ C(L)L^{-\frac j4}|x_jy_j-\dot x_j\dot y_j|
\cr&\qquad\qquad
+ C(L)A^{-1}\Big[\|W_j-\dot W_j\|_{h,T_j}+\d_{j,1}\e_1(|x_1-\dot x_1|
+|y_1-\dot y_1|)\Big]
\\\cr
\lb{aaR}
&|\tilde \RR_j(x_j,y_j,W_j)-\tilde \RR_j(\dot x_j,\dot y_j,\dot W_j)|
\le c (L^{-\th} +A^{-\h})\d_{j,1}\e_1(|x_1-\dot x_1|
+|y_1-\dot y_1|)
\cr&\qquad
+ C(A,L)\Big[\e_j^2|x_j-\dot x_j|
+\e_j^2|y_j-\dot y_j|+\e_j\|W_j-\dot W_j\|_{h,T_j}\Big]
\eal
The RG equations that we are describing here are well defined, and so
considered,  in ``infinite volume'', \ie for any $j\ge 1$ and 
with polymer activities defined 
on $\ZZZ^2$. 
Theorem \ref{t.sm} (which is about the RG on finite lattice $\L$ and
with $j=1, \ldots, R$) 
will be  a corollary of the
following result - see the remark after the proof. Hypotheses 
slightly more general than what we need, since
$W_1$ is not assumed to be 0. 
\begin{theorem}\lb{t.6} 
Consider the solution of \pref{lk8} with
initial data $(x_1, y_1, W_1)$. If $L$ and $A$ are large enough,  
if  $|y_1|\le \e_1$ for  a small $\e_1\=\e_1(A,L)$ and if 
$\|W_1\|_{h,T_1}$ is small enough w.r.t $|y_1|$,  
there exist $x_1=\Sigma(y_1)$  
such that  $|x_1|\le 2\e_1$ and 
\be
x_j=|q_j|+O(j^{-\frac32})\;,\qquad
y_j=q_j+O(j^{-\frac32})\;,\qquad
\|W_j\|_{h,T_j}=O(j^{-3})\;,
\ee
where $q_j$ is the same defined in \pref{dq} with $q_1:=y_1$.
\end{theorem}
\bpr Without loss of generality, assume $y_1$ positive.  
The strategy - partially 
inspired to \cite{[BS]} - is to mimic the treatment of the continuous
version of \pref{lk8}. 
We look for a solution $(x_j, y_j, W_j)$ 
of \pref{lk8} into the form 
$x_j=q_j+u_j$, $y_j=q_j+ v_j$,  where
$u_j$, $v_j$, $W_j$ are unknowns with initial data
$u_1=x_1-y_1$, $v_1=0$ and $W_1$.
By the identity  $q_{j+1}-q_j=-q_j q_{j+1}$, 
the  equations for $u_j$ and $v_j$ are   
\bal\lb{lk8b}
&u_{j+1}-u_j=-2v_jq_j 
+ \UU_j(u_j, v_j, W_j)
\cr
&v_{j+1}-v_j=-u_jq_j -v_jq_j 
+\VV_j(u_j, v_j, W_j)
\eal
for $\UU_j(u_j, v_j, W_j):=-v_j^2-q_j^2 q_{j+1} +\tilde \FF_j(x_j,y_j,W_j)$
and $\VV_j(u_j, v_j, W_j):= -u_jv_j-q_j^2 q_{j+1}
+\tilde \MM_j(x_j,y_j,W_j)$. In order to diagonalize the linear part, 
introduce the {\it stable direction}, 
$w^+_j:=u_j+2v_j$, and the {\it unstable direction},
$w^-_j:=u_j-v_j$, that have initial data $w^+_1=w^-_1=x_1-y_1$ 
and satisfy
\bal\lb{lk9}
&w^+_{j+1}-w^+_j=(-2 q_{j+1}+q^2_{j+1})w^+_j 
+ \WW^+_j(u_j, v_j, W_j)
\cr
&w^-_{j+1}-w^-_j=q_jw^-_j 
+ \WW^-_j(u_j, v_j, W_j)
\eal
for $\WW^+_j:=\UU_j+2\VV_j+[2q_j-q_{j+1}]q_{j+1} w^+_j$ and $\WW^-_j:=\UU_j-\VV_j$
(changing the linear order of the former equation into 
$(-2 q_{j+1}+q^2_{j+1})w^+_j$, up to a correction $O(q^2_j)w^+_j$ 
that is absorbed into
$\WW^+_j$, will make easier the next step). 
Since $q_j$
is strictly positive, $w^+_j$ is driven to zero by the linear term in
any case; whereas  $w^-_j$ converges to
zero only for a special initial $w^+_1=w^-_1=x_1-y_1$, to be found.
With some simple algebra, turn \pref{lk9},
together to the condition  $w^-_\io=0$, into a form that resembles the integral 
equation for the continuous flow:
\bal\lb{lk10}
&w^+_{j}=\lft(\frac{q_{j}}{q_1}\rgt)^2w^+_{1}
+\sum_{s=1}^{j-1}\lft(\frac{q_{j}}{q_{s+1}}\rgt)^2
\WW^+_s(u_j, v_j, W_j)
\cr
&w^-_{j}=-\sum_{s\ge j}\frac{q_{s+1}}{q_j}
\WW^-_s(u_j, v_j, W_j)
\cr
&W_j=\LL_{j-1,1} W_1+\sum_{s=1}^{j-1}\LL_{j-1,s+1}
\WW^0_s(u_j, v_j, W_j)
\eal
for $\LL_{n,m}:=\LL_{n}\circ \LL_{n-1}\circ\cdots
\circ\LL_{m}$ and $\WW^0_j(u_j, v_j, W_j):=\tilde \RR_j(x_j,y_j,K_j)$.
(As  in standard conventions, when a counter  
ranges over an empty set, the corresponding summation is zero, 
the corresponding chain of convolutions is the identity.) 

Introduce a vector
notation, $\underline w=(w^+_j,w^-_j,W_j)_{j\ge 1}$,  and the norm
$$
\|\underline w\|=\sup_{j\ge 1} 
\max\{(\t h_j)^{-1}|w^+_j|,\;2(\t h_j)^{-1}|w^-_j|,
\;(\t h_j)^{-2}\|W_j\|_{h,T_j}\}\;,
$$
where $h_j$ is the sequence
$h_j=y_1[1+y_1 (j-1)]^{-\frac{3}{2}}$ and $\t>0$. 
It is easy to see that 
$\WWW=\{\underline w\;:\; \|\underline w\|\le 1\}$ is a Banach
space.
Then define the operator 
$\TT=(\TT_j^+,\TT_j^-, \TT^0_j)_j$ such that 
\pref{lk10} reads $\underline w=\TT \underline w$, \ie
\be\lb{lk11}
w^+_{j}=\TT^+_j(\underline w)\;, \qquad
w^-_{j}=\TT^-_j(\underline w)\;, \qquad
W_{j}=\TT^0_j(\underline w)\;.
\ee
We shall prove that, if $\t$ is
small enough (independently of $A$ and $L$),  
$\TT$ is a contraction of $\WWW$; then 
$\underline w=\TT \underline w$ has  a unique solution in $\WWW$.

Define $p_j:=2^{-(j-1)}$ and the operators 
$$
T^+_j(\WW)
:=\sum_{s=1}^{j-1}\lft(\frac{q_{j}}{q_{s+1}}\rgt)^2 \WW_s\;,
\qquad
T^-_j(\WW):=-\sum_{s\ge j}\frac{q_{s+1}}{q_j}\WW_s\;.
$$
For $\a=\pm$, 
\bal\lb{bT+}
|T^\a_j (h^2)|\le C h_j \;, \qquad 
|T^\a_j (pq)|\le C  h_j \;.
\eal
By definition of $\WW^\a_j$, for $\a=\pm$, if $L$ is large enough,
\bal
&|\WW^\a_j(u_j, v_j, W_j)|
\le C h_j^2 [\t^2+q_1] +C(L)\lft[p_j q_j^2 + 
A^{-1} 
h^2_j\rgt]+  C  L^{-\frac14}p_j q_j\;;
\eal
therefore, using \pref{bT+}, 
\bal\lb{e11}
|T_j^\a(\WW^\a)|
&\le  C h_j \Big[\t^2+q_1+L^{-\frac14} +  C(L)(A^{-1}+q_1)\Big]\;.
\eal
Now consider two vectors in $\WWW$, 
$\underline w=(u_j,v_j, W_j)_j$ and 
$\underline {\dot w}=(\dot u_j,\dot v_j, \dot W_j)_j$; 
then 
\bal
&|\WW^\a_j(u_j, v_j, W_j)- \WW^\a_j(\dot u_j, \dot v_j,\dot {W}_j)|
\cr
&\qquad\qquad
\le
C\|\underline w -\underline {\dot w}\|\t
\lft[ h^2_j [\t +q_1]
+C(L)A^{-1} h^2_j
+  L^{-\frac14}p_jq_j\rgt]\;;
\eal
hence, by \pref{bT+},
\bal\lb{e12}
|T_j^\a(\WW^\a)-T_j^\a(\dot \WW^\a)|&\le
C \|\underline w -\underline {\dot w}\|\t h_j
\lft[\t +q_1+L^{-\frac14} +C(L) A^{-1}\rgt]\;.
\eal
Finally, from \pref{e11} and \pref{e12}, 
assuming  $\t$ small enough  
so that $16C\t\le1$, $L$ large enough so that $ 16CL^{-\frac14}\le \t$,
$A$ large enough so that $16 C(L)A^{-1}C\le \t$ and finally
$q_1=|y_1|$ small enough so that $16C(L)q_1C\le \t$,  
\bal\lb{prt+}
|T_j^\a(\WW^\a)|
\le \frac{\t h_j}{2} 
\qquad
|T_j^\a(\WW^\a)-T_j^\a(\dot \WW^\a)|
\le\|\underline w -\underline {\dot w}\|\frac {\t h_j}{4}\;.
\eal
Now consider  the third of \pref{lk10}. For a polymer activity
$\WW_j$, define
$$
T^0_j(\WW)=\sum_{s=1}^{j-1} \LL_{j-1, s+1} \WW_s\;.
$$
Suppose that $\|\WW_j\|_{h,T_j}\le h^2_j$; 
if $(L^{-\th} + A^{-\h})$ is small enough 
(\cfr Lemma \ref{t3.2}), 
\bal\lb{bT0}
\|(T_0 Q)_j\|_j\le C h_j^2\;.
\eal 
By \pref{aaR}
\bal
\|\WW^0_j(u_j, v_j, W_j)\|_{h,T_{j+1}}&\le 
C\Big[L^{-\th}+A^{-\h}+ C(A,L)q_1\Big]h_j^2\;,
\eal
and then 
\bal
\|T_j^0(\WW^0)\|_{h,T_{j}}&\le 
C\Big[L^{-\th}+A^{-\h}+ C(A,L)q_1\Big]h_j^2\;,
\eal
Besides,  given   two vectors in $\WWW$, 
$\underline w=(u_j,v_j, W_j)_j$ and 
$\underline {\dot w}=(\dot u_j,\dot v_j, \dot W_j)_j$,  
\bal
&\|\WW^0_j(u_j, v_j, W_j)- 
\WW^0_j(\dot u_j, \dot v_j,\dot W_j)\|_{h,T_{j+1}}
\cr
&\qquad\qquad
\le \t \|\underline w-\underline{\dot w}\|
 C\lft[L^{-\th}+A^{-\h} + C(A,L) q_1\rgt]h^2_j\;.
\eal
and using  \pref{bT0} again, 
\bal
\|T_j^0(\WW^0)- T_j^0(\dot{\WW}^0)\|_{h,T_{j}}
\le \t \|\underline w-\underline{\dot w}\|
 C\lft[L^{-\th}+A^{-\h} + C(A,L) q_1\rgt]h^2_j\;.
\eal
For $L$ and $A$ large enough, $4C[L^{-\th}+A^{-\h}]\le \t^2$; 
then, for  $q_1$ small enough, $4C(A,L,\a)q_1C<\t^2$; then 
\bal\lb{prt0}
\|T_j^0(\WW^0)\|_{h,T_{j}}&\le 
(\t h_j)^2\;,\qquad
\|T_j^0(\WW^0)- T_j^0(\dot{\WW}^0)\|_{h,T_{j}}
\le \frac{(\t h_j)^2}{2} \|\underline w-\underline{\dot w}\|\;.
\eal
Finally, consider the operator $\TT$. By \pref{prt+} and \pref{prt0}, 
if $\underline  w\in \WWW$, 
\bal
|\TT_j^+(\underline w)|\le \lft(\frac{q_j}{q_1}\rgt)^2 |w^-_1|
&+|T_j^+(\WW^+)|\le \t h_j\;,
\quad
|\TT_j^-(\underline w)|\le 
|T_j^-(\WW^-)|\le \frac{\t h_j}{2}\;,
\notag\\
&\|\TT_j^0(\underline w)\|_{h,T_j}\le 
\|T_j^0(\WW^0)\|_{h,T_j}\le (\t h_j)^2\;;
\eal
and, if $\underline  w,  \dot{\underline w}\in \WWW$,
with similar derivation 
\bal
&|\TT_j^+(\underline w)-\TT_j^+(\dot{\underline w})|
\le \frac{\t h_j}{2}\|\underline  w- \dot{\underline w}\|\;,
\quad
|\TT_j^-(\underline w)-\TT_j^-(\dot{\underline w})|
\le\frac{\t h_j}{4}\|\underline  w- \dot{\underline w}\|\;,
\notag\\
&\qquad\qquad\qquad\|\TT_j^0(\underline w)-\TT_j^0(\dot{\underline w})\|_{h,T_j}
\le \frac{(\t h_j)^2}{2}\;.
\eal
Hence $\TT$ is a contraction on $\WWW$ with norm $1/2$. 
The Theorem is proven.
\qedd
\epr
\brm
Using the same choice $\Si(y_1)$ for $x_1$ found above in the
infinite-volume case, the very same bounds for the decay of $x_j, y_j, W_j$ 
are valid also for the finite-volume RG flow. 
In fact,  the flows of  $x_j$ and $y_j$ are unchanged, 
for $\tilde F_j$ and $\tilde \MM_j$
are built out  of $W_j$'s living on small sets only. The flow of
$W_j$,  instead, is changed, but only when the support of $W_j$
is a polymer that wraps around the torus $\L$ - 
a non-small at any scale $j=1,2,\ldots, R-1$: 
since $W_j$ is anyways a stable (\ie contracting) direction of 
the flow, this change does not require a different initial datum
$(x_1,y_1)$. Therefore,  Theorem \ref{t.6}
indirectly implies Theorem \ref{t.sm}.
\erm
\appendix
\section{Functional Integral Formula}
\subsection{Siegert-Kac transformation}\lb{a1}
We begin with stressing two properties of the lattice Yukawa potential 
in two dimensions, \pref{yuk}.
First, the {\it self-energy} of a particle, 
\bal\lb{as2}
W_\L(0;m)
&=
\frac{1}{|\L|}\sum_{p\in \L^*}
{1\over
m^2-\hat\D(p)}
\eal
is positive divergent. Next,  if we define 
the {\it normalized interaction}
$W_\L(x|0;m):=W_\L(x;m)-W_\L(0;m)$, we find  
\be\lb{as1}
W_\L(x|0):=\lim_{m\to 0}W_\L(x|0;m)=
\frac{1}{|\L|}\sum_{p\in \L^*\bs \{0\}}
{e^{ipx}-1\over
-\hat\D(p)}\;;
\ee 
then, in the infinite volume, 
$W(x|0):= \lim_{\L\to \io}W_\L(x|0)$ 
has  large $|x|$ behavior
\be\lb{as3}
W(x|0)=-{1\over 2\p} \ln |x|+O(1)\;,
\ee
%
which is what is called the Coulomb potential in two-dimensions.
Now consider \pref{eneb}
written in terms of  normalized interactions and self-energies, 
\be\lb{A1}
H_\L(\o;m)
={1\over 2}\sum_{i,j=1}^n\s_i\s_j W_\L(x_i-x_j|0;m)
+{1\over 2}\lft(\sum_{i=1}^n\s_i\rgt)^2W_\L(0;m)\;.
\ee
As consequence of \pref{as1} and \pref{as3}, 
in the limit $m\to 0$, 
the latter term of \pref{A1} assign an infinite energy, \ie a zero
probabilistic weight, to configurations of $\O_n$ that are not
globally neutral. For neutral configurations, instead, the latter term
is zero and the particle interacts through the  normalized potential.
This justifies the use of the Yukawa potential 
as regularization of the Coulomb one.
Notice that  $W_\L(x;m)$ 
is positive definite and thus 
$e^{-H_\L(\o;m)}\le 1$; then, in the formula for the partition
function,  the sum on $n$ can be taken before the limit 
$m\to 0$,  for the resulting series is convergent  
uniformly in $m$:
\be\lb{gf2}
Z_\L(\b)=\lim_{m\to 0}\sum_{n\ge 0}
{z^n\over n!}
\sum_{\o\in \O_n}
e^{-\b H_\L(\o;m)}\;.
\ee
Now we introduce the Siegert-Kac construction.
Since $W_\L(x;m)$ is strictly-positive definite, introduce at any
site $x\in \L$ a random variable, or 'field', $\f_x$,
with 
Gaussian measure $dP_{R,0}(\f;m)$ determined by 
$$
\int\!dP_{R,0}(\f;m)\;
\f_x=0\;,\qquad 
\int\!dP_{R,0}(\f;m)\;
\f_x\f_y=W_{\L}(x-y;m)\;.
$$
Accordingly, Boltzmann weights of a given configuration in $\O_n$ 
can be re-written as characteristic functions of the Gaussian measure   
$$
\int\!dP_{R,0}\lft(\frac{\f}{\sqrt\b};m\rgt)\;
\exp\Big(
i\sum_{j=1}^n \s_j\f_{x_j}\Big)=e^{-\b H_\L(\o;m)}\;.
$$
 Notice that at this point of the 
analysis the Gaussian measure is
finite-dimensional and non-degenerate; 
then its density has an explicit formula. For RG analysis purposes, 
we want to consider a part of the Gaussian density as an integrand: 
for any $s\in [0,1/2]$, if
$\a^2:=\b(1-s)$ and replacing $m$ with $\frac{m}{\sqrt{1-s}}$,
\be\lb{A4}
dP_{R,0}\lft(\frac{\f}{\sqrt\b};\frac{m}{\sqrt{1-s}}\rgt)
=dP_{R,0}\lft(\frac{\f}{\a};m\rgt) 
\exp\Big\{{s\over 2\a^2}\sum_{x\in \L\atop \m\in\hat e}(\dpr^\m\f_x)^2\Big\}
\NN^{1/2}_\L(s;m) 
\ee
where $\NN_\L$ takes into account the different normalization of the two
measures
\be
\NN_\L(s;m)=\prod_{k\in \L^*}
{m^2-(1-s)\hat\D(k)\over m^2-\hat\D(k)}\;.
\ee
With some algebra, the functional
  integral representation of   
the partition function of the Coulomb Gas is 
\be\lb{fi}
Z_\L(\b)=\lim_{m\to 0}
\int\!dP_{R,0}(\f;m)\; e^{\VV(\f)}
\ee
for $\VV$ given by \pref{pot}. To obtain \pref{fif} we need a
decomposition of the fields.
\subsection{Multiscale decomposition.}
Suppose there exist positive-definite functions
$\G_{0}(x;m), \G_1(x;m), \ldots,\G_{R-1}(x;m)$ and  $\G_{\ge R}(x;m)$ 
such that
\be\lb{dec}
W_\L(x;m)=\sum_{j=0}^{R-1}\G_{j}(x;m)+ \G_{\ge R}(x;m)\;.
\ee
By standard theory of (finite
dimensional) Gaussian processes, $\f$ can be decomposed into
$R+1$ random variables, 
$$
\f_x=\z^{(R)}_x+\z^{(R-1)}_x+\cdots+\z^{(0)}_x
$$ 
that are independent and Gaussian:
\be\lb{fact} 
dP_{R,0}(\f;m)=dP_{\ge R}(\z^{(R)};m)\;
dP_{R-1}(\z^{(R-1)};m)\;\cdots dP_{0}(\z^{(R-1)};m)
\ee
for $dP_j$  and $dP_{\ge R}$ Gaussian measures with
covariances $\G_j$ and $\G_{\ge R}$, respectively. 

Decomposition \pref{dec} can be obtained in many ways. Here 
we take advantage of a similar decomposition 
for $\bar W(x;m)$, the massive covariance
for the {\it infinite} lattice, $\ZZZ^2$, derived in \cite{[BGM]}, 
but implemented on a finer set of scales (\cfr Sec. \ref{d}). 
Suppose $L=\g^M$, for $\g$  and $M$ positive integers; $\g$ odd. The
decomposition of the covariance that we shall consider 
is done on scales $\g^h$, for integer
$h>0$.  Define, for $j=0,\ldots,R-1$, 
\be
I_{j,j+1}:=\{j+m\log_L\g\;|\;m=0,1\ldots,M-1\}\;;
\ee
also, define $Q_h:=[-\p\g^h,\p\g^h]$ and $Q:=Q_0$.
\begin{theorem} There exists a positive-definite $\bar \G_{\ge R}(x;m)$
and, for any $j=0,1,\ldots, R-1$ and any $s\in I_{j,j+1}$, 
a positive-definite $C_{s}(x;m)$ 
such that
\be\lb{fdc}
\bar W(x;m)=\sum_{j=0}^{R-1}\bar \G_{j}(x;m)+ \bar\G_{\ge R}(x;m)\;,
\quad
\bar \G_{j}(x;m)=\sum_{s\in I_{j,j+1}} C_{s}(x;m)\;,
\ee
with $C_s(x;m)=0$ if $|x|\=\max\{x_0,x_1\}> \g^{sM+1}/2$. Besides, for any 
$h/M\in I_{j,j+1}$, 
there exists $F_{h}(p;m)$, 
defined for $p\in Q_h$,  such that, if 
$h=sM$, 
\bal\lb{A10}
&C_{s}(x;m)=\int_{Q}{d^2p\over (2\p)^2}\;
{e^{ipx}\over
  m^2-\hat\D(p)}\lft[F_{h}(\g^hp;m)-F_{h+1}(\g^{h+1}p;m)\rgt]\;, 
\cr\cr
&\bar \G_{\ge R}(x;m)=\int_{Q}{d^2p\over (2\p)^2}\;
{e^{ipx}\over
  m^2-\hat\D(p)}F_{MR}(\g^{MR}p;m)\;.
\eal
\end{theorem}
The proof (without the finer scales, \ie 
for $\g$ replaced by $12 L$) is in \cite{[BGM]}. 
Notice that, as opposed to the full
interactions $W_\L(x;m)$ or $\bar W(x;m)$ that are 
{\it strictly} positive-definite, 
the $C_s$'s might be positive-definite only,  
i.e. the corresponding Gaussian measure
might be degenerate. Anyways notice that, for example, if
$C_s=g_s\circ g_s$ for  positive-definite $g_s$, and if $\EEE_C$ 
is the expectation w.r.t. the Gaussian measure 
with covariance $C(x-y)$, 
then, for any integrable function $F$, 
\be\lb{dG}
\EEE_{C_s}[F(z)]=\EEE_I[F(g_s\circ z)]
\ee
where $I(x-y):=\d_{x,y}$ and
hence {\it strictly} positive-definite (see App. of \cite{[Di]}).

There is a standard way to construct the periodic 
$W_\L(x;m)$ out of its
infinite volume version $\bar W(x;m)$,  
\be\lb{cd}
W_\L(x;m)=\sum_{y\in \ZZZ^2}\bar W(x+L^Ry;m)\;.
\ee
By finite support of  $C_s$, 
$\bar \G_j(x;m)=0$ if $|x|>{L^{j+1}/2}$; 
therefore by simple arguments 
\pref{fdc}, through \pref{cd}, implies \pref{dec} for 
\be
\G_j(x;m)=\bar\G_j(x;m)\;,\quad
\G_{\ge R}(x;m)={1\over |\L|}\sum_{p\in\L^*}
{e^{ipx}\over
  m^2-\hat\D(p)} F_{MR}(L^Rp;m)\;.
\ee
\cite{[BGM]} provides also some properties of  
$F_h(p;m)$'s that we shall need below. 
\begin{lemma} Given an integer $h\ge 0$, 
for $n=1,2,\ldots,h$, there are complex functions 
$\hA_{h,n}(p,m)$, defined and differentiable for
$p\in Q_h$ and $m$ in a neighborhood of $0$,
such that 
$$
F_0(p;m)=1\;,\qquad
F_h(p;m)=\prod_{n=0}^{h-1}|\hA_{h,n}(p,m)|^2\;.
$$
Besides,  there exists a constant
$c\=c(\g)$  such that 
\be\lb{bgm1}
|\hA_{h,n}(p;m)|\le 1\;,
\ee
\be\lb{bgm2}
\hA_{h,n}(0;0)=1\;,\qquad
|1-\hA_{h,n}(p;0)|
\le c |p|\g^{-(h-n)}\;,
\ee
\be\lb{bgm3}
|\hA_{h,n}(p;m)|\le \frac{c}
{1+(\g^{-(h-n-1)}p)^4}\;,
\ee
\be\lb{bgm4}
|\hA_{h,n}(p;0)-\hA_{h+1,n+1}(p;0)|
\le 2 c \g^{-n}\;.
\ee
\end{lemma}
W.r.t. \cite{[BGM]}, here we abridged the
notation $\hA^{m^2}_{\e_h,h-n}(R_{h-n})(p)$ into
$\hA_{h,n}(p;\g^{-h}m)$. Therefore the above 
estimates \pref{bgm1}, \pref{bgm2}, \pref{bgm3} \pref{bgm4} correspond
to formula (3.12),
formula (6.17), Lemma 5.4 and Lemma 6.7  of \cite{[BGM]},
respectively; 
to optimize the constants, 
read off that paper the case $n=1$ only; the other values of $n$ 
can be obtained by scaling (3.26)
of \cite{[BGM]}.

A direct consequence of \pref{bgm1}, \pref{bgm2} and \pref{bgm3} is that 
\be\lb{cutb}
0\le 1-F_h(p;0)= c |p|\;,\qquad |F_h(p;m)|\le \frac{c}{1+p^{4}}\;.
\ee
Now we can complete the proof of \pref{fif}.
Define  $dP_{R-1,0}(\f;m)$ 
as the  Gaussian measure corresponding to the
covariance
$\G_{R-1,0}(x;m)=\G_{R-1}(x;m)+\G_{R-2}(x;m)+\cdots+\G_{0}(x;m)$, 
and neglect the $m$ dependence in  measures when $m=0$, i.e.
$$
dP_{R-1,0}(\f):=dP_{R-1,0}(\f;0)\;, \qquad
dP_{j}(\f):=dP_{j}(\f;0)\;. 
$$
\pref{fif} is a consequence of the following result.
\begin{lemma} 
\be\lb{nuo}
\lim_{m\to 0}
\int\!dP_{R,0}(\f;m)\; e^{\VV(\f)}
=
\lim_{m\to 0}
\int\!dP_{\ge R}(\z;m)
\int\!dP_{R-1,0}(\f)\; e^{\VV(\z+\f)}\;.
\ee
\end{lemma}
\bpr
It is an explicit check. For masses $m$ and $m'$, define the Yukawa
potential 
\bal
&T_\L(x;m,m'):=\frac{1}{|\L|}\sum_{p\in \L^*}
e^{ipx}\frac{(1-s)\lft[\hGG_{R-1,0}(p;m')+\hGG_{\ge R}(p;m)\rgt]}
{1+s\lft[\hGG_{R-1,0}(p;m')+\hGG_{\ge R}(p;m)\rgt] \hD(p)}\;.
\eal
In particular, notice that 
$W_\L(x;\frac{m}{\sqrt{1-s}})=T_\L(x;m,m)$.
Consider $T_\L(x;m,0)$ instead: it is well-defined and
positive-definite; $T_\L(0;m,0)$ is positive 
divergent for $m\to 0$; and
$T_\L(x|0):=\lim_{m\to 0}[T_\L(x;m,0)-T_\L(0;m,0)]=W_\L(x|0)$. 
This means that there is no change in \pref{gf2}
if we replace $W_\L(x;\frac{m}{\sqrt{1-s}})$ with $T_\L(x;m,0)$. 
The r.h.s. member of \pref{nuo} stems from repeating 
all the derivations of Section \ref{a1} for the latter potential. \qedd
\epr

\subsection{Explicit Computations}
In order to prove \pref{p3} and \pref{r0} we need explicit
computations that involve the covariances. For  $p\in \RRR^2$, 
consider the differentiable
function $u(p)=\lim_{h\to \io}F_h(p;0)$; and define the covariance
\be\lb{rmk}
\tilde C(x):=\int\!{d^2p\over (2\p)^2}\;
e^{ixp}{u(p)-u(\g p)\over p^2}\;.
\ee
By construction, 
$u(0)=1$ and, uniformly in
$\th$, 
$\lim_{\r\to \io}u(\r\cos\th,\r\sin\th)=0$
\begin{lemma} For any $j=0,1,\ldots,R-1$ and $s\in I_{j,j+1}$, if
  $h=sM$, 
\be\lb{as00}
|C_s(x;0)-\tilde C(\g^{-h}x)|\le c \g^{-\frac {h}{4}}\;,
\qquad
|\dpr^\m C_s(x;0)-\g^{-h}\tilde C^{,\m}(\g^{-h}x)|
\le c \g^{-\frac {5h}{4}}\;,
\ee
where the upper label $,\m$ indicates the continuous derivative (as opposed
to $\dpr^\m$ that is the lattice one).
\end{lemma}
As consequence, $\tilde C(x)=\lim_{h\to \io} C_s(\g^h x;0)$, and
$\tilde C(x)=0$ if $|x|\ge \g/2$.

\bpr We discuss the details for the former inequality only; 
for the latter the argument is similar. 
Consider the formula for the Fourier transform of the covariance
$C_h$ given in \pref{A10} for $m=0$: 
the idea is to replace the $h$-dependent 
cutoff function $F_h(p;0)$ with $u(p)$; 
and to replace $-\hat\D(p)$ with
$p^2$ and the interval of integration $Q$ with $\RRR^2$.
We have to prove that the  errors so generated are   $O(\g^{-\th h})$ 
for some $\th>0$.
If $p\in Q$, $-\hat\D(p)\ge (4p^2/\p^2)$ and
$|p^2+\hat\D(p)|\le (\sqrt 2/3)|p|^3$; then, by the second of
\pref{cutb}, 
\be\lb{err}
\lft|{1\over -\hat\D(p)}-{1\over p^2}\rgt|
F_h(\g^hp;0)
\le \frac{C}{ |p|}\frac{1} {1+|\g^{h}p|^4}\;.
\ee
The  replacement of  $\hD(p)$ with $p^2$ in $C_s$ gives an 
error  not larger than the integral of \pref{err} over $\RRR^2$, which
is  O($\g^{-h}$). Add and subtract $u(\g^hp)$ and $u(\g^{h+1}p)$ in place 
of $F_h(\g^hp;0)$ and $F_{h+1}(\g^{h+1}p;0)$, respectively, to get
\bal\lb{as0}
C_s(x)&=\int_Q\!{d^2p\over (2\p)^2}\;
e^{ixp}{u(\g^hp)-u(\g^{h+1}p)\over p^2}+
\int_Q{d^2p\over (2\p)^2}\;
e^{ixp}{F_{h}(\g^hp;0)-u(\g^hp)\over p^2}
\notag\\
&-
\int_Q{d^2p\over (2\p)^2}\;
e^{ixp}{F_{h+1}(\g^{h+1}p;0)-u(\g^{h+1}p)\over p^2}+ {\rm O}(\g^{-h})\;.
\eal
Now assume that (proof is below)  
\be\lb{as}
\int_{Q_h} {d^2p\over (2\p)^2}\;
{|F_{h}(p;0)-u(p)|\over p^2}\le c(\g) \g^{-\frac{h}{4}}\;;
\ee
\pref{as0} and \pref{as} directly give \pref{as00}. 
We are only left with proving assumption \pref{as}, that is 
a consequence of 
\be\lb{asb}
\int_{Q_{h'}} {d^2p\over (2\p)^2}\;
{|F_{h}(p;0)-F_{h+1}(p;0)|\over p^2}\le c(\g) \g^{-\frac{h}{4}}\;,
\ee
for any $h>h'>0$. Define 
\bal
&O_h(p):=
\prod_{n=1}^{h}|\hA_{h,n}(p;0)|^2
-\prod_{n=0}^{h}|\hA_{h+1,n+1}(p;0)|^2 
\cr
&=\sum_{m=1}^h
\lft[|\hA_{h,m}(p;0)|^2-|\hA_{h+1,m+1}(p;0)|^2\rgt]
\prod_{n=1}^{m-1}|\hA_{h,n}(p;0)|^2 
\prod_{n=m+1}^h|\hA_{h+1,n+1}(p;0)|^2 
\cr
&+
 \lft[1-|\hA_{h+1,1}(p;0)|^2\rgt]
\prod_{n=1}^h|\hA_{h+1,n+1}(p;0)|^2 
:=\sum_{m=0}^{h} O_{h,m}(p) \;;
\eal
then, using \pref{bgm1}, \pref{bgm2}, \pref{bgm3} and \pref{bgm4}, for $0\le m\le h$,
$$
|O_{h,m}(p)|\le  C(\g)
\frac{\min\{\g^{-m},|p|\g^{-(h-m)}\}}{1+p^4}\;;
$$
therefore 
$$
\sum_{m=1}^{h+1}\int \frac{d^2p}{(2\p)^2}\;
\frac{|O_{h,m}(p)|}{p^2}\le
C(\g) \g^{-\frac{h}{2}} h 
$$ 
(when $m\le h/2$, use the bound $|p|\g^{-(h-m)}$; when  $m> h/2$, use
the bound  $|p|\g^{-(h-m)}$ for the integration over $|p|\le \g^{-h/2}$
and the bound $\g^{-m}$ for the integration over $\g^{-h/2}\le
|p|\le1$ and $|p|\ge 1$).
This proves \pref{asb}. \qedd
\epr
\subsubsection{Proof of \pref{p3}}
Integrating in polar coordinates, only using that 
$u(0)=1$ and $
\lim_{\r\to \io}u(\r\cos\th, \r\sin\th)=0$, 
\be\lb{p3b}
\tilde C(0)= \int\!{d^2p\over (2\p)^2}\;
{u(p)-u(\g p)\over p^2}=\frac{1}{2\p}\ln \g\;;
\ee
\pref{p3} follows.
\subsubsection{Proof of Lemma \ref{t3.0}}\lb{aec}
A preliminary result is that, for large $|x|$, 
\bal\lb{smpt}
\tilde \G_{\io, 0}(x|0):=\int\!\frac{d^2p}{(2\p)^2}
\;(e^{ipx}-1)\frac{u(p)}{p^2}=-{1\over 2\p} \ln|x|+c+o(1)\;.
\eal
This formula is an easy consequence of the decomposition of the 
integral into three parts: for $I_1=\{|p|\le |x|^{-1}\}$ and
$I_2=\{|p|> |x|^{-1}\}$, the above integral is equal to    
$$
\int_{I_1}\!\frac{d^2p}{(2\p)^2}
\;(e^{ipx}-1)\frac{u(p)}{p^2}+
\int_{I_2}\!\frac{d^2p}{(2\p)^2}
\;e^{ipx}\frac{u(p)}{p^2}-
\int_{I_2}\!\frac{d^2p}{(2\p)^2}
\;\frac{u(p)}{p^2}\;;
$$
the only term that is divergent for large $|x|$ is the third, and its
leading term  can be
explicitly computed. Now consider  the coefficient $a_j$.
Let $\tilde w_{b,j}$ and $\tilde \G_j(0|y)$ be the same function defined
in Sec.\ref{s4.2}, but with $\tilde C(\g^{-h}x)$ in place of $C_s(x)$. 
It is not difficult to see, also using \pref{as00},  
that an equivalent formula for the
coefficient $a_j$ is, up to  $O(L^{-\frac{j}{4}})$ errors, 
\bal\lb{aup}
&{\a^2\over 2} \int\!d^2y\; y^2
\lft[
\tilde w_{b,j}(y)\lft(e^{-\a^2\tilde \G_j(0|y)}-1\rgt)
+
e^{-\a^2\tilde \G_j(0)}\lft(e^{\a^2\tilde \G_j(y)}-1\rgt)L^{-4j}\rgt]
\notag\\
&={\a^2\over 2} \int\!d^2y\; y^2
\lft[\sum_{n=0}^j R^{(j)}_{n}(y)
-\sum_{n=0}^{j-1} R^{(j-1)}_{n}(y)\rgt]
\eal
for 
$$
R^{(j-1)}_{n}(y):=e^{-\a^2\tilde \G_{j-1,n+1}(0|y)}
e^{-\a^2\tilde \G_n(0)}\lft(e^{\a^2\tilde \G_n(y)}-1\rgt)L^{-4n}\;.
$$
From now on we only consider the case $\a^2=8\p$.
Since $R^{(j-1)}_{n}(y)=L^4 R^{(j)}_{n+1}(yL)$, \pref{aup} becomes
\bal
{\a^2\over 2} \int\!d^2y\; y^2
R^{(j)}_{0}(y)
&={\a^2\over 2} \int\!d^2y\; y^2
e^{-\a^2\tilde \G_{\io,1}(0|y)}
e^{-\a^2\tilde \G_{0}(0)}
\lft(e^{-\a^2\tilde \G_{0}(y)}-1\rgt)
+O(L^{-j})
\notag\\
&=\frac{\a^2}{2} \int\!\frac{d^2y}{y^2}\; \lft[w(y)-w(y L^{-1})\rgt]
+O(L^{-j})
\eal
for 
$w(y)=y^4e^{-\a^2\tilde \G_{\io,0}(0|y)}$ and the new error
$O(L^{-j})$ in the first line 
is due to the replacement of $\tilde \G_{j,1}(0|y)$ with 
$\tilde \G_{\io,1}(0|y)$. The last integral can be computed
in polar coordinates only using the fact that  $w(0)=0$ and, by
\pref{smpt}, 
$\lim_{y\to\io}w(y)=e^c$. This proves the first of \pref{r0}.

Now consider the coefficient $b_j$ (still $\a^2=8\p$). With the same
argument used for $a_j$, an equivalent formula for $b_j$ is, up to
 an error term 
$O(L^{-\frac {j}{4}})$,
\bal\lb{aba}
\frac{\a^2}{2}\sum_{\m=\hat e}\int\! d^2y\;
\lft[(\tilde \G_{j,0}^{,\m})^2(y)-(\tilde
  \G_{j-1,0}^{,\m})^2(y)\rgt]\;.
\eal
As $\tilde \G_{j-1,0}^{,\m}(y)=L \tilde \G_{j,1}^{,\m}(yL)$,
\pref{aba} becomes, up another error term  $O(L^{-j})$,
\bal
&\frac{\a^2}{2}\sum_{\m=\hat e}
\int\! d^2y\;
\lft[2\tilde \G_{\io,1}^{,\m}(y)\tilde \G_{0}^{,\m}(y)
+(\tilde\G_{0}^{,\m})^2(y)\rgt]
\cr
&=\frac{\a^2}{2}\sum_{\m=\hat e}
\int\! d^2y\;
\lft[(\tilde \G_{\io,0}^{,\m})^2(y)-(\tilde
  \G_{\io,1}^{,\m})^2(y)\rgt] 
\cr
&=
\frac{\a^2}{2}\int\!\frac{d^2p}{(2\p)^2}
\frac{[u(\r)]^2-[u(L\r)]^2}{p^2}\;.
\eal
This
proves the second of \pref{r0}.
\section{Charged Components}\lb{abCG}
The following constructions was introduced in  \cite{[DH]}.
By iteration on scale $j$, the polymer activities are showed to
satisfy, for any
$m\in \ZZZ$, 
\be\lb{gt}
K_j(\f+\frac{2m\p}{\a},X)=K_j(\f,X)\;.
\ee
Accordingly, given $X\in \PP_j^c$ and a field $\{\f_x\}_{x\in X}$, 
the function of real variable  $F(t):=K_j(\f+t,X)$ is smooth and 
periodic of period $2\p/a$: expand $F(t)$ in 
(absolutely convergent) Fourier series and, at $t=0$, 
obtain \pref{dec3} with charged components  
\be
\hK_j(q,\f,X):={\a\over 2\p}\int_{0}^{2\p\over\a}ds\ 
K_j(\f+t,X) e^{-iq\a s }\;.
\ee 
Besides, by properties of the norms and 
since $G_j(\f,X)$ only depends 
upon the derivatives of $\f$,  
\be\lb{cin}
\|\hK_j(q,\f,X)\|_{h,T_j(\f,X)}
\le \|K_j(X)\|_{h,T_j(X)}  G_j(\f,X)\;.
\ee
\section{Lowest Orders Computation}\lb{sc}
\subsection{First Order}\lb{sc1} 
Let $B\in \BB_j$. An explicit computation of 
$\EEE_j\lft[V_j(\f,B)\rgt]$ yields
\bal\lb{pt1}
&\frac{s_j}{2}\sum_{x\in B\atop \m\in\hat e}(\dpr^\m \f')_x^2
-|B|
\frac{s_j}{2}\sum_{\m\in\hat e}(\dpr^\m \dpr^\m\G_j)(0)
+
z_j L^{-2j} e^{-\frac{\a^2}{2}\G_j(0)}
\sum_{x\in B\atop \s=\pm} e^{i\s\a\f'_x} 
\cr\cr
&\qquad= 
-\frac{s_j}{2}|B| \sum_{\m\in \hat e}(\dpr^\m\dpr^\m\G_j)(0)+
V_{j+1}(s_j, L^2e^{-\frac{\a^2}{2}\G_j(0)}z_j,\f',B)\;. 
\eal
\subsection{Second Order}\lb{sc2}  
Notice the following ``cancellations'':
$$
\sum_y(\dpr^\m\dpr^\n\G_j(y))=
-k^\m k^\n\hat\G_j(k)\Big|_{k=0}=0\;;
$$
$$
\sum_y e^{-\a^2 \G_j(0)}\lft(e^{\a^2 \G_j(y)}-1\rgt)y^\m y^\n=\d^{\m\n}
\sum_y e^{-\a^2 \G_j(0)}\lft(e^{\a^2 \G_j(y)}-1\rgt){|y|^2\over 2}\;.
$$

Let  $D=\bar B$, and $B\in \BB_j$. An explicit computation yields: 
\bal\lb{dF}
{1\over 2}E^T_j\lft[V_j(\f,B);V_j(\f,D^*)\rgt]
=&s^2_jF_{a,j+1}(\f',B)+z^2_jF_{b,j+1}(\f',B)
\cr\cr
&+z^2_jF_{c,j+1}(\f',B)+z_js_jF_{d,j+1}(\f',B)
\eal
where we have defined:
\bal
&F_{a,j+1}(\f',B)
:=
-\frac{1}{2}\sum_{y\in \ZZZ^2}\sum_{\m\in\hat e\atop\n\in\hat e}
(\dpr^\m\dpr^\n \G_j)(y)
\sum_{x\in B} (\dpr^\m\f'_x)\big[(\dpr^\n\f'_{x+y})
-(\dpr^\n\f'_{x})\big]
\notag\\
&\qquad\qquad
+|B|
{1\over 4}\sum_{y\in \ZZZ^2}\sum_{\m\in\hat e\atop\n\in\hat e}
(\dpr^\m\dpr^\n\G_{j})(y)(\dpr^\m\dpr^\n\G_{j})(y|0)
\eal
\bal
&F_{b,j+1}(\f',B)
:=
\sum_{y\in \ZZZ^2}\frac{1}{2}
e^{-{\a^2}\G_j(0)}\lft( e^{{\a^2}\G_j(y)}-1\rgt)L^{-4j}
\cdot\notag\\
&\qquad\qquad\qquad\qquad\cdot
\sum_{x\in B\atop\s=\pm}
\Bigg[e^{i\s\a(\f'_x- \f'_{x+y})}-1 +\frac{\a^2}{2}
\sum_{\m\in\hat e\atop\n\in\hat e}
(\dpr^\m\f'_x)(\dpr^\n\f'_x) y^\m y^\n\Bigg]
\notag\\
&\qquad+\sum_{y}
e^{-\a^2\G_j(0)}\lft(e^{\a^2\G_j(y)}-1\rgt)L^{-4j}|B|
\notag\\
&\qquad
-\sum_{y\in \ZZZ^2}
\frac{\a^2}{2}
e^{-{\a^2}\G_j(0)}|y|^2\lft( e^{{\a^2}\G_j(y)}-1\rgt)L^{-4j}
\frac{1}{2}\sum_{x\in B}
\sum_{\m\in\hat e}
(\dpr^\m\f'_x)^2
\eal
\bal
&F_{c,j+1}(\f',B)
:=
\frac{L^{-4j}}{2}\sum_{y\in \ZZZ^2}
e^{-{\a^2}\G_j(0)}\lft(
e^{-{\a^2}\G_j(y)}-1\rgt)
\sum_{x\in B}\sum_{\e}e^{i\s\a(\f'_x+\f'_{x+y})}
\eal
\bal
&F_{d,j+1}(\f',B)
:=
\frac{L^{-2j}}{2}\sum_{y\in \ZZZ^2\atop \n\in\hat e}
e^{-\frac{\a^2}{2}\G_{j}(0)}
(\dpr^\n \G_j)(y)
\cdot
\notag\\
&\qquad\qquad\qquad\qquad\qquad\qquad\cdot
\sum_{x\in B\atop \s=\pm}i\s\a
\lft[e^{i\s\a\f'_x} (\dpr^\n\f'_{x+y})
-e^{i\s\a\f'_{x+y}} (\dpr^\n\f'_x)\rgt]
\notag\\
&\qquad
-\frac{L^{-2j}}{4}\sum_{y\in \ZZZ^2\atop \n\in\hat e}
e^{-\frac{\a^2}{2}\G_{j}(0)}
(\dpr^\n \G_j)^2(y)\a^2
\sum_{x\in B\atop \s=\pm}
\lft(e^{i\s\a\f'_{x+y}}-e^{i\s\a\f'_x} \rgt) 
\notag\\
&\qquad
-\frac{L^{-2j}}{2}\sum_{y\in \ZZZ^2\atop \n\in\hat e}
e^{-\frac{\a^2}{2}\G_{j}(0)}
(\dpr^\n \G_j)^2(y)\a^2
\sum_{x\in B\atop \s=\pm}
e^{i\s\a\f'_x}\;.
\eal
Accordingly,
by the definitions of $W_{m,j}$ in \pref{dfW}, 
\bal
&W_{a,j+1}(\f',B)+L^{-2j}\mce_{3,j}|B|
=F_{a,j+1}(\f',B)
+\EEE_j[W_{a,j}(\f',B)] 
\notag\\\cr
&W_{b,j+1}(\f',B)+L^{-2j} \mce_{4,j}
|B|+V_{j+1}(-a_j,0,\f',B)
=F_{b,j+1}(\f',B)+\EEE_j[W_{b,j}(\f,B)]
\notag\\\cr
&W_{c,j+1}(\f',B)=F_{c,j+1}(\f',B)+\EEE_j[W_{c,j}(\f,B)]
\notag\\\cr
&W_{d,j+1}(\f',B)+V_{j+1}(0,-L^2e^{-\frac{\a^2}{2}\G_j(0)}b_j,\f',B)
=F_{d,j+1}(\f',B)+\EEE_j[W_{d,j}(\f,B)]
\cr\eal
where $\mce_{3,j}$ and $\mce_{4,j}$ are defined in \pref{dfm2}.
\subsection{Estimates for $W_j$}\lb{sC.3}
Let $B\in \BB_{j}$.  If $|y|\le
L^{n+1}/2$, uniformly in $j$
$$
|\G_{j-1,n+1}(0)-\G_{j-1,n+1}(y)|\le C\;,
$$
$$
|\G_{j-1,n+1}(0)+\G_{j-1,n+1}(y)-(j-n-1){1\over \p}\ln L|\le 
C\;.
$$
Accordingly,
\bal\lb{c13}
L^{-j} \sum_{y\in \ZZZ^2}|w^{\m\n}_{a,j}(y)|\; |y|\le C(L)\;,
\qquad
L^{-j} \sum_{y\in \ZZZ^2}|w^{\m\n}_{b,j}(y)|\; |y|^3\le C(L)\;,
\eal
\bal\lb{c14}
L^{2j}\sum_{y\in \ZZZ^2}|w_{c,j}(y)|\le C(L)\;,
\qquad
L^{j} \sum_{y\in \ZZZ^2}|w^{\m}_{d,j}(y)|\;\le C(L)\;,
\cr
L^{j} \sum_{y\in \ZZZ^2}|w_{e,j}(y)|\; |y|\le C(L)\;.
\qquad\qquad\qquad
\eal
The above inequalities yield
\bal\lb{dWa}
&\|W_{a,j}(s,z,\f,B)\|_{h,T_j(\f,B)}
\notag\\
&\qquad\qquad
\le
s^2\sum_{\m,\n,\r\in\hat e\atop}
\sum_{y\in \ZZZ^2} |w_{a,j}^{\m\n}(y)|\; |y^\r|
|B|
\sup_{x,z\in B}\|(\dpr^\m\f_x)(\dpr^\r \dpr^\n\f_{z})\|_{h,T_j(\f,B)}
\notag\\
&\qquad
\le
s^2 C(L) \lft(1+\max_{n=1,2}\|\nabla_j^n\f\|^2_{L^\io(B^*)}\rgt)
\eal
\bal
&\|W_{b,j}(s,z,\f,B)\|_{h,T_j(\f,B)}
\le z^2 C(\a,L)\sum_{y\in \ZZZ^2}|w_{b,j}(y)|\;|y|^3
\;\cdot
\notag\\
&\quad\qquad\qquad\qquad\cdot
|B|\sum_{\m,\n,\r=\hat e}
\sup_{x,z\in B}\Bigg[\|(\dpr^{\m}\f_z)^3
+(\dpr^\r\dpr^{\m}\f_x)(\dpr^{\n}\f_z)\|_{h,T_j(\f,B)}\Bigg]
\notag\\
&\qquad\qquad\qquad
\le z^2 C(\a,L)
\lft(1+\max_{n=1,2}\|\nabla_j^n\f\|^2_{L^\io(B^*)}\rgt)
\eal
\bal
&\|W_{c,j}(s,z,\f,B)\|_{h,T_j(\f,B)}
\le z^2C(\a)\sum_{y\in \ZZZ^2} 
|w_{c,j}(y)|\; |B|
\le z^2 C(L,\a)
\\
\cr
&\|W_{d,j}(s,z,\f,B)\|_{h,T_j(\f,B)}
\le |zs| C(\a)
\sum_{\m}\sum_{y\in \ZZZ^2} 
\lft[|w^\m_{d,j}(y)|+|w_{e,j}(y)||y|\rgt]
\cdot\cr
&\qquad\cdot
|B|
\sup_{z\in B^*}\|(\dpr^\m\f_{z})\|_{h,T_j(\f,B)}
\cr
&\qquad\le 
|zs|C(\a,L)\lft(1+\max_{n=1,2}\|\nabla_j^n\f\|^2_{L^\io(B^*)}\rgt)
\eal
All together these bounds give \pref{bww}.
\section{Proof of Lemma \ref{l6.53}.}\lb{d}
The goal in to implement the idea of finer decomposition of the
covariances, \cite{[Br2]}, in order to have control over the different
field regulators needed for the Coulomb Gas.
By \pref{fdc}, decompose the covariances
and   the field
\be\lb{fx}
\G_j(x)=\sum_{s\in I_{j,j+1}} C_s(x)\;,
\qquad
\z^{(j)}_x=\sum_{s\in I_{j,j+1}} \x^{(s)}_x\;,
\ee
so that $\x^{(s)}_x$ is a Gaussian random field with covariance 
$E_s[\x^{(s)}_x\x^{(s)}_y]=C_s(x-y)$, with range
$L^{s+\log_L\g}/2=L^j\g^{m+1}/2$ and typical momentum
$L^{-s}=L^{-j}\g^{-m}$ .
 The notation for the fields is now
\be\lb{ff}
\f^{(s)}_x =\f^{(s')}_x + \x^{(s)}_x  
\ee
for $s'=s+\log_L\g$. 
It is clear how to extend previous definitions to have $s$-blocks,
$s$-polymers, $L_s^2$ norms and $G_s$ field regulators. We use 
$\overline X_s$ to indicate the
closure of the set $X$ on scale $s$ (hence, 
for any $j-$polymer,  $X$ $\overline X_j\= X$ and 
$\overline X_{j+1}\=\overline X$).  
Given $c_4>0$, define (\cfr \pref{6.69})
\be\lb{gg}
\ln g_s(\x,\overline X_s)
=c_4 \kappa_L\sum_{n=0,1,2} W_s(\nabla^n_s\x,\overline X_s)^2\;.
\ee
%
\begin{lemma}\lb{l.int}
There exists a choice of the constants $c_1$, $c_3$, $c_4$ and $\g$
such that:
\bd
\item{1)} for any $X\in \PP_s^c$, 
\be\lb{ex6.54}
G_s(\f^{(s)},\overline X_s)
\le G_{s'}(\f^{(s')}, \overline{X}_{s'}) g_s(\x^{(s)},\overline X_{s})\;;
\ee
\item{2)} for any $X\in \SS_s$, 
\be\lb{ex6.128}
G_s(\f^{(s)},\overline X)
\le G_{s'}(\f^{(s')},\overline X_{s'})^{2/3} g_s(\x^{(s)},\overline X_s)\;.
\ee
\ed
\end{lemma}
\bpr
Only in this proof, let us shorten the notations
$\f^{(s)}$, $\f^{(s')}$, $\x^{(s)}$, $\overline X_s$ 
and $\overline X_{s'}$
into $\f$, $\f'$, $\x$, $X$ and $\overline X$,  respectively.
Since $\f=\f'+\x$, for any $0<\a<1$, 
by discreet  partial integration (see (6.117) of \cite{[Br]})  
\bea\lb{6.117}
\|\nabla_s \f\|^2_{L^2_s(X)}
&\le&
\|\nabla_s \f'\|^2_{L^2_s(X)}
+
\a\|\nabla_s \f'\|^2_{L^2_s(\dpr X)}
+
\a\|\nabla^2_s \f'\|^2_{L^2_s(X)}
\cr\cr
&&
+\|\nabla_s\x\|^2_{L^2_s(X)}
+
\a^{-1}\|\x\|^2_{L^2_s(\dpr X)}
+
\a^{-1}\|\x\|^2_{L^2_s(X)}\;;
\eea
whereas, by squaring  triangular inequalities
\bea
\lb{6.112}
\|\nabla_s\f\|^2_{L^2_s(\dpr X)}
&\le& 
2\|\nabla_s\f'\|^2_{L^2_{s}(\dpr X)}
+
2\|\nabla_s\x\|^2_{L^2_{s}(\dpr X)}\;,
\\ \cr
\lb{6.115}
W_s(\nabla^2_s\f,X)^2&\le& 
2W_s(\nabla^2_s\f',X)^2
+2W_s(\nabla^2_s\x,X)^2\;.
\eea
Collecting the last three formulas, by (6.110) of \cite{[Br]}, 
\bea\lb{6.125}
\ln G_s(\f,X) 
&\le&
c_1\kappa_L\|\nabla_s \f'\|^2_{L^2_s(X)}+
f_3\kappa_L\|\nabla_s \f'\|^2_{L^2_s(\dpr X)}+
f_1\kappa_LW_s(\nabla^2_s\f',X)^2 
\cr\cr
&&+ f_4 \kappa_L \sum_{n=0,1,2} W_s(\nabla^n_s\x,X)^2
\eea
for $f_3=\a c_1+2c_3$, $f_1=(4\a+2)c_1<6c_1$
and $f_4=4c_1\a^{-1}+ 8c_3+4c_1$.
In order to pass to scale $s'$, use the inequality
\be\lb{6.105}
\|\nabla_s \f'\|^2_{L^2_s(\dpr X)}
\le
\|\nabla_s \f'\|^2_{L^2_s(\dpr \overline X)}
+c\|\nabla_s \f'\|^2_{L^2_s(\overline X\bs X)}
+cW_s(\nabla^2_s\f',\overline X\bs X)^2  
\ee
for a certain constant $c>0$
(for the proof, \cfr the proof of 
(6.105) of \cite{[Br]}).
If $\a$ and $c_3$ are so small that 
$cf_3\le c_1$, 
%
\bea
\ln G_s(\f,X)
&\le&
c_1\kappa_L\|\nabla_s \f'\|^2_{L^2_s(\bar X)}
+f_3\kappa_L\|\nabla_s \f'\|^2_{L^2_s(\dpr \overline X)}
+6c_1\kappa_LW_s(\nabla^2_s\f',\overline X)^2
\cr\cr
&&
+ 
f_4\kappa_L \sum_{n=0,1,2} W_s(\nabla^n_s\x,X)^2
\eea
Finally, to restore the old constant $c_3$ and $c_1$ in the 2nd and
3rd terms, observe that these two terms are irrelevant,  i.e. 
passing to scale $s'$ gives $\g^{-1}$'s factors:
\bea
&&\|\nabla_s \f'\|^2_{L^2_s(\dpr \overline X)}\le
\g^{-1} \|\nabla_{s'} \f'\|^2_{L^2_{s'}(\dpr \overline X)}
\cr\cr
&&
W_s(\nabla^2_s\f',\overline X)^2\le
\g^{-2}W_{s'}(\nabla^2_{s'}\f',\overline X)^2\;;
\eea 
the first term is, instead, scale invariant (and the fourth 
is relevant but there is no need to scale it).
Therefore, for $c_4\ge f_4$, 
if $\g$ is so large that $f_3\g^{-1}<c_3$ and $6\g^{-2}<1$ 
\pref{ex6.54} is proven.

Then consider the case when $X\in \SS_s$.
If $\g>8$, it is possible to find the polymer $TX$, a translation of $X$,
such that $TX$ is disjoint from $X$, but $(TX)^*\cap X^*\neq \emptyset$
and $TX\subset \overline X$ ; by (6.129) of \cite{[Br]},
setting $X_T=TX\cup X$ and for a  $c>0$,
\be\lb{6.129}
{3\over 2}\|\nabla_s \f'\|^2_{L^2_s(X)}
\le \|\nabla_s \f'\|^2_{L^2_s(X_T)}
+c W_s(\nabla ^2\f';X_T)^2\;;
\ee
besides, as  $X$ and $TX$ have disjoint boundaries,
\be\lb{6.129bis}
\|\nabla_s \f'\|^2_{L^2_s(\dpr X)}\le 
\|\nabla_s \f'\|^2_{L^2_s(\dpr X_T)}\;,\qquad
W_s(\nabla ^2_s\f',X)^2\le
W_s(\nabla ^2_s\f',X_T)^2\;.
\ee
Multiply both sides of \pref{6.125} times $3/2$; then use
\pref{6.129} and \pref{6.129bis} to obtain
\bea
{3 \over 2}\ln G_s(\f,X)
&\le&
c_1\kappa_L\|\nabla_s \f'\|^2_{L^2_s(X_T)}
+\frac{3 f_3}{2}\kappa_L\|\nabla_s \f'\|^2_{L^2_s(\dpr X_T)}
\cr\cr
&&
+t_1 \kappa_LW_s(\nabla^2_s\f',X_T)^2
+c_4\sum_{n=0,1,2} W_s(\nabla^n_s\x,X)^2
\eea
for   $t_1= 3f_1/2+cc_1$ and $c_4=3 f_4/2$.
This inequality is of the same form as \pref{6.125}, except for the
fact that $X$ has been replaced by $X_T$ in the first three terms of
the r.h.s,  and the constants are different. Therefore one can proceed
as done from \pref{6.125} to obtain \pref{ex6.54}: since by construction
$\overline X_T=\overline X$, \pref{ex6.128} follows 
for $\a$ and $c_3$ small enough, and $\g$ large enough.\qedd
\epr
To conclude the proof of Lemma \ref{l6.53}, 
we need two results of \cite{[Br]}.
\begin{lemma}
If $\kappa_L$ is small enough (w.r.t $\g^{-2}$), for any $X\in \PP_s$
\be\lb{6.53quater}
\EEE_s\lft[g_s(\x^{(s)},X)\rgt]\le e^{c \kappa_L |X|_s}\;,
\ee
where $|X|_s$ is the number of $s-$blocks in $X$. 
\end{lemma}
\begin{lemma}
For $X\in \PP_j$
\be\lb{6.127}
\lft(1+\max_{n=1,2}\|\nabla^n_{j+1}\f^{(j+1)}\|_{L^\io(X^*)}\rgt)^3
\le
{c\over \kappa_L^{3/2}} \lft[G_{j+1}(\f^{(j+1)},\bar X)\rgt]^{1/3}\;.
\ee 
\end{lemma}
The former is Lemma 6.31 of \cite{[Br]} and its hypothesis is
fulfilled if $M$ is large enough; the latter  
corresponds to  formula 
(6.127) of \cite{[Br]}.
Use \pref{ex6.54} iteratively, then  
\pref{6.53quater}, 
to find, for any $X\in \PP_j$,
\bal\lb{aa}
\EEE_j\lft[G_j(\f^{(j)},X)\rgt]
&\le G_{j+1}(\f^{(j+1)},\bar X)
\prod_{s\in I_{j,j+1}}\EEE_s\lft[g_s(\x^{(s)},\bar X_s)\rgt]
\cr
&\le G_{j+1}(\f^{(j+1)},\bar X)
e^{c\kappa_L \sum_{s\in I_{j,j+1}}|\bar X_s|_s} \;.
\eal
This, for $\g$ large enough proves \pref{6.54} 
for $|\bar X_s|_s\le |X|_j$ and 
$c \kappa_L=C/\ln L= C(\log_\g e)/M$. In the same fashion, using  
by \pref{ex6.128} iteratively, and then \pref{6.53quater}, \pref{6.127}, 
for any $X\in\SS_j$ (so that $\bar X_s\in \SS_s$ for any $s$),
\bal
&\lft(1+\max_{n=1,2}\|\nabla^n_{j+1}\f^{(j+1)}\|_{L^\io(X^*)}\rgt)^3
\EEE_j\lft[G_j(\f^{(j)},X)\rgt]
\cr
&
\le\lft(1+\max_{n=1,2}\|\nabla^n_{j+1}\f^{(j+1)}\|_{L^\io(X^*)}\rgt)^3
 G^{2/3}_{j+1}(\f^{(j+1)},\bar X)
\prod_{s\in I_{j,j+1}}\EEE_s\lft[g_s(\x^{(s)},\bar X_s)\rgt]
\cr
&
\le 
{c\over \kappa_L^{3/2}} G_{j+1}(\f^{(j+1)},\bar X)
e^{c\kappa_L \sum_{s\in I_{j,j+1}}|\bar X_s|_s}  \;.
\eal
This proves \pref{6.58}. (When $G_{j}(\f^{(j)},X)$ is replaced by 
$\sup_{t\in[0,1]} G_{j}(t\f^{(j+1)}+\z,X)$, only little changes 
are required.) 
\section{Proof of Theorems \ref{l6.12},  \ref{l13} and \ref{l13b}}
\lb{A6}
For $F\in \NN_j(X)$ and $m$ a positive integer, define
\bal\lb{sn+}
&\|F(\f)\|_{h,T^{\ge m}_j(\f,X)}
:=\sum_{n\ge m}{h^n\over n!}\|F(\f)\|_{T^n_j(\f,X)}\;
\eal
(it corresponds to \pref{sn} when $m=0$). 
\begin{lemma}\lb{lf1}
Let   $F\in \NN_j(X)$
and $X\in \PP_j$:
\bd
\item{a)}
if $X\in \SS_j$ and $x_0\in X$, 
for $(\d\f)_x:=\f_x-\f_{x_0}$ and $\r=5L^{-1}$,  
\be\lb{f1}
\|F(\d\f)\|_{h,T_{j+1}(\f,X)}\le 
\|F(\x)\|_{\r h,T_j(\x,X)}\Big|_{\x_x=\d\f_x}\;;
\ee
\item{b)}
given $\ps\in \CC^2_j(X)$, if $\D:=\|\ps\|_{\CC^2_j(X)}$ is finite,
\be\lb{f2}
\|F(\f+\ps)\|_{h,T_{j}(\f,X)}\le \|F(\f)\|_{h+\D,T_{j}(\f,X)}\;;
\ee
\item{c)}
for $m=0,2$ and any $h>0$,  
\be\lb{f3}
\|{\rm Rem}_{m}\; F(\x)\|_{h,T_{j}(\x,X)}
\le 
(1+h^{-1}\|\x\|_{\CC^2_j(X)})^m
\sup_{t\in [0,1]} \|F(t\x)\|_{h,T^{\ge m}_{j}(t\x,X)}\;.
\ee
where the Taylor remainder is in the field $\x$.
\ed
\end{lemma}
\bpr
\pref{f1} follows from 
the identity 
$$
\sum_{x\in X^*} f_x {\dpr F(\d\f)\over \dpr\f_x}=
\lft.\sum_{x\in X^*} (\d f)_x {\dpr F\over \dpr\f_x}(\x)\rgt|_{\x=\d\f}
$$
and the fact that, for  $X$ small,   
$\|\d f\|_{\CC^2_j(X)}\le 5 L^{-1} \|f\|_{\CC^2_{j+1}(X)}$.
\pref{f2} is a direct consequence of Taylor series.
Consider \pref{f3} for $m=2$ - the case $m=0$ can be proved 
with similar arguments -  and 
set $O(\x):={\rm Rem}_{2}F(\x)$;  
then using test functions $f_1,f_2,\ldots$ such that 
$\|f_j\|_{\CC^2_{j}(X)}\le 1$, for $n=0,1,2$
\bal
&{h^n\over n!}|D^n_{\x} O(\x)\cdot (f_1,\ldots,f_n)|
\notag\\
&\qquad\le{h^n\over n!}\int_0^1dt\; {(1-t)^{2-n}\over (2-n)!} \lft|
D^3_{\x} F(t\x)\cdot (f_1,\ldots,f_n,\x,\ldots,\x)\rgt|
\notag\\
&\qquad\le
\frac{3!}{(3-n)!\ n!}\big[h^{-1}\|\x\|_{\CC^2_{j}(X)}\big]^{3-n}
\;{h^3\over 3!} \sup_{t\in [0,1]}\;\|F(t\x)\|_{h,T^3_j(t\x,X)}\;,
\eal
whereas, for $n\ge 3$, 
\bal
{h^n\over n!}|D^n_\x O(\x)\cdot (f_1,\ldots,f_n)|
\le 
{h^n\over n!}\|F(\x)\|_{h,T^n_j(\x,X)}\;. 
\eal
The result follows by first summing the terms with 
$n=0,1,2,3$ with the binomial theorem,
and then summing the terms with $n>3$. \qedd
\epr

\subsection{Proof of Theorem \ref{l13}}
Consider the measure $\EEE_j$ and an integrable function
$F(\f)$. Factorize the covariance into $\G_j= g_j\circ
g_j$ and call  $\EEE_I$ the Gaussian expectation with 
covariance $I=(\d_{i,j})$;
under the imaginary translation $\z_x\to \z_x+i(g_jf)_x$ where 
$f$ is any test function  with finite support, 
$$
\EEE_j[F(\f)]=\EEE_I[F(\f'+(g_j\z))]=
e^{{1\over 2} (f,\G_jf)}\;
\EEE_j\lft[e^{-i(\z,f)} F(\f+i (\G_j f))\rgt]\;.
$$
Now apply this identity with  $F(\f)=\hK_j(q,\f,X)$:  
calling $\ps_x:=(\G_j f)_x$ and  
$(\d\ps)_x:=(\G_j f)_x-(\G_j f)_{x_0}$, by \pref{dec4}, 
\bal
\EEE_j\lft[\hK_j(q,\f,X)\rgt]
&= e^{{1\over 2}(f, \G_j f)}\;
\EEE_j\lft[e^{-i(\z,f)} \hK_j(q,\f+i \ps,X)\rgt]
\cr\cr
&= e^{{1\over 2}(f, \G_j f)-\a q (\d_{x_0}, \G_j f)}\;
\EEE_j\lft[\hK_j(q,\f+i \d\ps,X)\rgt]
\eal
where $(\d_{x_0})_x:=\d_{x,x_0}$. 
Finally,
\bal\lb{op}
&\|\EEE_j\lft[\hK_j(q,\f,X)\rgt]\|_{h,T_{j+1}(\f',X)}
\cr
&\qquad
\le e^{{1\over 2}(f, \G_j f)-\a q (\d_{x_0}, \G_j f)}\;
\EEE_j\lft[\|\hK_j(q,\f+i \d\ps,X)\|_{h,T_{j+1}(\f',X)}\rgt]\;.
\eal
With \cite{[DH]}, choose $f_x=\a {\rm sgn}(q)\d_{x,x_0}$ 
(see the Remark below)  and obtain
\bal\lb{ct1}
&\|\EEE_j\lft[\hK_j(q,\f, X)\rgt]\|_{h,T_{j+1}(\f', X)}
\cr
&\qquad\qquad\le e^{-(|q|-{1\over 2})\a^2\G_j(0)}\;
\EEE_j\lft[\|\hK_j(q,\f+i \d\ps,X)\|_{h,T_{j+1}(\f',X)}\rgt]\;.
\eal
This inequality is an important result of \cite{[DH]}.
Now consider the expectation on the r.h.s. of \pref{ct1}:
for  $\r:=5 L^{-1}$, $\h:=\z+i\d\ps$, $\D:=\|\d\ps\|_{\CC^2_j(X)}$,
since $\|e^{iq \f_{x_0}}\|_{h,T_{j+1}(\f',X)}$ is less than
$e^{h|q|\a}$, by
\pref{dec4}, \pref{f1}, \pref{f2}, \pref{cin},
and for $L$ so large that $\r h+\D\le h$, 
\bal\lb{ct2}
&\|\hK_j(q,\f+i\d\ps,X)\|_{h,T_{j+1}(\f',X)}
\le e^{h|q|\a}
\|\hK_j(q,\d\f'+\h,X)\|_{h,T_{j+1}(\f',X)} 
\cr
&\qquad\qquad\qquad\qquad\qquad
\le e^{h|q|\a}\|\hK_j(q,\x+\h,X)\|_{\r h,T_j(\x, X)}\Big|_{\x:=\d\f'}
\cr
&\qquad\qquad\qquad\qquad\qquad
\le e^{h|q|\a}\|\hK_j(q,\x+\z,X)\|_{\r h+\D,T_j(\x,X)}\Big|_{\x:=\d\f'}
\cr
&\qquad\qquad\qquad\qquad\qquad
\le e^{h|q|\a}\|K_j(X)\|_{h,T_j(X)} G_j(\f,X)\;.
\eal
(The final inequality is a consequence of the fact that
$G_j(\f,X)$ depends on the derivatives of $\f$, and then 
$G_j(\x+\z,X)=G_j(\f,X)$.)
Finally, plugging \pref{ct2} into \pref{ct1}, 
for any $\th >0$ and  $L$ large enough
\bal\lb{ct3}
&\|\EEE_j\lft[\hK_j(q,\f,X)\rgt]\|_{h,T_{j+1}(\f',X)}
\cr
&\qquad\qquad\le e^{c(\a)|q|} L^{-(2|q|-1){\a^2\over 4\p}}
\|K_j\|_{h,T_j} \lft({A\over 2}\rgt)^{-|X|_j} G_{j+1}(\f',\bar X)\;.
\eal
This inequality proves Theorem \ref{l13}.
\brm
In order to minimize  the prefactor in  the r.h.s. of 
\pref{op}, one should rather choose 
$f_x=\a q\d_{x,x_0}$; then, the prefactor in \pref{ct1} would be  
$ e^{-{q^2\over 2}\a^2\G_j(0)}$, as in heuristic arguments. 
Though this choice conflicts with the condition  
$\r h+\D\le h$, for the corresponding $\D$ grows 
with $q$, whereas the radius of analyticity,  
$h$, is assumed independent of  $q$. 
\erm
\subsection{Proof of Theorem \ref{l6.12}}
With Taylor remainder in  $\d\f'$, by 
\pref{f1}, \pref{f3}, 
\bal
&\|{\rm Rem}_{2}\hK_j(0,\d\f'+\z,X)\|_{h,T_{j+1}(\f', X)}
\le\|{\rm Rem}_{2}\; 
\hK_j(0,\x+\z,X)\|_{\r h,T_j(\x,X)}\Big|_{\x=\d\f'}
\cr
&\qquad
\le
\lft(1+(\r h)^{-1}\|\x\|_{\CC^2_{j}(X)}\rgt)^3
\sup_{t\in[0,1]}\|\hK_j(0,t\x+\z,X)\|_{\r h,T^{\ge 3}_j(t\x,X)}\Big|_{\x=\d\f'}
\notag\\
&\qquad
\le 
C\lft(1+L\|\x\|_{\CC^2_{j}(X)}\rgt)^3
\r^3\sup_{t\in[0,1]}\|\hK_j(0;t\x+\z,X)\|_{h,T_j(t\x,X)}\Big|_{\x=\d\f'}
\notag\\
&\qquad
\le 
C'\lft(1+L\|\x\|_{\CC^2_{j}(X)}\rgt)^3
L^{-3}\|K_j(X)\|_{h,T_j(X)} \sup_{t\in [0,1]}G_j(t\f'+\z,X)\;.
\eal
(We used that, as $\r=5 L^{-1}<1$, then
$\|\cdot\|_{\r h,T^{\ge3}_j}\le \r^3\|\cdot\|_{h,T_j}$.)
As  $X\in \SS_j$, 
\be\lb{lng}
L\|\d\f'\|_{\CC^2_j(X)}
\le C \max_{p=1,2}\|\nabla^p_{j+1}\f'\|_{L^\io(X^*)}\;.
\ee
Therefore, by \pref{dec4}, \pref{6.58}
\bal
&\|{\rm Rem}_{2}\EEE_j
\lft[\hK_j(0,\f,X)\rgt]\|_{h,T_{j+1}(\f',X)}
\cr
&\qquad\qquad
\le C{L^{-3}\over \kappa_L^{3/2}}
\|K_j(X)\|_{h,T_j(X)}\lft(\frac A2\rgt)^{-|X|_j} 
G_{j+1}(\f',\bar X)\;,
\eal
which proves Theorem \ref{l6.12}.
\subsection{Proof of Theorem \ref{l13b}}
Set $\h:=\z+i\d\ps$. 
By \pref{f1}, \pref{f3}, \pref{f2}, and since, by
definition of $h$, $\frac h2+\D\le h$, 
\bal\lb{ct5}
&\|{\rm Rem}_{0}\hK_j(q,\d\f'+\h,X)\|_{h,T_{j+1}(\f',X)}
\le
\|{\rm Rem}_{0}\hK_j(q,\x+\h,X)\|_{\r h,T_j(\x,X)}
\Big|_{\x=\d\f'}
\cr
&\qquad\qquad
\le
C\lft(1+L\|\x\|_{\CC^2_j(X)}\rgt)
\sup_{0\le t\le 1}\|\hK_j(q,t\x+\h,X)\|_{\r h,T^{\ge1}_j(\x,X)}
\Big|_{\x=\d\f'}
\cr
&\qquad\qquad
\le
C'\lft(1+L\|\x\|_{\CC^2_j(X)}\rgt) L^{-1}
\sup_{0\le t\le 1}\|\hK_j(q,t\x+\h,X)\|_{\frac h2,T_j(\x,X)}
\cr
&\qquad\qquad
\le
C'\lft(1+L\|\x\|_{\CC^2_j(X)}\rgt) L^{-1}
\sup_{0\le t\le 1}\|\hK_j(q,t\x+\z, X)\|_{h,T_j(\x,X)}
\cr
&\qquad\qquad
\le
C'\lft(1+L\|\x\|_{\CC^2_j(X)}\rgt) L^{-1}
\|K_j(X)\|_{h,T_j}
\sup_{0\le t\le 1} G_j(t\f'+\z,X)\;.
\eal
In analogy with the derivation
of  \pref{ct1}, 
\bal
&\|\EEE_j\lft[{\rm Rem}_{0}\hK_j(q,\f, X)\rgt]\|_{h,T_{j+1}(\f', X)}
\cr
&\qquad\le e^{-(|q|-{1\over 2})\a^2\G_j(0)}\;
\EEE_j\lft[\|{\rm Rem}_{0}\hK_j(q,\f+i \d\ps,X)\|_{h,T_{j+1}(\f',X)}\rgt]\;.
\eal
Then, continuing the chain of inequalities by  \pref{lng} and
\pref{ct5}, \pref{6.58}, 
\bal
&
\|\EEE_j\lft[{\rm Rem}_{0}\hK_j(q,\f, X)\rgt]\|_{h,T_{j+1}(\f', X)}
\cr
&\le e^{-(|q|-{1\over 2})\a^2\G_j(0)}\;
\EEE_j\lft[\|{\rm Rem}_{0}\hK_j(q,\f+i
  \d\ps,X)\|_{h,T_{j+1}(\f',X)}\rgt]
\cr
&\le
e^{-(|q|-{1\over 2})\a^2\G_j(0)}
e^{h|q|\a}
\EEE_j\lft[\|{\rm Rem}_{0}\hK_j(q,\d\f'+\h,X)\|_{h,T_{j+1}(\f',X)}\rgt]
\cr
&\le C e^{-(|q|-{1\over 2})\a^2\G_j(0)}
e^{h|q|\a}
\frac{L^{-1}}{\kappa_L^{3/2}}  \|K_j\|_{h,T_j}
\lft(\frac{A}{2}\rgt)^{-|X|_j}G_{j+1}(\f',\bar X)\;,
\eal
which proves Theorem \ref{l13b}. 
\bac 
It is a pleasure to thank David Brydges for 
patient guidance during the study of \cite{[Br]},  
clarifying discussions on  
\cite{[DH]} and  \cite{[BS]}, 
many important suggestions - including a
crucial one labeled as reference 
\cite{[Br2]};  and, most of all, 
for constant support during the (long)
writing process. 
{\it This work is supported by the Giorgio and Elena Petronio
  Fellowship Fund.}
\eac

\end{document}